\documentclass[11pt,a4paper,english]{article}
\pdfoutput=1 
\usepackage{jheppub}
\usepackage{graphicx}
\usepackage{amssymb,latexsym}
\usepackage{amsmath}
\usepackage{epsf}
\usepackage{pdfsync}
\usepackage{epsfig}
\usepackage[parfill]{parskip}
\usepackage{color}
\usepackage[matrix,arrow,color]{xy}
\usepackage{wrapfig}
\usepackage{extarrows}

\usepackage{mathrsfs}

\ProcessKeyvalOptions{Hyp}

\input{xy}
\xyoption{all}
\UseCrayolaColors

\input{epsf}

\newcommand{\ba}{\begin{eqnarray*}}
\newcommand{\ea}{\end{eqnarray*}}
\newcommand{\ban}{\begin{eqnarray}}
\newcommand{\ean}{\end{eqnarray}}

\makeatletter
\newcommand{\xRightarrow}[2][]{\ext@arrow 0359\Rightarrowfill@{#1}{#2}}
\makeatother

\newcommand{\Tr}{{\rm Tr\,}}

\newcommand{\IQ}{\mathbb{Q}}

\newcommand{\bbE}{\mathbb{E}}

\newcommand{\cN}{{\cal N}}
\newcommand{\cM}{{\cal M}}
\newcommand{\cS}{{\cal S}}
\newcommand{\cB}{{\cal B}}

\newcommand{\cH}{{\cal H}}
\newcommand{\cA}{{\cal A}}

\newcommand{\cC}{{\cal C}}
\newcommand{\cQ}{{\cal Q}}

\newcommand{\cZ}{{\cal Z}}
\newcommand{\cL}{{\cal L}}
\newcommand{\cF}{{\cal F}}

\newcommand{\cP}{{\cal P}}

\newcommand{\cY}{{\cal Y}}

\newcommand{\sfA}{{\mathsf{A}}}
\newcommand{\sfB}{{\mathsf{B}}}
\newcommand{\sfe}{{\mathsf{e}}}
\newcommand{\sfa}{{\mathsf{a}}}
\newcommand{\sfb}{{\mathsf{b}}}

\newcommand{\sfQ}{{\mathsf{Q}}}
\newcommand{\sfD}{{\mathsf{D}}}

\newcommand{\sfR}{{\mathsf{R}}}
\newcommand{\sfX}{{\mathsf{X}}}
\newcommand{\sfY}{{\mathsf{Y}}}

\newcommand{\sfW}{{\mathsf{W}}}
\newcommand{\sfL}{{\mathsf{L}}}
\newcommand{\sfy}{{\mathsf{y}}}
\newcommand{\sfx}{{\mathsf{x}}}
\newcommand{\sfz}{{\mathsf{z}}}
\newcommand{\sfu}{{\mathsf{u}}}
\newcommand{\sfv}{{\mathsf{v}}}

\newcommand{\scrM}{{\mathscr{M}}}

\newcommand{\scrT}{{\mathscr{T}}}
\newcommand{\scrL}{{\mathscr{L}}}

\newcommand{\re}{{\rm Re \,}}
\newcommand{\im}{{\rm Im \,}}

\def \min{\rm{min}}

\newcommand{\mbf}[1]{{\boldsymbol {#1} }}
\newcommand{\complex}{{\mathbb C}} 
\newcommand{\zed}{{\mathbb Z}} 
\newcommand{\real}{{\mathbb R}} 
\newcommand{\torus}{{\mathbb T}}

\def\e{{\,\rm e}\,}

\def\ii{{\,{\rm i}\,}}
\def\dd{{\rm d}}

\newcommand{\mut}{\mathsf{mut}}

\def\beq{\begin{equation}}
\def\bee{\begin{equation}}
\def\eeq{\end{equation}}
\def\bea{\begin{eqnarray}}
\def\eea{\end{eqnarray}}
\def\bd{\begin{displaymath}}
\def\ed{\end{displaymath}}

\newcommand{\Cint}{\int\kern-10.5pt-\kern7pt}

\newcommand{\PP}{{\mathbb{P}}}

\newcommand{\be}{\begin{equation}}
\newcommand{\ee}{\end{equation}}

\newcommand\fverbit{\egroup\item[\fbox{\unhbox\pippobox}]}
\newbox\pippobox

{

\def\be{\begin{equation}}
\def\ee{\end{equation}}
\def\bea{\begin{eqnarray}}
\def\eea{\end{eqnarray}}

\title{Line defects and (framed) BPS quivers}

\author{Michele Cirafici}

\affiliation{Center for Mathematical Analysis, Geometry, and Dynamical Systems \\ Departamento de Matem\'atica and LARSyS \\ Instituto Superior T\'ecnico\\
Av. Rovisco Pais, 1049-001 Lisboa, Portugal}

\emailAdd{cirafici@math.ist.utl.pt}

\abstract{The BPS spectrum of certain $\cN=2$ supersymmetric field theories can be determined algebraically by studying the representation theory of BPS quivers. We introduce methods based on BPS quivers to study line defects. The presence of a line defect opens up a new BPS sector: framed BPS states can be bound to the defect. The defect can be geometrically described in terms of laminations on a curve. To a lamination we associate certain elements of the Leavitt path algebra of the BPS quiver and use them to compute the framed BPS spectrum. We also provide an alternative characterization of line defects by introducing framed BPS quivers. Using the theory of (quantum) cluster algebras, we derive an algorithm to compute the framed BPS spectra of new defects from known ones. Line defects are generated from a framed BPS quiver by applying certain sequences of mutation operations. Framed BPS quivers also behave nicely under a set of ``cut and join" rules, which can be used to study how $\cN=2$ systems with defects couple to produce more complicated ones. We illustrate our formalism with several examples.}

\begin{document}

\maketitle

\section{Introduction}

The physics of the BPS sector of supersymmetric field theories is deeply intertwined with geometric and algebraic structures. The low energy description of the theory is governed by the geometry of the moduli space of quantum vacua. Its structure is directly determined by the amount of supersymmetry preserved. In the case of $\cN=2$ field theories, the constraints imposed by supersymmetry are enough to solve completely for the Wilsonian low energy effective action in the Coulomb branch \cite{Seiberg:1994rs}. The solution is completely encoded into a family or Riemann surfaces $\Sigma_u$ fibered over the moduli space of quantum vacua. 

At each point in the moduli space there is a tower of BPS states, characterized by certain parameters which depend explicitly on the neighbourhood of the moduli space one is considering, and a corresponding set of single particle Hilbert spaces $\cH_u$. Moving along the Coulomb branch, these parameters may change and a process may become energetically possible in which several BPS states form a bound state, or a BPS particle decays into its elementary constituents. The locus where this can happen is called a wall of marginal stability. Crossing a wall, a certain single particle Hilbert space may cease to have meaning, since multi-particle Hilbert spaces are described by continuous representations. As a consequence the degeneracies of BPS states, measured via the Witten indices over the single particle Hilbert spaces, jump at walls of marginal stability since, simply put, their corresponding Hilbert spaces are no more. This phenomenon is called wall-crossing.

These simple representation theory arguments reveal an enormously rich structure governing the behavior of BPS states in $\cN=2$ field theories. The BPS degeneracies, and their refined cousins such as the protected spin character, obey a universal wall-crossing formula \cite{Kontsevich:2008fj}. This formula originates in the theory of Donaldson-Thomas invariants, certain integrals over the appropriate physical moduli spaces which model mathematically the BPS states. In the context of $\cN=2$ field theories, when reduced to three dimensions, it is the physical condition that the exact quantum metric on the moduli space of vacua is hyperK\"ahler, which means that all the massive BPS states has been accounted for and duly integrated out to give the exact Wilsonian action \cite{Gaiotto:2008cd}.  

The string theory engineering of these field theories provides a different and insightful perspective on such structures. A special class of field theories, the theories of class $\cS [\mathfrak g]$ with $\mathfrak g$ a Lie algebra of ADE type,  can be obtained from the low energy limit of the effective theory of multiple $M5$ branes wrapping a curve $\cC$. The most studied case correspond to $\mathfrak g = \mathfrak{su} (2)$ with only two $M5$ branes \cite{Gaiotto:2009we,Gaiotto:2009hg,Dimofte:2009tm}. In this case the field theory moduli space is identified with the moduli space of solutions of Hitchin equations on $\cC$. The BPS spectrum is captured by the study of an ideal triangulation of $\cC$ derived from the Hitchin system.

A different and more algebraic perspective can be gained via the use of quivers. A BPS quiver is a certain oriented graph whose vertices are in correspondence with elements of a basis of charges of BPS states, while the arrows are dictated by the Dirac-Schwinger-Zwanziger pairing between the charges \cite{Cecotti:2011rv,Alim:2011ae,Alim:2011kw,Cecotti:2012se}. These quivers can be obtained by studying D-branes on Calabi-Yau threefolds; for a restricted sample of the vast literature, see \cite{Douglas:1996sw,Douglas:2000ah,Douglas:2000qw,Fiol:2000wx,Denef:2002ru}. For theories of class $\cS [\mathfrak{su} (2)]$ the nodes of the quiver are identified with the edges of the ideal triangulation of $\cC$, while the arrows are determined by the orientation of the triangles. Alternatively BPS quivers can be constructed using the $4d/2d$ correspondence of \cite{Cecotti:2010fi}. The BPS spectrum is determined via the theory of the generalized Donaldson-Thomas invariants associated with the quiver \cite{Kontsevich:2008fj}. The wall-crossing formula is then interpreted as a consequence of the quantum dilogarithm identities associated with the quiver. Using the interplay between geometry and representation theory, several BPS quivers have been constructed and their BPS spectra understood for a variety of superconformal field theories \cite{Cecotti:2011gu,DelZotto:2011an,Cecotti:2012jx,Cecotti:2013lda,Cecotti:2013sza}, theories with non-simply laced gauge groups \cite{Cecotti:2012gh} and more exotic models coupled to half-hyper multiplets \cite{Cecotti:2012sf}. Quiver methods are also useful to study higher rank theories \cite{Alim:2011kw,Xie:2012gd,Galakhov:2013oja}.

All of these structures are vastly enriched by the presence of defects. Defects are modifications in the fundamental definition of a theory. From the functional perspective they are incorporated in the Feynman path integral by integrating only over those field configurations which have a prescribed behavior. Similarly in canonical quantization they lead to new Hilbert spaces of states, different from the original theory. In four dimensional quantum field theory they can be ordered by increasing co-dimension: domain walls, surface defects and line defects. In this paper we will discuss line defects in four dimensional $\cN=2$ quantum field theories. Line defects can be used to distinguish physically inequivalent theories  \cite{Gaiotto:2010be,Aharony:2013hda,Gadde:2013wq,Gukov:2013zka}. They can be conjecturally characterized by the spectrum of BPS states which can bound them, called framed BPS states \cite{Gaiotto:2010be}. Framed BPS states have an interesting dynamics of their own and exhibit wall crossing phenomena \cite{Lee:2011ph,Gaiotto:2010be}. From the perspective of \cite{Gaiotto:2009we} line defects can be described in terms of certain lines on the curve $\cC$ \cite{Drukker:2009tz,Alday:2009fs,Drukker:2010jp,Gaiotto:2010be,Xie:2013lca}. Furthermore they play an important role in connection with the geometry of Hitchin moduli space when the theory is compactified to three dimensions \cite{Gaiotto:2010be,Ito:2011ea}.

The purpose of this paper is to adapt the method of BPS quivers to study line defects. We will do this in two ways. Firstly we will give a prescription to associate to a line defect a certain series of paths on the quiver, with specific rules to evaluate the framed BPS spectrum. Then we will introduce framed BPS quivers, which contain an extra set of nodes connected with the nodes of the original BPS quiver via a series of rules determined by the defect. In the first setup we devise a dictionary between the traffic rules used in \cite{Gaiotto:2010be} to compute framed BPS spectra and appropriate objects associated with the extended BPS quiver;  here ``extended" means that the BPS quiver is appropriately modified to include the information about the boundary of the curve $\cC$. We identify these objects with elements of a certain algebra of paths associated with the extended quiver, which is a slight generalization of the so-called Leavitt path algebra. The results of \cite{Gaiotto:2010be} are then reproduced by mapping elements of these algebras into specific matrix products or traces thereof.

Framed BPS quivers have the advantage of having clear transformations under quiver mutations. We show that by studying the quantum cluster algebra associated with the BPS quiver, we can compute the framed BPS spectra of certain line operators from known ones by using certain sequences of quiver mutations. More precisely the new line defect is identified with a new framed quiver obtained via mutations from a given one: there exists certain sequences of mutations which leave the underlying unframed BPS quiver invariant, while acting nontrivially on the framing and thus generating new defects. The new framed BPS spectrum is obtained by acting with an appropriate operator obtained by studying the action of quiver mutations on the coefficients of a cluster algebra. We will find that line defects in supersymmetric field theories come in families, or ``mutation orbits". This construction is based on the relation between Donaldson-Thomas theory and cluster algebras \cite{keller,nagao}.

BPS quivers can also be used to construct different supersymmetric theories via graphical manipulations on the quivers \cite{Alim:2011ae,Cecotti:2011rv}. Interestingly we find that a similar set of ``cut and join" rules can be applied to framed BPS quivers, corresponding to the case where the theory is modified by the presence of a line defect. We show some examples of elementary gluing operations acting on framed quivers obtained from surfaces with boundaries, now defined with a line defect.

As a more ``philosophical" remark: the present paper is part of a more broad project, aiming to understand generalizations of Donaldson-Thomas theory in the presence of defects. In a companion work \cite{Cirafici:2013nja} we have introduced divisor defects on Calabi-Yau threefolds and conjectured that the appropriate moduli problem involves sheaves with a certain parabolic structure. In simple examples, quivers make an appearance and allow to compute Donaldson-Thomas type of invariants by instanton counting techniques. In this paper we also find that ordinary Donaldson-Thomas theory for class $\cS [\mathfrak{g} , \cC]$ theories has to be consistently modified when including one dimensional defects. Still, a more general and precise perspective is lacking.

This paper is organized as follows. In Section \ref{susygauge} we collect some generalities about supersymmetric field theories, their BPS quivers and their relations with Hitchin systems and Donaldson-Thomas theory. In Section \ref{linedef} we review some aspects of line defects and introduce the necessary ingredients we will need. In Section \ref{extended} we introduce extended BPS quivers and argue that  line defects correspond to certain elements of the so-called Leavitt path algebra associated with the extended quiver. We outline an algorithm which computes the framed BPS spectrum from the aforementioned elements. In Section \ref{framedBPS} we introduce framed BPS quiver, to provide an alternative description of line defects. Framed BPS quivers transform nicely under quiver mutations; we use this fact to derive another algorithm, which allows one to generate framed BPS spectra of new line defects from known ones. We use various ingredients from the theory of generalized Donaldson-Thomas invariants, namely quantum cluster algebras and quantum dilogarithm identities. Sections \ref{ADtheories} and \ref{SU2gauge} contain several examples which we use to illustrate the formalism. Next in Section \ref{cutjoin} we discuss how the formalism of framed BPS quivers can be used to construct line defects in field theories via a series of ``cut and join" quiver rules. Physically these rules can be traced to certain operations, such as coupling or decoupling of BPS states or gauging flavor symmetries, done when the theory is defined with a defect. We summarize our findings  in Section \ref{discussion}. The Appendix \ref{selffolded} contains some technical results on the shear coordinates for triangulations with self-folded triangles. 

\section{$\cN=2 $ supersymmetric field theories and their BPS spectra} \label{susygauge}

\subsection{Generalities}

In this paper we will consider $\cN=2$ supersymmetric field theories in four dimensions. These theories have a moduli space of vacua and we will denote with $\cB$ their Coulomb branch. The low energy Wilsonian description is encoded in a certain function $\cF$, the prepotential, which determines the quantum metric on $\cB$. In general $\cF$ is a sum of a classical part, a perturbative one-loop term and an infinite series of instanton corrections. At a generic point $u \in \cB$ the theory is abelian and has a symplectic lattice of charges $\Gamma_u$ of rank $2 r + f$, where $f$ is the rank of the flavor symmetry group and the gauge symmetry is broken to $U(1)^r$. The lattice is endowed with a nondegenerate symplectic integer-valued pairing
\begin{equation}
\langle \ \ , \ \ \rangle \ : \ \Gamma \times \Gamma \longrightarrow \zed \ .
\end{equation}
Flavor charges are in the radical of the symplectic pairing. Because of $\cN=2$ supersymmetry, the central charge operator of the supersymmetry algebra is represented by a function
\begin{equation} \label{Zstab}
\cZ_{\gamma} (u) : \Gamma \longrightarrow \complex \ ,
\end{equation}
which depends holomorphically on $u \in \cB$. 
Supersymmetry implies that the Wilsonian low energy theory can be described in the Coulomb branch in terms of an auxiliary family of non compact Riemann surfaces $\Sigma_u$ fibered over $\cB$  \cite{Seiberg:1994rs};  the lattice of electric and magnetic charges is identified with the homology group $H_1 (\Sigma_u ; \zed)$ and the central charge is represented by the integral of a certain differential $\lambda_u$, the Seiberg-Witten differential, over 1-cycles of $\Sigma_u$
\begin{equation}
Z_{\gamma} (u) = \frac{1}{\pi} \oint_{\gamma} \ \lambda_u \ .
\end{equation}

The Wilsonian low energy action describes the propagating degrees of freedom which are light near the low energy scale. The full BPS spectrum will also contain heavier massive states. Some of these can become light at some point in the moduli space and appear in the Wilsonian description, others will be generically integrated out. At a generic point in the Coulomb branch, the single particle BPS Hilbert space is graded by the lattice of charges $\cH_u = \bigoplus_{\gamma} \ \cH_{u,\gamma}$. The degeneracy of BPS states can be computed via the second helicity supertrace
\begin{equation}
\Omega (\gamma , u) = \frac12  \Tr_{\cH_{u,\gamma}}  (2 \, J_3)^2 (-1)^{2 J_3} \ ,
\end{equation} 
where $J_3$ is an angular momentum generator for the spatial $\mathfrak{so} (3)$ symmetry. Note that CPT implies the property $\Omega (\gamma , u)  = \Omega (-\gamma , u) $.  The above BPS degeneracy has a far reaching generalization in the Protected Spin Character \cite{Gaiotto:2010be}
\begin{equation} \label{PSC}
\left( q - q^{-1} \right) \Omega (\gamma , u ; q) = \Tr_{\cH_{u,\gamma}} (2 \, J_3)  (-1)^{2 J_3} (-q)^{2 J_3 + 2 I_3} \ ,
\end{equation} 
where $I_3$ is a $\mathfrak{su} (2)_R$ generator and $q$ is a quantum parameter. The above indices vanish on long representations of the supersymmetry algebra. If we denote with $\cH_{u,\gamma}'$ the Hilbert space with the half-hypermultiplet representation factored out, the Protected Spin Character has the more compact form
\begin{equation}
\Omega (\gamma , u ; q) =  \Tr_{\cH_{u,\gamma}'}  (-1)^{2 J_3} (-q)^{2 J_3 + 2 I_3}
\end{equation}
In particular for a hypermultiplet $\Omega (\gamma , u ; q) = 1$ while for a vector multiplet $\Omega (\gamma , u ; q) = q + \frac{1}{q}$. The BPS degeneracy is recovered in the $q \longrightarrow -1$ limit.

As we move along the Coulomb branch $\cB$ certain BPS states may become unstable, leave the single particle spectrum, and join the continuum of multi-particle states . This process (and the inverse where new states appear) is known as wall-crossing and governs the dependence of the BPS indices on the point $u \in \cB$. Both $\Omega (\gamma , u)$ and $\Omega (\gamma , u ; q)$ are piecewise constant functions of $u$ which may jump at walls of marginal stability, when a state enters or leaves the single particle spectrum. The formation of bound states is governed by conservation of energy, or for a BPS state by the central charge. For example, if two BPS particles $\gamma_1$ and $\gamma_2$ form a BPS bound state $\gamma_1 + \gamma_2$, then the binding energy is
\begin{equation}
E_{bin} =  \vert \cZ_{\gamma_1 + \gamma_2} (u) \vert - \vert \cZ_{\gamma_1} (u) \vert - \vert \cZ_{\gamma_2} (u) \vert \le 0 \ .
\end{equation}
When the two central charges become aligned  the binding energy vanishes. This condition defines a wall of marginal stability
\begin{equation}
MS (\gamma_1 , \gamma_2) = \left\{ u \in \cB \ \vert \ \cZ_{\gamma_1} (u) \parallel \cZ_{\gamma_2} (u) \right\} \ .
\end{equation}
Walls of marginal stability divide the moduli space in chambers, and the jumps in the indices are governed by the wall-crossing formula and its quantum generalization \cite{Kontsevich:2008fj}. In principle this formula allows for a computation of the full spectrum of the theory over all the moduli space, once the spectrum is known in a chamber.

\subsection{BPS quivers} \label{BPSquivers}

The information about the BPS spectrum can be nicely packaged into a so called \textit{BPS quiver}, as reviewed in \cite{Cecotti:2012se}. A quiver $\sfQ = (\sfQ_0 , \sfQ_1)$ is an algebraic structure defined by a set of nodes $v \in \sfQ_0$ and a set of arrows $\left( v \xrightarrow{\sfa} w \right) \in \sfQ_1$ between the vertices $v,w \in \sfQ_0$. We will denote by $\sfe_i$ the trivial path of length zero at the vertex $i \in \sfQ_0$. The BPS quiver is constructed accordingly to the following rules. At any point of the Coulomb branch $\cB$, $\cZ_{\gamma} (u)$ is a complex number, which can be represented as a vector, a BPS ray, on the central charge plane. We will assume that these vectors do not form a dense set. Assume we are given an integral positive basis of charges $\{ \gamma_i \}$ for the lattice $\Gamma$. To find such a basis we need to fix conventions on what we call a particle and what we call an anti-particle. This is done by dividing arbitrarily the central charge plane in two half-planes, and declaring that one of them, call it $\cH$, contains the particle states. The set of particle states defines a positive cone $\Gamma_+ = \bigoplus_i \, \zed_+ \, \gamma_i$ inside $\Gamma$. In the following we will often confuse the central charge of a state with the actual state, and speak freely of ``ordering" of the charges $\gamma_i$ when talking about the BPS rays. For each element of the basis $\{ \gamma_i \}$ we draw a node of the quiver. Then for each pair of charges $\gamma_i $ and $\gamma_j$ such that $\langle \gamma_i , \gamma_j \rangle > 0$ one draws $\langle \gamma_i , \gamma_j \rangle$ arrows $ i \longrightarrow j$. If the pairing is negative, one draws the arrows in the opposite sense, $i \longleftarrow j$. In other words the adjacency matrix of the quiver is $B_{ij} = \langle \gamma_i , \gamma_j \rangle$. Note that because of the antisymmetry of the pairing the resulting quiver is 2-acyclic: all the arrows between two nodes point in the same direction and there are no arrows which start and end at the same node. This construction is conventional and drawing the arrows with the opposite orientation results in the opposite quiver $\sfQ^{op}$ \footnote{One can find different conventions in the literature. Any statement about a quiver can be rephrased for the opposite quiver by appropriately changing the conventions; for example going from one quiver to the opposite switches the mutation operators $\mu_{k,\pm}$ and their generalizations, to be introduced in the following.}. 

Every BPS state in a given chamber can be realized as a bound state of the particles corresponding to the nodes of the quiver, by the assumption that these form a positive basis. The dynamics of the BPS states is governed by a supersymmetric quantum mechanics where the scalar fields are bi-fundamental Higgs fields given by the representations of the BPS quiver. A quiver representation $\sfR$ is a collection of vector spaces $\mathsf{V}_v$ associated with the vertices $v \in \sfQ_0$ and a collection of linear maps $\mathsf{B}_{\sfa} \, : \, \mathsf{V}_v \longrightarrow \mathsf{V}_w$ for each arrow $\sfa \in \sfQ_1$. Each representation $\sfR$ is characterized by a dimension vector $\dim \sfR$ whose elements are the dimensions of the representation spaces $\dim \mathsf{V}_v$ for all the nodes $v \in \sfQ_0$. A distinguished set of representations are the simple ones $\sfD_v$, such that $\mathsf{V}_v = \complex$ and all the others $\mathsf{V}_w = 0$. A representation of dimension $\dim \sfR$ is associated with a state of charge $\gamma = \sum_{i \in \sfQ_0} \dim \mathsf{V}_i \, \gamma_i$.

The supersymmetric quantum mechanics has four supercharges. Its vacua preserve all the supercharges and therefore correspond to four dimensional BPS states. Vacua are field configurations which obey the D-term conditions and the F-term conditions, which are critical points of a superpotential $\sfW$. Alternatively one can trade the D-term conditions for a stability condition, at the price of modding out by complexified gauge transformations (change of basis in the representation spaces). Ground states correspond to stable quiver representations. The equations of motion derived from the superpotential correspond to a set of relations $\langle \, \partial_{\sfa} \sfW \, \vert \, \sfa \in \sfQ_1 \rangle$ associated with the arrows of the quiver (in a more mathematical language we regard $\sfW$ as a sum of cyclic monomials and we say that the BPS quiver is a bound quiver). The notion of stability is induced by the central charge function $\cZ_\gamma (u)$. The construction is quite generic and holds for an abelian category $\cA$, in this case the category of bound quiver representations $\mathsf{Rep} (\sfQ , \sfW)$ (another notable example is the category of coherent sheaves $\mathsf{coh} (X)$ of a smooth projective variety $X$). In physically interesting situations there is typically an homomorphism from the Grothendieck group $K (\cA) $ to the lattice of charges $\Gamma$. In our case this is the identification of the simple quiver representations with a basis of BPS states (in the case of $\mathsf{coh} (X)$ the map is given by the Chern character). 

Then the function (\ref{Zstab}) is a stability condition, and a certain state $\gamma_r$ described by a representation $\sfR \in \mathsf{Rep} (\sfQ , \sfW)$ is $\cZ$-stable (semi-stable) if $\arg \cZ_{\gamma_s} < \arg \cZ_{\gamma_r}$ ($\arg \cZ_{\gamma_s} \le \arg \cZ_{\gamma_r}$) for each sub-state $\gamma_s$ associated with a sub-representation $\mathsf{S}$ other than $\mathsf{R}$ and zero. If the state is stable, then the moduli space 
\begin{equation}
\scrM_{\gamma} = \left\{ \sfR \in \mathsf{Rep} \left( \sfQ , \sfW \right) \  \vert \ \partial \, \sfW = 0  \text{ \ and $\sfR$ is \ } \cZ-\text{stable} \right\} / \prod_i Gl (n_i , \complex) \ ,
\end{equation}
is not empty, and its cohomology determines the spin of the BPS state (via the Lefschetz action). In particular if $\scrM_{\gamma}$ is zero dimensional the state $\gamma$ is an hypermultiplet, while if it is one-dimensional it is a vector multiplet. Simple representations are automatically stable. Physically this implies that in all the BPS chambers there are at least $|\sfQ_0|$ stable states.

An efficient algorithm to compute the spectrum is the mutation method \cite{Alim:2011ae,Alim:2011kw,Gaiotto:2010be}. In the mutation method one changes the particle anti-particle assignment \textit{at the same} point in the moduli space by rotating the plane $\cH$. As one rotates the plane $\cH$ across a BPS ray, the corresponding particle leaves $\cH$ while its antiparticle enters. Therefore the basis of charges, and consequently the quiver, changes. The locus where this happens is called a wall of second kind \cite{Kontsevich:2008fj} or BPS wall \cite{Gaiotto:2010be}. Now one has to study the representation theory of the new quiver with the new ordering for the central charges. However since no wall of marginal stability was crossed, the BPS spectrum must be the same. The new quiver quantum mechanics gives therefore a Seiberg-dual description of the same physics. The algebraic operation which give rise to the new quiver is called a mutation, and if we label by $k$ the node corresponding to the particle which exits the $\cH$ plane, in terms of the adjacency matrix it is given by
\begin{equation} \label{mutB}
B'_{ij} = \left\{ \begin{matrix} - B_{ij} & \text{if} \ i=k  \ \text{or} \ j=k \\ B_{ij} + \text{sgn} (B_{ik}) [ B_{ik} \, B_{kj} ]_+ & \text{otherwise} \end{matrix} \right. \ ,
\end{equation}
where we have introduce the notation $[x]_+ = \text{max} \{  0, x \}$. The mutation on the quiver at node $k$ has the effect of changing the basis of charges as follows. If the node $k$ is the one with the biggest phase of the central charge, that is the corresponding particle is the first one to disappear from the $\cH$ plane as this is rotated clockwise, we have the mutation $\mu_{k,+}$ whose effect is 
\begin{equation} \label{mu+}
\mu_{k,+} (\gamma_i) = \left\{ \begin{matrix} - \gamma_k & \text{if} \ i = k \\ 
\gamma_i + [\langle \gamma_i , \gamma_k \rangle]_+ \, \gamma_k & i \neq k  
\end{matrix} \right.  \ .
\end{equation}
Similarly we can rotate the plane $\cH$ counterclockwise; in this case the first state to exit the plane is the one with the smallest central charge and the basis undergoes a mutation $\mu_{k,-}$
\begin{equation} \label{mu-}
\mu_{k,-} (\gamma_i) = \left\{ \begin{matrix} - \gamma_k & \text{if} \ i = k \\ 
\gamma_i - [\langle \gamma_i , \gamma_k \rangle]_- \, \gamma_k & i \neq k  
\end{matrix} \right. \ ,
\end{equation}
where $[x]_- = \text{min} \{  0, x \}$. These rules extend to the superpotential $\sfW$, although we will not need them for the moment. It is natural to introduce a discrete ``time" variable to keep track of a sequence of quiver mutations. The quiver at time zero $t_0$ is the original quiver $\sfQ (t_0) = \sfQ$. After a mutation $\mu_k \sfQ (t_0) = \sfQ (t_1)$ where $\sfQ (t_1)$ is the quiver with adjacency matrix $B (t_1) = \mu_k \, B (t_0)$. Note that at each time $t_i$ one can choose which node to mutate among the set $\sfQ_0$, which is ``time independent".

To generate the full spectrum one continues with a $\pi$ rotation of the half-plane $\cH$. Each node at which the quiver is mutated automatically correspond to a stable representation, since nodes are associated with simple  representations, and is therefore part of the stable spectrum. Moreover, since simple representations have no moduli, it corresponds to a BPS hypermultiplet. Once a full $\pi$-rotation is completed, the full particle spectrum within the chamber will have been generated, and the anti-particle spectrum is simply the CPT conjugate. There might be obstructions to complete the full half-rotation; this is typically due to a higher spin state which appears as an accumulation ray in the central charge plane. Note that we are using different conventions respect to \cite{Alim:2011ae,Alim:2011kw}, whose quivers have $i \xrightarrow{n} j$ if $\langle \gamma_i , \gamma_j \rangle = -n$. Our conventions for the clockwise and counterclockwise mutations $\mu_{k,\pm}$ are those of Section 5.2 of \cite{Gaiotto:2010be}, since we are going to compare our formalism with their results.

The order of the mutation sequence depends on the ordering of the central charges of the BPS states corresponding to the nodes of the quiver. A randomly chosen ordering does not necessarily correspond to a physically realized regime of the theory. In other words, there might be kinematically determined chambers which cannot be reached dynamically. A field theory in which any ordering of the central charge phases of the nodes is physically accessible is called a \textit{complete} theory. Complete theories were classified in \cite{Cecotti:2011rv} and are associated with finite mutation type quivers; that is quivers which, upon all possible mutations, only generate a finite number of distinct quiver topologies. These quivers are the ones arising from an ideal triangulation of a curve, as we will explain in the next section, plus a finite number of other cases. These are the nine cases associated with Dynkin diagrams of type $E$ (finite, affine and elliptic), two Derksen-Owen quivers and quivers with at most two nodes.

\subsection{BPS quivers and triangulated curves} \label{BPStriang}

For a large class of theories the BPS quiver can be read directly from a geometric engineering of the field theory \cite{Cecotti:2011rv,Alim:2011ae,Alim:2011kw}. The field theory can be realized for example via type II compactified on a local Calabi-Yau threefold, or as low energy description of a multiple M5 brane configuration in M-theory. In both cases string theory provides an ``internal" description of the theory via a certain curve $\cC$, on which the threefold is fibered or the M5 branes wrapped. Theories which arise from the low energy dynamics of multiple $M5$ branes wrapping a curve $\cC$ are known as theories of class $\cS$ \cite{Gaiotto:2009hg}. They are characterized by an ADE Lie algebra $\mathfrak g$, the curve $\cC$ and eventually some extra data $D$ associated with the punctures of $\cC$; they are customarily denoted with $\cS [\, \mathfrak g , \cC , D]$ or simply $\cS$. We will consider theories of class $\cS$ which are obtained by wrapping two M5 branes over the curve $\cC$, for which $\mathfrak g = \mathfrak{su} (2)$. 

The effective theory of the $M5$ branes compactified on  $\real^3 \times S^1_R \times \cC$ is a sigma model with target a certain manifold $\cM$. This manifold is identified with the Hitchin moduli space $\cM$. Indeed for the system to have four dimensional $\cN=2$ supersymmetry on $\real^3 \times S^1_R$, the internal fields $(A,\varphi)$ on $\cC$ have to satisfy Hitchin equations
\begin{eqnarray}
F_A + R \, [ \, \varphi , \overline{\varphi} \, ] \, =0 \ , \\
\overline{\partial}_A \, \varphi \, = 0 \ , \\
\partial_A \, \overline{\varphi} \, =0 \ ,
\end{eqnarray}
with prescribed singularities along the punctures of $\cC$. The manifold $\cM$ is hyperK\"ahler and is a fibration over $\cB$ whose fibers are compact tori.

The Seiberg-Witten curve $\Sigma$ which describes the IR regime of the four dimensional gauge theory is a double cover of $\cC$
\begin{equation}
\Sigma = \{ \lambda \, \vert \, \det (\lambda - \varphi) = 0 \}  \ .
\end{equation}
In the case $\mathfrak g = \mathfrak{su} (2)$ the Seiberg-Witten differential $\lambda$ is the square root of a certain meromorphic differential. The boundary condition on $\varphi$ implies that the quadratic differential $\lambda^2$ has double poles at ordinary punctures which are associated to the mass parameters of the field theory (and are indeed first order poles of the Seiberg-Witten differential whose residue is a mass parameter) and poles of order $p_i =  c_i+2 \ge 3$ at irregular punctures. This description provides a geometric picture for the BPS states via an ideal triangulation of $\cC$. This triangulation is obtained by studying constant phase $\theta$ flows on $\cC$, that is curves defined by the condition that their tangent vector $\partial_t$ obeys
\begin{equation}
\lambda \ \partial_t \in \e^{\ii \theta} \real^* \ .
\end{equation}
For irregular punctures the flow exhibits Stokes phenomena \cite{Gaiotto:2009hg,Alim:2011ae}. A number of Stokes rays emanates from the singularity, dividing a neighbourhood into Stokes sectors. Solutions of the Hitchin equations in those sectors differ by the appropriate Stokes factors. These phenomena are dealt with by cutting out a small disk around the irregular puncture and replacing it with an ideal boundary with $c_i$ marked points. Each marked point on the new ideal boundary is a representative of a trajectory within a Stokes sector. There are then two types of flow lines, the separating and the generic trajectories. Separating trajectories connect a zero of $\lambda^2$ with an ordinary puncture or a marked point. On the other hand a generic trajectory have both endpoints at a ordinary puncture or at a marked point, whose neighborhoods act as basin of attractions. There are also  curves separating two zeros of $\lambda^2$; these trajectories are called finite and correspond to BPS hypermultiplets. Generic trajectories come in families. For each family we take a representative and those arcs determine an ideal triangulation of $\cC$. By construction, each triangle contains exactly one zero of $\lambda^2$. In general this triangulation allows for self-folded triangles. As the angle $\theta$ changes, the triangulation varies continuously until the angle corresponding to a BPS state is reached, signaling the appearance of a finite trajectory. Crossing this angle, the triangulation undergoes a flip. In other words flip of the triangulations detect BPS hypermultiplets.

In summary we have a curve  $ \cC_{g,n,b,\{ c_i \} }$, or simply $\cC$, of genus $g$, $n$ ordinary punctures, $b$ boundaries with $c_i$, $i =1,\dots,b$ marked points on each boundary component. The flow problem describes an ideal triangulation $T (\cC_{g,n,b,\{ c_i \} })$ or simply $T$, that is a collection of curves, identified up to isotopy, mutually and self non-intersecting except at end points, which end at the punctures or at the marked points, and finally cannot be contracted to a puncture or to a boundary segment. The number of these curves, called arcs, is
\begin{equation}
D  = 6 \, g - 6 + 3 \, n + \sum_i (c_i + 3) \ .
\end{equation}
To this triangulation we can associate a BPS quiver as follows \cite{FT1}. To each arc of the triangulation which is not a boundary segment we associate a node of the quiver. The numbers of arrows between the nodes is determined by the triangulation. To do so we need to define the  map $\mathrm{ext}_T ( \bullet )$ which acts on the arcs $i$ of the triangulation $T$ as follows: if $i$ is the internal arc of a self-folded triangle, then $\mathrm{ext}_T ( i )$ is the external arc of the same triangle; otherwise $\mathrm{ext}_T ( i )  = i$. Then for each triangle $\triangle \in T$ which is not self-folded, we define the $D \times D$ integer matrix $B^{\triangle}$ whose entries are
\begin{equation}
B^{\triangle}_{ij}  = \left\{\ \begin{matrix} +1 &  \mathrm{ext}_T ( i )  \ \text{and} \  \mathrm{ext}_T ( j ) \ \text{are both sides of} \ \triangle \ \text{and}  \\ &  \mathrm{ext}_T ( i ) \ \text{precedes}  \ \mathrm{ext}_T ( j ) \ \text{in counterclockwise order} \ ,
\\
-1 &  \mathrm{ext}_T ( i )  \ \text{and} \  \mathrm{ext}_T ( j ) \ \text{are both sides of} \ \triangle \ \text{and}  \\ &  \mathrm{ext}_T ( i ) \ \text{precedes}  \ \mathrm{ext}_T ( j ) \ \text{in clockwise order} \ , \\
0 & \text{otherwise} \ .
\end{matrix}
\right.
\end{equation} 
Then the matrix $B = \sum_{\triangle \in T} B^{\triangle} $, where the sum is over all the non self-folded triangles, is the adjacency matrix of the BPS quiver. Note that its entries can be $0, \pm 1, \pm 2$. A flip of the ideal triangulation corresponds to a mutation of the BPS quiver, signaling the appearance of a BPS state.

We fix the conventions as in Section 7.1.1 of \cite{Gaiotto:2009hg}, where if the edge $E$ precedes $E'$ going counterclockwise
\begin{equation}
\langle E , E' \rangle = \langle \gamma_E , \gamma_{E'} \rangle \ ,
\end{equation}
and the pairing is positive. The pairing defined as such, coincides with the pairing on $\Gamma$, and determines the arrow structure of the quiver. Other conventions are possible and used in the literature.

From the triangulation, or equivalently the associated BPS quiver, we can read several properties of the associated quantum field theory \cite{Cecotti:2011rv}. For example the information about the number of flavor charges is encoded in the triangulation as
\begin{equation} \label{flavornum}
f  = n + \sum_{c_i \ \mathrm{even}} 1 \ .
\end{equation}
Triangulations which only have ordinary punctures lead to the so-called Gaiotto theories (more precisely theories with $g>0$ and regular punctures and theories with $g=0$ and at least $3$ regular punctures) which are conformal in the UV. Other conformal theories are the $ADE$ Argyres-Douglas theories, whose BPS quiver has an element in its mutation class equal to a Dynkin diagram of type $ADE$. The Argyres-Douglas theories of type $A_{p-5}$ arise from triangulations of a sphere with a single irregular puncture of order $p \ge 7$, while the theories of type $D_{p-2}$ correspond to the sphere with an ordinary puncture and an irregular puncture of order $p$. Other superconformal theories correspond to the elliptic quivers of type $E$ and the quiver $X_7$ from the Derksen-Owen classification. All the other quantum field theories arising from triangulations of a Riemann surface are asymptotically free in the UV.

\subsection{Generalized Donaldson-Thomas theory} \label{genDT}

The stable spectrum of BPS states has a very rich structure. For theories which admit a BPS quiver, the BPS degeneracies and the wall-crossing properties of the BPS states are elegantly encoded in the generalized Donaldson-Thomas invariants introduced by Kontsevich and Soibelman \cite{Kontsevich:2008fj}. To the lattice of charges $\Gamma$ we associate the quantum torus $\torus_{\Gamma}$ generated by formal variables $\sfX_{\gamma}$ which obey the $q$-commutation relations
\begin{equation}
\sfX_{\gamma_i} \ \sfX_{\gamma_j} = q^{\langle \gamma_i , \gamma_j \rangle} \ \sfX_{\gamma_i + \gamma_j} \ .
\end{equation}
Generalized Donaldson-Thomas invariants appear in a certain ordered product of \textit{quantum dilogarithms}. The quantum dilogarithm function is defined as
\begin{equation}
\bbE (\sfX) = 1 + \frac{q^{1/2}}{q-1} \ \sfX + \cdots + \frac{q^{n^2/2}}{(q^n-1) \, (q^{n-1}-1) \, \cdots \, (q-1)} \ \sfX^n + \cdots \ .
\end{equation} 
We will typically use the notation $\bbE (\gamma)$ as a shorthand for $\bbE (\sfX_{\gamma})$. BPS degeneracies are typically defined as certain integrals over the relevant moduli spaces and with an appropriate measure. In the case of a quiver with a potential $(\sfQ , \sfW)$ the appropriate moduli space is the moduli space $\mathsf{mod} \, \sfA$ of left-modules of the path algebra $\sfA = \complex \sfQ / \partial \, \sfW$. The path algebra is defined via the composition of paths on the quiver (the product being zero if two arrows do not compose) modulo the relations derived from the superpotential. Sometimes the name noncommutative Donaldson-Thomas invariants is used when the BPS degeneracies are associated with a quiver. Donaldson-Thomas invariants $\Omega (\gamma)$ are weighted Euler characteristics of these moduli spaces. In certain chambers they can be computed explicitly via techniques of equivariant localization \cite{Cirafici:2010bd,Cirafici:2011cd}. We will not discuss these matters in detail and refer the reader to the review \cite{Cirafici:2012qc}.

Fixing a stability condition $\cZ$ for $\sfA$ gives an ordering of the central charge phases corresponding to the stable BPS states. Assume they are ordered as $(\gamma_1 , \cdots , \gamma_N)$ with increasing phase. In this chamber we construct the ordered product of quantum dilogarithms
\begin{equation} \label{NCDT}
\bbE (\gamma_N)^{\Omega (\gamma_N)} \cdots \bbE (\gamma_1)^{\Omega (\gamma_1)}  \ .
\end{equation}
This is the Kontsevich-Soibelman operator of $(\sfQ , \sfW)$. The BPS degeneracies, the  Donaldson-Thomas invariants, appear as exponents of the quantum dilogarithms. If the quiver has a minimal chamber with only hypermultiplets, they are all unity. Remarkably the product (\ref{NCDT}) is \textit{independent} on the stability condition $\cZ$; this is the key property underlying the wall crossing behavior of the BPS degeneracies. As we move within the Coulomb branch $\cB$ the central charges of the BPS states vary smoothly. When we cross a wall of marginal stability the ordering of the central charge phases changes. As a consequence the ordering of the factors and the BPS degeneracies in (\ref{NCDT}) change in such a way that the product is invariant.

As an example consider the Argyres-Douglas theory $A_2$. In this case the BPS quiver is
\begin{equation}
\xymatrix@C=10mm{  
 \ \gamma_1 \ \bullet \ \ar[rr] & & \ \bullet \ \gamma_2
} \ .
\end{equation}
This theory has two chambers. When $\arg \cZ_{\gamma_1} > \arg \cZ_{\gamma_2}$ the spectrum consists only of $\{ \gamma_1 , \gamma_2 \}$. On the other hand in the chamber where  $\arg \cZ_{\gamma_2} > \arg \cZ_{\gamma_1}$ a bound state becomes stable and the spectrum is $\{ \gamma_1 ,  \gamma_1 + \gamma_2 , \gamma_2 \}$. The invariance of the product (\ref{NCDT}) is expressed by the famous pentagon identity
\begin{equation} \label{pentagon}
\bbE (\sfX_{\gamma_1}) \ \bbE (\sfX_{\gamma_2}) = \bbE (\sfX_{\gamma_2}) \ \bbE (\sfX_{\gamma_1 + \gamma_2}) \ \bbE (\sfX_{\gamma_1}) \ .
\end{equation}

The same story holds for chambers with an infinite number of states. The simplest example is $SU(2)$ super Yang--Mills, whose BPS quiver is
\begin{equation}
\xymatrix@C=10mm{  
 \ \gamma_1 \ \bullet \ \ar@<-0.5ex>[r] \ar@<0.5ex>[r]  & \ \bullet \ \gamma_2 
} \ .
\end{equation}
The lattice $\Gamma$ has rank two, and here $\gamma_1 = (2,-1)$, $\gamma_2 = (0,1)$ such that $\langle \gamma_1 , \gamma_2 \rangle = + 2$ (our quiver is the opposite to that of \cite{Alim:2011kw}). To compute the spectrum we consider first a chamber where $\arg \cZ_{\gamma_1} > \arg \cZ_{\gamma_2}$. In this case we apply the mutation operator $\mu_{k,+}$ going clockwise in the central charge plane: firstly mutating at $\gamma_1$ and then at $\gamma_2$. In doing this no new BPS state is generated and after a full half rotation of the plane $\cH$ the quiver goes back to itself but now written in terms of the antiparticles. This is the strong coupling chamber. If on the other hand we consider the weak coupling chamber, where $\arg \cZ_{\gamma_2} > \arg \cZ_{\gamma_1}$, we generate the weak coupling spectrum by sequences of mutations which alternate between the nodes; one can start  with $\mu_{2,+}$, and move clockwise in the central charge plane, or with $\mu_{1,-}$ and move counterclockwise. Both sequences are needed, since the weak coupling spectrum contains a vector multiplet, which is an obstruction to complete the full half rotation in either sense. Invariance of the product (\ref{NCDT}) respect to the choice of the stability  condition implies the $SU(2)$ wall-crossing formula 
\begin{eqnarray}
\bbE (\sfX_{\gamma_1}) \ \bbE (\sfX_{\gamma_2}) =  && \bbE (\sfX_{\gamma_2}) \ \bbE (\sfX_{\gamma_1 + 2 \gamma_2}) \ \bbE (\sfX_{2 \gamma_1 + 3 \gamma_2}) \cdots \ \bbE^{\text{vect}} (\sfX_{\gamma_1 + \gamma_2}) \ \cdots \cr && \cdots \bbE (\sfX_{3 \gamma_1 + 2 \gamma_2}) \  \bbE (\sfX_{2 \gamma_1 + \gamma_2})  \ \bbE (\sfX_{\gamma_1}) \ ,
\end{eqnarray}
where $\bbE^{\text{vect}} (\sfX_{\gamma_1 + \gamma_2}) = \bbE (-q^{-1} \, \sfX_{\gamma_1+ \gamma_2})^{-1} \, \bbE (- q \, \sfX_{\gamma_1+ \gamma_2})^{-1} $ corresponds to the W-boson in the weak coupling chamber.

\section{Line defects} \label{linedef}

The goal of this paper is to adapt the formalism of BPS quivers to study line defects. We will now discuss some generalities about line defects in supersymmetric field theories following \cite{Gaiotto:2010be} and in particular the concept of framed BPS degeneracies, which correspond to BPS states bound with the defect. For theories of class $\cS [\mathfrak{su} (2) , \cC]$ line defects can be characterized as certain laminations on $\cC$.

\subsection{Line defects and framed BPS states}

We will only consider straight lines in $\real^{3,1}$ which can be described as point defects in space located, say, at the origin of $\real^3$. We think of a line defect $L$ as a way of modifying the path integral by integrating only on those configurations which obey specified boundary conditions on a neighborhood of the defect. Similarly the Hilbert space of the theory is modified to $\cH_{L}$, which now decomposes according to representations of the  subalgebra of the super-Poincar\'e algebra left unbroken by the defect. As locally a neighborhood of a straight line defect in Minkowski space is conformal to $AdS_2 \times S^2$, an equivalent way of defining a line defect is in the ultraviolet as a boundary condition of a superconformal fixed point theory on $AdS_2 \times S^2$ \cite{Kapustin:2005py,Kapustin:2006pk}. The line defects we will consider are parametrized by a phase $\zeta$ which determines the subalgebra of the supersymmetry algebra which is preserved by the defect.

A good IR model of the defect is to consider the theory as if in the presence of an infinitely heavy dyon with charge $\gamma$ located at the position of the defect. From this perspective it is natural to ask if BPS particles can form supersymmetric bound states with the defect. It turns out that this is true and the new bound states are called \textit{framed BPS states} \cite{Gaiotto:2010be}. Moreover a stronger statement is conjectured to hold, the defect is entirely characterized by its framed BPS spectrum. Framed BPS states saturate the energy bound
\begin{equation}
E + \re (Z_\gamma (u) / \zeta) = 0 \ .
\end{equation}
The Hilbert space is graded by the charges  as
\begin{equation}
\cH_{L,\zeta,u}^{BPS} = \bigoplus_{\gamma \in \Gamma_L} \cH_{L,\zeta,u,\gamma}^{BPS} \ ,
\end{equation}
where $\Gamma_L$ is the lattice of electric and magnetic charges in the presence of the line defect (which takes into account that the electric and magnetic charges as measured at infinity will be shifted by the charge of the defect); for simplicity we will usually omit the dependence on the Coulomb branch point $u \in \cB$ and on the phase $\zeta$. If we imagine the defect as being a infinitely heavy dyon, then framed BPS states  look like a halo of particles of charge $\gamma_h$ bound to a core particle of charge $\gamma_c$. Above the energy threshold 
\begin{equation}
E = -\re (Z_{\gamma_c} (u)) + |Z_{\gamma_h} (u)| \ ,
\end{equation}
the halo is not bound to the line defect and forms a continuum of ordinary BPS states, free to wander around. Ordinary BPS states and framed BPS states are separated in energy by a tower of excited framed BPS states. The halo bound state radius is 
\begin{equation}
r_{halo} = \frac{\langle \gamma_h , \gamma_c \rangle}{ 2 \, \im (\cZ_{\gamma_h} (u) / \zeta)} \ .
\end{equation}
In particular note that the halo is stable if $\langle \gamma_h , \gamma_c \rangle \  \im (\cZ_{\gamma_h} (u) / \zeta) >0$. Therefore when there exists a halo charge $\gamma_h$ which satisfies
\begin{equation}
- \re (\cZ_{\gamma_h} (u) / \zeta) = |\cZ_{\gamma_h} (u)| \ ,
\end{equation}
the halo is free to join the continuum of ordinary BPS states. This condition defines the BPS walls
\begin{equation}
W (\gamma) = \{ (u,\zeta) \vert \cZ_{\gamma} (u) / \zeta \in \real_- \} \subset \cB \times \complex^* \ .
\end{equation}
Note that the BPS walls are precisely the walls of second kind, upon fixing $\zeta$. The BPS walls divide the space  $ \cB \times \complex^*$ in chambers, which we will generically denote by $c$.

Exactly as we have seen in the ordinary BPS case, we can associate to framed BPS states the framed Protected Spin Character (PSC)
\begin{equation} \label{framedPSC}
\underline{\overline{\Omega}} (u , L_{\zeta} , \gamma ; q) := \Tr_{ \cH_{L_\zeta, \gamma}^{BPS}} q^{2 J_3} (-q)^{2 I_3} \ ,
\end{equation}
and for $q=-1$ the framed BPS degeneracies
\begin{equation}
\underline{\overline{\Omega}} (u , L_\zeta , \gamma ) := \Tr_{ \cH_{u,L_\zeta,\gamma}^{BPS}} (-1)^{2 J_3} \ .
\end{equation}
Here $J_3$ is an $\mathfrak{so}(3)$ generator, while $I_3$ a generator of the $R$-symmetry group $\mathfrak{su}(2)_R$. 
It was conjectured in \cite{Gaiotto:2010be} that no exotic states (that is no states with non trivial $\mathfrak{su} (2)_R$ quantum numbers) contribute to (\ref{framedPSC}) or to (\ref{PSC}) at a generic smooth point in the Coulomb branch. This implies the \textit{strong positivity conjecture}, that the PSC is a positive integral linear combination of characters of irreducible representations and in particular at $q=1$ the PSC computes dimensions of vector spaces and is positive (since $I_3$ acts trivially).

Line operators form a semiring \cite{Kapustin:2007wm,Gaiotto:2010be,Saulina:2011qr,Moraru:2012nu}, with a rather rich structure. The sum operation between line defects is defined by stating that the Hilbert space of the theory with $L_1 + L_2$ is the direct sum of the two superselection sectors $\cH_{L_1}$ and $\cH_{L_2}$. In particular a line defect is called \textit{simple} if it is not the sum of other line defects. The product of line defects is more interesting, and defined via the path integral of the theory with the line defects inserted. The formalism that will be introduced in the following Sections can be used to study these structures, but we will not do so in this paper.

Since a line defect can be equivalently described by its framed BPS spectrum, it is natural to consider generating functions:
\begin{equation} \label{physgen}
F \left( u, L_\zeta , \{ X_{\gamma} \} ; q \right) = \sum_{\gamma} \underline{\overline{\Omega}} (u,L_\zeta , \gamma ; q) \ \sfX_{\gamma} \ .
\end{equation}
Crossing a BPS wall a physical line operator transforms by gaining or losing halos. Explicitly, if we consider crossing a wall associated with a single hypermultiplet of charge $\gamma_h$, then going from the region where $\im \cZ_{\gamma_h} (u) / \zeta <0 $ to the region where  $\im \cZ_{\gamma_h} (u) / \zeta >0 $, which we will label respectively with $ c^-$ and $c^+$, a line operator with charge $\gamma$ behaves as
\begin{eqnarray} 
F  \left( c^+, L , \zeta ,  \sfX_{\gamma} ; q \right) &=& \bbE (\sfX_{\gamma_h}) \ F  \left( c^-, L , \zeta ,  \sfX_{\gamma} ; q \right) \ \bbE(\sfX_{\gamma_h})^{-1} 
\\[4pt]
&\equiv&  \ \mathrm{Ad} (\bbE (\sfX_{\gamma_h})) \ F  \left( c^-, L , \zeta ,  \sfX_{\gamma} ; q \right) \ ,
\end{eqnarray}
and the line operator gains or loses a halo depending on the sign of $\langle \gamma_h , \gamma \rangle$.

It is difficult to compute the framed spectrum of a physical line operator from first principles. To overcome this problem one is led to define a \textit{strongly positive formal line operator} as collection of elements $F(c)$ defined in all BPS chamber $\{ c \}$ such that across BPS walls
\begin{equation}
F(c^+) =  \mathrm{Ad} (\bbE (\sfX_{\gamma_h})) \, F(c^-) \ ,
\end{equation}
and in each chamber
\begin{equation}
F(c) = \sum_{\gamma} P^c_{\gamma} (q) \ \sfX_{\gamma} \ ,
\end{equation}
where $P^c_{\gamma} (y)$ (which is a character of a $SU(2)$ representation) has only positive coefficients (\textit{strong positivity}). It is much easier to construct these objects and following  \cite{Gaiotto:2010be} we will assume that any strongly positive formal line operator is a generating function of framed BPS states for a certain physical line operator.

It is instructive to consider the specialization of these formulas for $q \longrightarrow -1$. In this limit the quantum torus algebra $\torus_{\Gamma}$ becomes the twisted algebra
\begin{equation}
\hat{x}_{\gamma} \, \hat{x}_{\gamma'} = (-1)^{\langle \gamma , \gamma' \rangle} \hat{x}_{\gamma + \gamma'} \ .
\end{equation}
Crossing a BPS wall can be seen as the action on the generating function
\begin{equation}
F \left( u, L_\zeta , \{ \hat{x}_{\gamma} \} ; q= -1 \right) = \sum_{\gamma} \underline{\overline{\Omega}} (u,L_\zeta , \gamma ; q= -1) \ \hat{x}_{\gamma}
\end{equation}
of the diffeomorphism
\begin{equation} \label{wchatx}
\hat{x}_{\gamma} \longrightarrow \hat{x}_{\gamma} \ (1 - \hat{x}_{\gamma_h})^{\langle \gamma_h , \gamma \rangle \Omega (\gamma_h)} \equiv \mathcal{K}_{\gamma_h}^{- \Omega(\gamma_h)} (\hat{x}_{\gamma}) \ .
\end{equation}

In this paper we will mainly take an algebraic perspective from the point of view of the BPS quivers. However the framed BPS degeneracies play a very important role in the geometrical description of the BPS sector. When  the theory is compactified on  $\real^3 \times S^1_R$, with the line defect wrapped around the circle, the path integral with the line defect insertion is a trace
\begin{equation} \label{deftraceL}
\langle L_{\zeta} \rangle = \Tr_{\cH_{u,L_{\zeta}}} \ (-1)^F \ \e^{- 2 \pi R H} \ \e^{\ii \theta \cdot \cQ} \ \sigma(\cQ) = \sum_{\gamma} \underline{\overline{\Omega}} (L_{\zeta} , \gamma) \ \mathcal{Y}_{\gamma} \ ,
\end{equation}
where $\theta$ parametrizes the holonomy of the electromagnetic field on $S^1_R$, $\cQ$ is the charge operator, $H$ the hamiltonian and $\sigma (\cQ)$ a certain sign \cite{Gaiotto:2010be}. In particular $\langle L_\zeta \rangle$ is an holomorphic function on $\cM$ which can be expanded into a set of universal functions $\cY_{\gamma}$ with coefficients given by the framed degeneracies. The functions $\mathcal{Y}_{\gamma}$ are a series of Darboux coordinates and satisfy the twisted group algebra
\begin{equation}
\mathcal{Y}_{\gamma} \ \mathcal{Y}_{\gamma'} = (-1)^{\langle \gamma , \gamma' \rangle} \ \mathcal{Y}_{\gamma + \gamma'} \ .
\end{equation}
When the theory is considered as a three dimensional sigma model $\real^3 \longrightarrow \cM$  with $\cN=4$ supersymmetry, the natural symplectic form constructed out of the $\mathcal{Y}_{\gamma}$ gives the exact hyperK\"ahler metric on the moduli space $\cM$ \cite{Gaiotto:2008cd}. One can be more general and define
\begin{equation} \label{deftraceL'}
\langle L_{\zeta} \rangle_y = \Tr_{\cH_{u,L_{\zeta}}} \ (-1)^F \ \e^{- 2 \pi R H} \  (-y)^{\mathcal{J}_3}  \ \e^{\ii \theta \cdot \cQ} \ \sigma(\cQ) \ .
\end{equation}
In the case $q=+1$ we have
\begin{equation}
\langle L_{\zeta} \rangle' = \Tr_{\cH_{u,L_{\zeta}}} \ (-1)^{2 I_3} \ \e^{- 2 \pi R H} \ \e^{\ii \theta \cdot \cQ} \ \sigma(\cQ) = \sum_{\gamma} \ \underline{\overline{\Omega}} (L_{\zeta} , \gamma , q=1) \ \tilde{\mathcal{Y}}_{\gamma} \ .
\end{equation}
Where now the coordinates obey
\begin{equation}
\tilde{\mathcal{Y}}_{\gamma} \ \tilde{\mathcal{Y}}_{\gamma'} = \tilde{\mathcal{Y}}_{\gamma + \gamma'} \ ,
\end{equation}
and span a certain hyperk\"ahler manifold conjecturally identified with $\cM$. The functions $\mathcal{Y}_{\gamma}$ are discontinuous across the BPS walls
\begin{equation} \label{wcY}
\mathcal{Y}_{\gamma} \longrightarrow   \mathcal{K}_{\gamma_h}^{\Omega(\gamma_h)} (\mathcal{Y}_{\gamma}) \ ,
\end{equation}
and the same property holds for the functions $\mathcal{\tilde{Y}}_{\gamma}$. Note that the transformation (\ref{wcY}) is precisely opposite to (\ref{wchatx}). As a consequence the indices $\langle L_{\zeta} \rangle$ and $\langle L_{\zeta} \rangle'$ are \textit{invariant} upon crossing the walls $W (\gamma_h)$. 

In the following we will introduce formal generating functions $\scrL$ which are similar to the $\langle L \rangle'$ but more algebraic in nature. The invariance under wall-crossing of $\langle L \rangle'$ will be a guidance to define the functions $\scrL$.

\subsection{Line defects and laminations}

For theories of class $\cS [\mathfrak{su}(2)]$ which arise by compactifying the superconformal $(2,0)$ theory on a surface $\cC$, line defects have an elegant description in terms of paths on $\cC$. In \cite{Gaiotto:2010be} these paths were conjecturally identified with \textit{laminations}, generalizing the results of \cite{Drukker:2009tz}. These are union of curves $\mathcal{P}_i$ on $\mathcal{C}$ (if $\cC$ has irregular punctures, each one is replaces with a boundary with marked points, corresponding to the Stokes sectors, as explained in \ref{BPStriang}) which are non self-intersecting and mutually non intersecting, considered up to isotopy. These curves can be either closed or open; open curves end on boundary arcs and are generic in the case where the quadratic differential on $\cC$ has irregular singularities. These curves are subjected to a series of conditions. In the case the curve $\cP$ is closed:
\begin{itemize}
\item if it surrounds a regular puncture $n$, it carries an integer weight  $\mathsf{wt} (\cP) \in \zed$ (positive or negative);
\item all other closed curves carry nontrivial irreducible representations of $SL(2,\complex)$.
\end{itemize}
On the other hand the curve $\cP$ can also be open, with both ends at a boundary segment. For example the boundary segment might be delimited by two triangulation edges which represent generic trajectories in two distinct Stokes sectors. Within a boundary segment there will be the intersection point between the boundary and a Stokes ray. Its specific location is immaterial. We call this point a \textit{special} point (not to be confused with a marked point from which an edge of the triangulation can depart\footnote{To compare with the language of \cite{Gaiotto:2010be} what we call special points correspond to Stokes rays emerging from irregular singularities, while what we call marked points correspond to anti-Stokes rays which are the vertices of the triangulation.}). Then
\begin{itemize}
\item an open curve $\mathcal{P}_i$ carries a positive integral weight $\mathsf{wt} (\cP_i) = k_i \in \zed_{>0}$, unless it can be retracted to a boundary segment containing precisely a special point. In the latter case we say that $\cP_i$ is a \textit{special curve} and its weight is allowed to be negative;
\item the sum of all weights $\sum_i \mathsf{wt} (\cP_i)$ over all the paths $\cP_i$ ending on the same boundary segment $b$ must vanish. 
\end{itemize}
Finally there are two isotopy conditions:
\begin{itemize}
\item two isotopic curves $\cP_i$ and $\cP_j$ with weights $\mathsf{wt} (\cP_i) = k_i$ and $\mathsf{wt} (\cP_j) = k_j$ can be replaced by a single curve $\cP_l$ with weight $\mathsf{wt} (\cP_l)= k_l = k_i  + k_j$;
\item a closed curve which is contractible or an open curve which can be contracted to a boundary segment without special points, are considered trivial and removed from the lamination.
\end{itemize}
Laminations defined according to these rules are similar, but slightly different, to the integral bounded measured laminations introduced by Fock and Goncharov \cite{FG1} in the study of higher Teichm\"uller theory. By partial abuse of language we will denote the space of laminations $\cL$ obeying these conditions with $\scrT ( \cC_{g,n,b, c_i} )$ or simply $\scrT (\cC)$. It will be useful in the following to consider a subspace of $\scrT (\cC)$, consisting of laminations $\cL$ where closed curves surrounding a regular puncture are not allowed (since in this case the formalism of Section \ref{framedBPS} cannot be applied); we will denote this space with  $\scrT_0 ( \cC_{g,n,b, c_i} )$ or simply $\scrT_0 (\cC)$.

Given a certain defect described by a lamination $\cL \in \scrT (\cC)$, the twisted trace on $\real^3 \times S^1_R$
\begin{equation}
\langle L_{\zeta} \rangle' = \sum_{\gamma} \ \underline{\overline{\Omega}} (L_{\zeta} , \gamma , q=1) \ \tilde{\mathcal{Y}}_{\gamma}
\end{equation}
can be computed via the \textit{traffic rules} algorithm \cite{Gaiotto:2010be}. One associates the coordinates $\tilde{\mathcal{Y}}_{\gamma}$ to each internal edge $E$ of the triangulation. Then one proceeds in giving local rules for the expansion of each curve in $\cL$ in the coordinates $\mathcal{Y}_{\gamma}$. The local rules associate to a curve in $\cL$ an $SL(2,\complex)$ matrix for each intersection between the curve and an internal edge, multiplied by certain  matrices if the path between two subsequent edges leans to the left or to the right.

In the following Section we will give an equivalent description of these rules in terms of certain paths on the BPS quiver in purely algebraic terms.

\section{BPS quivers and line defects} \label{extended}

In this paper we will establish two algebraic approaches to line defects. In the first approach, discussed in this Section, we associate the defect with a certain path on the quiver and give rules to compute its framed spectrum. This path is an element of the Leavitt path algebra associated to the extended BPS quiver. By mapping a Leavitt path algebra element into an appropriately defined product of $SL(2 , \complex)$ matrices, one recovers the framed BPS spectrum. In Section \ref{framedBPS} we will discuss a second approach, where we represent the defect via a framing of the BPS quiver associated with the lamination.

\subsection{Laminations and extended quivers}

As we have discussed, line defects correspond to collections of paths on triangulated curves $\cC$. Since the information on the triangulation is equivalently packaged into a BPS quiver, it is natural to look for a set of rules which translate the paths on the curve to paths on the quiver. A path on $\cC$ will cross various edges of a triangulation $T (\cC)$. Locally the sequence of edges it crosses is determined by its direction, left or right, after crossing a given edge. One can therefore expect that a path can be described by a sequence of nodes (which correspond to edges of $T (\cC)$) and arrows (whose orientation is determined by the relative orientation between two edges of a same triangle) of the BPS quiver $\sfQ$ constructed from $T (\cC)$. Indeed we will now show how to describe a lamination on $\cC$ in terms of algebraic data on the extended BPS quiver.

The extended quiver is simply defined by adding vertices associated to the boundary segments. These do not correspond to BPS states and are introduced only as a bookkeeping device to keep track of the boundaries. The extended quiver is defined precisely via the same set of rules as the BPS quiver. That is, the set of arrows between nodes is determined by the ordering of the edges of the triangles in $T (\cC)$. We will denote by $\mathsf{B}_0$ the set of nodes associated with the boundary segments and by $\mathsf{B}_1$ the set of arrows connecting this set with $\sfQ_0$ (that is the arrows ending on or departing from the boundary nodes). This set excludes arrows between the boundary nodes. The extended quiver is then the quiver $\widetilde{\sfQ}$ whose set of nodes is $\sfQ_0 \cup \mathsf{B}_0$ and whose set of arrows is $\sfQ_1 \cup \mathsf{B}_1$. Furthermore we will assume for simplicity that $T (\cC)$ does not contain self-folded triangles; a triangulation can always be reduced to this form by appropriately flipping a finite number of edges \cite{FT1}. 

Consider for example the Argyres-Douglas theory $A_2$. This theory is obtained by compactifying the $(2,0)$  superconformal theory on a sphere $\cC = \PP^1$ with a single irregular singularity of order $p=7$. Equivalently 5 Stokes line emerge from the singularity. The triangulation has therefore two edges, and 5 boundary segments. The extended quiver looks like 
\begin{equation}
\begin{matrix}
\xymatrix@C=10mm{ \circ   & & & & \circ \ar[dl] \\
&   \ \bullet \, \gamma_2 \ar[ul] \ar[dr] \  & & \ \bullet \, \gamma_1 \ar[ll]  \ar[dr] & &
  \\ \circ \ar[ur] & & \circ \ar[ur]  &  & \circ 
}
\end{matrix} \ ,
\end{equation}
where we have denoted with empty dots the nodes corresponding to boundary segments. We can easily generalize this result to generic Argyres-Douglas theories of type $A_n$. In these theories there is a canonical sink-source form of the BPS quiver. The associated extended BPS quiver is
\begin{equation}
\xymatrix@C=2.5mm{ \circ & &   & & \circ \ar[dl] & & & & \circ \ar[dl] & & \cdots  & & \circ \ar[dl] & & \circ  \ar[dl] \\
&  \ \bullet  \, \gamma_{i_1} \ar[ul] \ar[dr] & &  \ \bullet \, \gamma_{i_2} \ar[ll]  \ar[rr] &  & \ \bullet \, \gamma_{i_3} \ar[ul] \ar[dr] & & \ \bullet \, \gamma_{i_4} \ar[ll] & & \cdots & & \ \bullet \, \gamma_{i_{n-1}} \ar[rr]& & \ \bullet  \, \gamma_{i_{n}} \ar[ul] \ar[dr]  \\ 
   \circ \ar[ur] & & \circ  \ar[ur] &  & & & \circ \ar[ur] & &   \cdots &  &  \circ \ar[ur] & &  & & \circ 
}
\end{equation}

Before discussing the general case we shall now consider line operators on Argyres-Douglas theories of type $A_n$. We want to write laminations on $\cC$ as certain paths on the extended BPS quiver. Which kind of paths are allowed? For example it is easy to see that such a path must at least go through a node of the (non-extended) BPS quiver, in any direction, otherwise it would be contractible to a boundary segment without special points. Secondly since we only have open paths, each path must start at a boundary node and end at a (in this case different, since there are no regular punctures) boundary node. Note that, except for the beginning and ending nodes, all the nodes met by the path must be ordinary BPS nodes corresponding to edges of the triangulation. 

We have however to impose the condition that the sum of the weights of each path in the lamination vanish at each boundary segment. Recall that a boundary segment is divided in two Stokes sectors by a special point. The same condition can be restated as follow: firstly we remove any special curve surrounding the special point. Then the number of paths ending on the left of the special point must be equal to the number of paths ending on its right. To mimic this behavior we will draw \textit{shifted paths} on the quiver, which can start at the left or at the right of a boundary node. Note that this is only a graphically convenient way of drawing the paths to keep track of the number of intersections with the boundary segment at the left and at the right of the special point; but the paths themselves are \textit{always} to be understood as a sequence of  arrows and nodes on the extended BPS quiver. Finally we require the shifted paths to be non intersecting (with this we mean that it is possible to draw the shifted paths on the quiver in such a way that they do not intersect, eventually by rearranging the drawing of the quiver). 

\begin{figure}
\centering
\def\svgwidth{10cm}
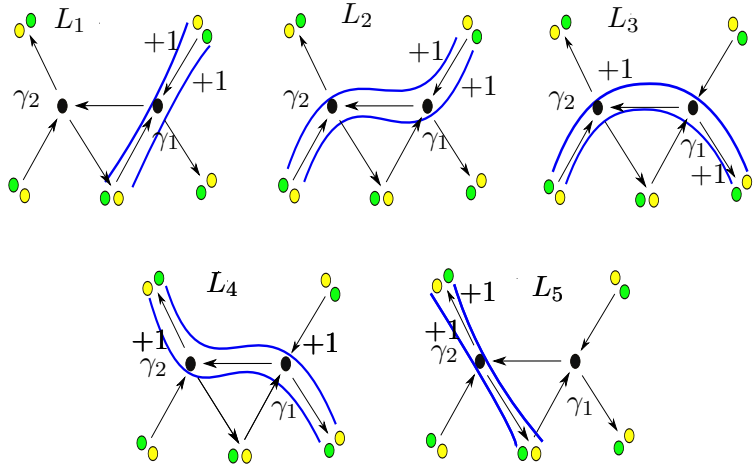
\caption{The five elementary operators in the Argyres-Douglas $A_2$ theory, drawn as paths on the extended BPS quiver}
\label{laminA2}
\end{figure}

To make it easier to draw the shifted paths, we will ``split" the boundary nodes in two, corresponding to the two Stokes sectors on the boundary segment separated by a special point. Indeed any boundary segment is delimited by two marked points (not necessarily distinct) from which two generic trajectories emanate, which can form two edges of the triangulation. Each generic trajectory is a representative of a WKB curve within a Stokes sector. Therefore any boundary segment will be split in two by the special point delimiting the two Stokes sectors. We will conventionally draw the two split nodes as a green one and a yellow one. This is just a convention to distinguish the Stokes sectors on the left and on the right of a special point lying on a boundary segment; sectors with the same color have \textit{not} to be identified. To each shifted path one can associate a sequence of nodes and arrows (pointing in either direction) in the extended quiver; in addition for the open paths one also has a starting (green or yellow) and an ending (yellow or green) point corresponding to boundary segments. We will continue to use the notation $\tilde{\sfQ}$ for the extended quiver with the split nodes, confident that no confusion should arise. In the following we will associated to each path thus constructed an appropriate product of matrices. These set of rules are enough to transpose the concept of lamination on a triangulated surface to a collection of (shifted, non intersecting) paths on an extended BPS quiver. For example the simple line operators in the Argyres-Douglas $A_2$ case (those operators which cannot be written as a sum of other line operators) are drawn in Figure \ref{laminA2}. We will make these concepts more precise in Section \ref{Leavitt} where we will describe the shifted paths as elements of certain algebras.

\begin{figure}[htbp]
\centering
\begin{minipage}[b]{0.4\textwidth}
\centering
\def\svgwidth{6cm}
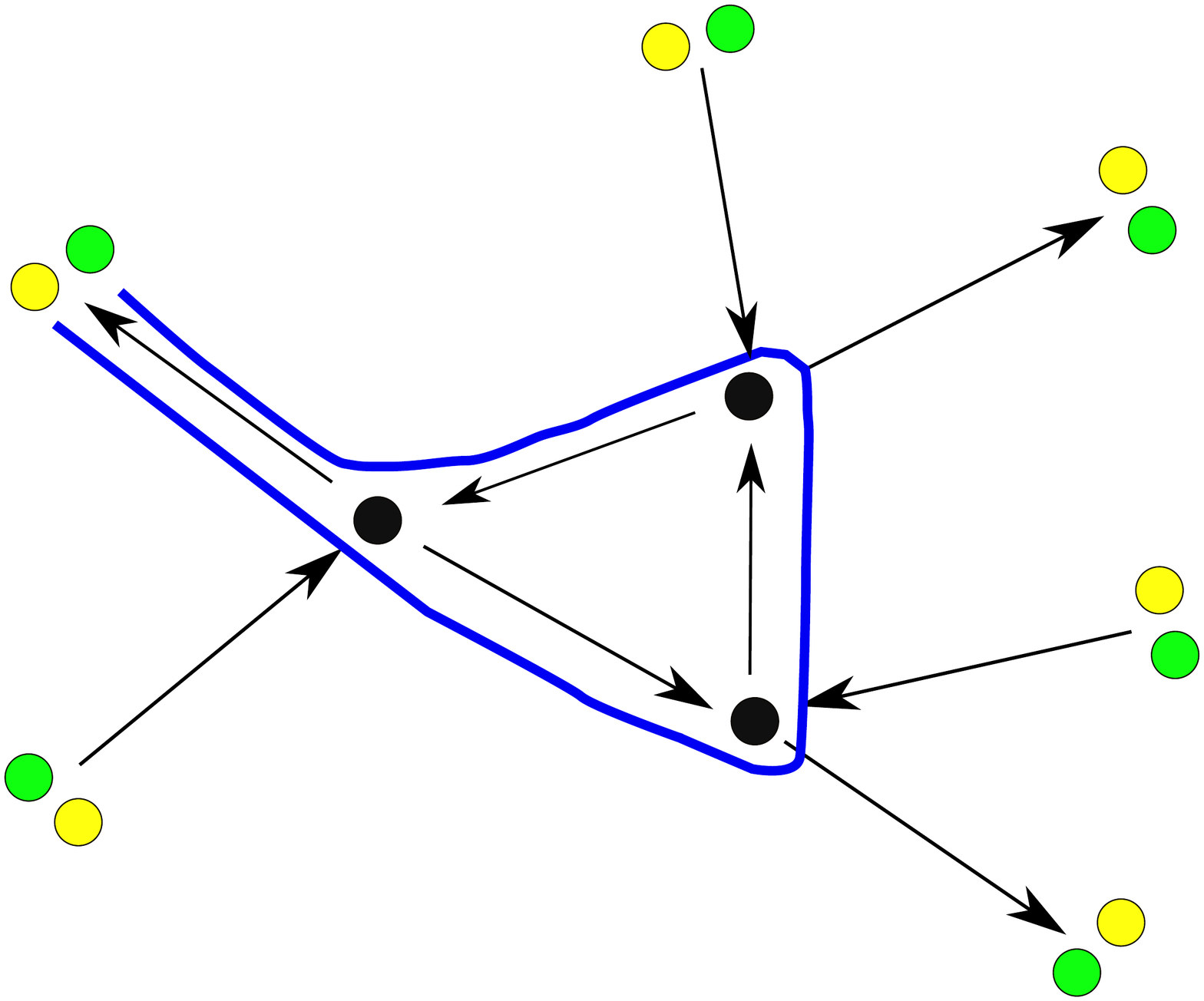
\caption{A non acceptable path in the $A_3$ Argyres-Douglas theory; the corresponding lamination is contractible}
\label{nogoodA3}
\end{minipage} \qquad
\begin{minipage}[b]{0.4\textwidth}
\centering
\def\svgwidth{6cm}
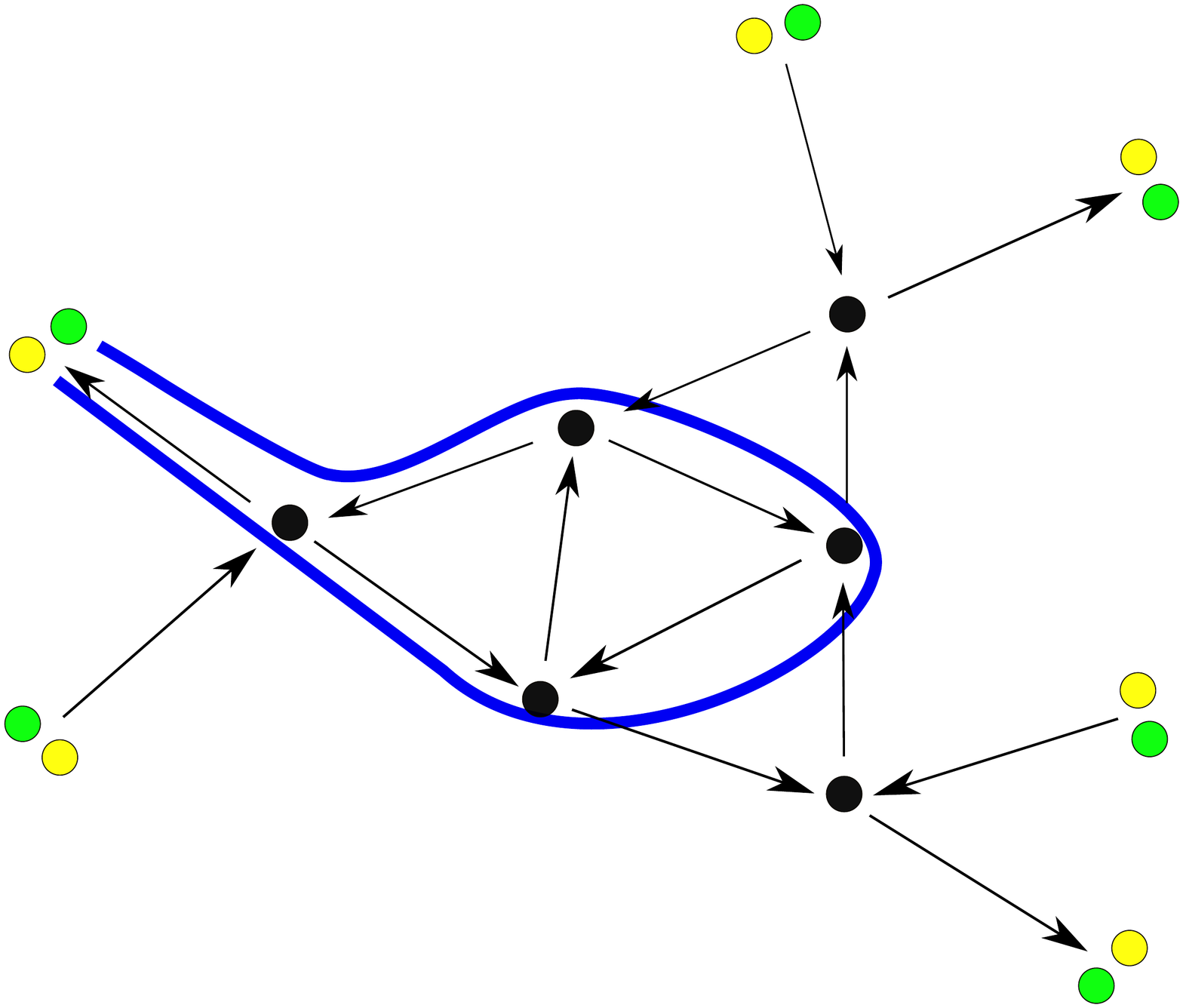
\caption{An acceptable path in the $D_6$ Argyres-Douglas theory. The corresponding open lamination is not contractible since it encircles a puncture.}
\label{D6loop}
\end{minipage}
\end{figure}

We can also have paths starting from a boundary node and ending at the same boundary node. This can happen if there is a loop inside the quiver. For example we can consider the $A_3$ theory which arises from the triangulation of a sphere with an irregular puncture from which six Stokes lines depart. The extended BPS quiver is depicted in Figure \ref{nogoodA3} along with a path. 
We seem to have found a paradox: the lamination on the curve corresponding to the path is clearly trivial; since there are no punctures in the interior, the path is contractible to the boundary segment, where it cancels the special curve. On the other hand the path on the quiver seems to be non trivial. The resolution of this paradox is that we have overlooked the rules to construct paths on the extended quiver. Consider a curve $\cP$ on $\cC$. As we follow it around, it crosses edges of the triangulation $T$ traveling from faces to faces. Consider for example to neighbouring faces $F_1$ and $F_2$ of the triangulation, sharing a common edge $E$. The path $\cP$ crossing $E$ coming from $F_1$, no matter which other edge of $F_1$ it had crossed before, it must now cross one of the two remaining edges of $F_2$. In other words, it is not allowed to ``go back" to $F_1$, simply because it would be isotopic to a path which does not crosses $E$ at all. A path $\cP$ going back to $F_1$ will have necessarily to cross an edge of $F_1$ at least for the third time. If we now recall that edges in the triangulation correspond to nodes of the quiver, we conclude that a path on the extended quiver is not allowed to touch {\it consecutively} more that two nodes of the same triangle representing a face of the triangulation. The last specification is important. Indeed a triangular loop in the BPS quiver needs not necessarily correspond to a face of the triangulation. It can also correspond to a regular puncture, out of which three edges emerge. Consider for example a theory of type $D$, which arise from a sphere with a regular puncture and an irregular one where the quadratic differential has a pole of order $8$. This theory has again $6$ Stokes lines emerging from the irregular puncture, but now the ideal triangulation consists of $6$ arcs. Take the triangulation showed in figure \ref{D6loop}. Because of the puncture the drawn path is non trivial and cannot be contracted to a boundary segment. We therefore conclude that a path on the extended quiver is not allowed to touch consecutively more than two nodes of the same triangular loop on the quiver, unless that triangle corresponds to a regular puncture (or a boundary component with marked points).

Similar arguments also hold for closed laminations. Recall that these carry an integral weight (if surrounding a regular puncture) or a representation of $SL(2 ,  \complex)$ (otherwise). Also closed laminations can be lifted to paths on the quiver, which now start from an internal node and end at the same node without touching the boundary nodes.

\subsection{Leavitt path algebras} \label{Leavitt}

To summarize given any triangulation $T$ of a bordered Riemann surface $\cC$ we can construct the corresponding extended BPS quiver $\tilde{\sfQ}$. Then any curve in a lamination $\cL$ on $\cC$ lifts to a certain path on the extended quiver. We have been loosely talking about ``paths" on the extended quiver. For example the paths in $L_1$ in Figure \ref{laminA2} are given by the formal strings of elements
\begin{eqnarray}
\sfe_{G} \, \sfa_{G \gamma_1} \, \sfe_{\gamma_1} \sfa_{Y \, \gamma_1}^{-1} \, \sfe_{Y} \ , \\
\sfe_{Y} \, \sfa_{Y \gamma_1} \, \sfe_{\gamma_1} \sfa_{G \, \gamma_1}^{-1} \, \sfe_{G} \ ,
\end{eqnarray}
in the notation of Section \ref{BPSquivers}. These paths are \textit{not} elements of the path algebra $\sfA_{\tilde{\sfQ}} = \complex \tilde{\sfQ}$ of the extended quiver $\tilde{\sfQ}$. Indeed both paths are identically zero in the path algebra, since due to the direction of the arrows there is no concatenation possible. Rather these paths are elements of the algebra obtained from $\sfA_{\tilde{\sfQ}}$ by adding the formal inverses to the arrows (its ``localization"). We will denote the algebra with $\sfL \sfA_{\tilde{\sfQ}}$. This algebra is rather similar to what is called Leavitt path algebra in the literature \cite{abrams} and reviewed for example in \cite{goodearl}. Ordinary Leavitt path algebras are not defined in the extended quiver and in particular do not have ``colored" nodes. 

Recall that the path algebra $\sfA_{\tilde{\sfQ}} = \complex \tilde{\sfQ}$ is defined via concatenation of paths. If $s,t \, : \, \tilde{\sfQ}_1 \longrightarrow \tilde{\sfQ}_0$ are the functions which to an arrow in $\tilde{\sfQ}_1$ associate its starting and terminating vertices, then the product of two paths $\mathsf{p_1} = \mathsf{a_1 \cdots a_n}$ and $\mathsf{p_2} = \mathsf{a_{n+1} \cdots a_{m}}$ is defined as $\mathsf{p_1 p_2} = \mathsf{a_1 \cdots a_n a_{n+1} \cdots a_m}$ if $t (\mathsf{a_n}) = s (\mathsf{a_{n+1}})$ and zero otherwise (we write path composition from the left to the right). Consider now the set of formal inverse arrows $\{ \sfa^{-1} \, \vert \, \sfa \in \tilde{\sfQ}_1 \}$, sometimes called \textit{ghost paths}. The Leavitt path algebra $\sfL \sfA_{\tilde{\sfQ}}$ of $\sfA_{\tilde{\sfQ}}$ is the algebra with generators $\{ \sfe_i , \sfa , \sfa^{-1} \, \vert \, i \in \tilde{\sfQ}_0 , \sfa \in \tilde{\sfQ}_1 \}$ and with coefficients in $\complex$, such that the following relations hold
\begin{itemize} 
\item $\sfe_i \, \sfe_j = \sfe_i \, \delta_{ij}$ for all $i \in \tilde{\sfQ}_0$,
\item $\sfa \, \sfe_{t(\sfa)} = \sfa = \sfe_{s (\sfa)} \sfa$ and $\sfa^{-1} \, \sfe_{s(\sfa)} = \sfa^{-1} = \sfe_{t (\sfa)} \sfa^{-1}$ for all $\sfa \in \tilde{\sfQ}_1$,
\item $\sfa^{-1} \, \sfb = \delta_{\sfa, \sfb} \, \sfe_{t(\sfa)}$ for all $\sfa , \sfb \in \tilde{\sfQ}_1$,
\item $\sum_{\sfa \in \tilde{\sfQ}_1 \, \vert \, s(\sfa) = i } \, \sfa \, \sfa^{-1} = \sfe_i$ for every vertex $i \in \tilde{\sfQ}$ which is not a sink (since for a sink the map $s$ is trivial)\footnote{In the mathematical literature one also requires that there are not infinitely many edges emitted from $i$, but since we are only considering finite quivers this condition would be redundant.}.
\end{itemize}
In plain words we now have a formal tool to compose arrows even if they are oriented in the wrong way. Our Leavitt path algebra is slightly different from the above definition as we double the vertex set in $\sfB_0$. Since the above defining relations extend immediately (by assuming that the two trivial paths of different colors at each boundary node are orthogonal), we will continue to call our algebras $\sfL \sfA_{\tilde{\sfQ}}$, Leavitt path algebras. 

To summarize to a collection of curves $\cP_i$, $i=1,\dots,n$ on $\cC$ forming a lamination $\cL \in \scrT (\cC)$ we can associate a collection of elements of the Leavitt path algebra $\sfL \in \sfL \sfA_{\tilde{\sfQ}}^n$ (although we will often omit to specify the length $n$ of the vectors, hoping that no confusion will arise). To be more precise, since each curve $\cP_i$ comes with a weight $\mathsf{wt} (\cP_i)$ we are really talking about collections of the form $(\mathsf{wt} (\cL) , \sfL)$. We will however not use  this notation in general to avoid cluttering the formulas, and mostly consider curves with weight one. This connection between Teichm\"uller theory and Leavitt path algebra does not seem to have been noticed in the literature, and we believe deserve further investigation. Now we will show how to extract the framed BPS spectrum from elements of Leavitt path algebras.

Note however that, while given $\cL \in \scrT (\cC)$ we can easily write down the corresponding $\sfL \in \sfL \sfA_{\tilde{\sfQ}}$, the correspondence is not one to one. In general we do not have a set of algebraic necessary and sufficient conditions to guarantee that a Leavitt path algebra element corresponds to a physical line defect. 
 
\subsection{Matrix formulae from paths} \label{matrix}

The quantum theories we are studying come equipped with a certain lattice of charges $\Gamma$. As we have seen it is natural to consider the quantum torus $\torus_{\Gamma}$ associated with this lattice, which is generated by the formal variables $\sfX_{\gamma}$. We can consider a classical commutative (formal) limit $q \longrightarrow +1$ of these variables and introduce variables $x_{\gamma}$, which are element of the group algebra of $\Gamma$, such that
\begin{equation}
x_{\gamma_i} \ x_{\gamma_j} = x_{\gamma_i + \gamma_j} \ .
\end{equation}
We think of the set $\{ x_{\gamma} \}$ as formal variables, which could be identified with the functions $\{Ê\tilde{\cY}_{\gamma} \}$ of Section \ref{linedef} or with the  quiver variables $\{ \tilde{\sfy}_{\gamma} \}$ of Section \ref{framedBPS}. We will describe now a natural set of rules to expand the vacuum expectation value of a line operator onto these variables and thus obtain the framed BPS spectrum. These rules are a direct translation of the traffic rules of \cite{Gaiotto:2009hg,Gaiotto:2010be,FG1}; from our point of view they allow to associate to an element $\mathsf{l} \in \sfL \sfA_{\tilde{\sfQ}}$ a certain product of $SL (2 ,\complex)$ matrices. Consider an element of the algebra $\sfL \sfA_{\tilde{\sfQ}}$ of the extended BPS quiver which correspond to a lamination $\cL \in \scrT (\cC)$. In general $\cL = \left( \cP_1 , \dots , \cP_n \right)$ will be a collection of curves $\cP_i$ on $\cC$ and the corresponding $\sfL = \left( \mathsf{I}_1 , \dots , \mathsf{I}_n \right)$ is a collection of Leavitt paths $\mathsf{I}_i$ in $\sfL \sfA_{\tilde{\sfQ}}$. For simplicity we will write explicitly the trivial paths $\sfe_{\gamma_i}$ associated with a basis of BPS states $\{ \gamma_i \}$. This is not necessary if one keeps carefully track of the functions $s , t \, ; \, \tilde{\sfQ}_1 \longrightarrow \tilde{\sfQ}_0$, but by writing down the trivial paths explicitly we find a set of matrix rules notationally less involved. Quite generically we can consider a certain node lying on the path and then continue following the path in any direction we like. The path will cross other nodes moving along the arrows of the quiver, in the same or in the opposite direction, corresponding to elements $\sfa \in \sfQ_1 \cup \sfB_1$ or their formal inverses $\sfa^{-1}$. We define the map
\begin{eqnarray}
\mathsf{m} \ : \ \sfL \sfA_{\tilde{\sfQ}} & \longrightarrow & SL (2 , \complex) \cr
\sfe_{\gamma} & \longrightarrow & M_{\gamma} = \left( \begin{matrix} \sqrt{x_{\gamma}} & 0 \\ 0 & \frac{1}{\sqrt{x_{\gamma}}} \end{matrix} \right) \ , \nonumber\\[4pt] \sfa & \longrightarrow & D =  \left( \begin{matrix}1 & 0 \\ 1 & 1 \end{matrix} \right) \ , \nonumber\\[4pt] \sfa^{-1} & \longrightarrow & U = \left( \begin{matrix} 1 & 1 \\ 0 & 1 \end{matrix}  \right) \ .
\end{eqnarray}
Furthermore for a boundary node $\sfe_i$, $i \in \sfB_0$ of either color, $\mathsf{m} (\sfe_i) = \mathrm{id}$, is the identity matrix (formally, the boundary nodes have $\gamma=0$). In particular under composition of paths, $\mathsf{m}$ sends the Leavitt product into ordinary matrix multiplication. Note that this is not a representation of the algebra.

To any path on the extended quiver we associate a rational function of the variables $\{ x_{\gamma} \}$ via the map $\mathsf{m}$ as follows. For every node corresponding to a charge $\gamma_1$ we write down the matrix $M_{\gamma_1}$. Then following the path to the next node $\gamma_2$, we write down a matrix $U$ (for upstream) if the path direction is opposite to the arrow between $\gamma_1$ and $\gamma_2$, or $D$ (for downstream)  otherwise. A path going through the nodes $\gamma_1$, $\gamma_2$ and $\gamma_3$ (in this order) will correspond to a matrix of the form
\begin{equation}
\cdots \ M_{\gamma_1} \ D \ M_{\gamma_2} \ U \ M_{\gamma_3} \ \cdots
\end{equation}
if $\langle \gamma_1 , \gamma_2 \rangle > 0$ and $\langle \gamma_2 , \gamma_3 \rangle < 0$. Now there are two possibilities, the path is open or it is closed. If it is closed, then $\mathsf{p} = \sfe_{\gamma} \cdots \sfe_{\gamma}$ for some $\gamma$. In this case we omit one of the $\sfe_{\gamma}$, or more precisely use $\sfe_i \sfe_i = \sfe_i $ for every $i \in \tilde{\sfQ}_0$, and take the trace of the product of matrices (which is already cyclic). This corresponds to the fundamental representation of $SL (2,\complex)$. If the path is open we have to specify boundary condition at its endpoints. In this case we simply associate with a boundary node the identity matrix and with the arrow between the boundary node and an internal node, a matrix $U$ or $D$ according to the aforementioned rules. In this way we have associated to the open path an $SL(2 , \complex)$ matrix. Finally, depending on the boundary endpoints of the path, we pick one element of this matrix by using the projector $P_{ij}$ which extract the $(i,j)$th element of an $SL(2,\complex)$ matrix. Here the first index $i$ is associated with the starting node of a path, while the second index $j$ corresponds to the ending node. According to the path we are considering, each index can be green ($G$) or yellow ($Y$). We pick conventions in which a $Y$ entry corresponds to the first row or column of the $SL(2,\complex)$ matrix while a $G$ entry to the second row or column. Note that the condition that the weights of the lamination vanish at each boundary segment implies that for each path what we call starting point and ending point is just a matter of conventions. Let us apply now these rules to the paths of Figure \ref{laminA2}. It is easy to see that the Leavitt path algebra elements associated with those paths are
\begin{eqnarray}
 \sfL_1 &=& \left( \sfe_{G} \, \sfa_{G \gamma_1} \, \sfe_{\gamma_1} \, \sfa_{\gamma_1 Y}^{-1} \, \sfe_Y  \, , \,   \sfe_{Y} \, \sfa_{Y \gamma_1} \, \sfe_{\gamma_1} \, \sfa_{\gamma_1 G}^{-1} \, \sfe_G \right) \ ,
\cr
\sfL_2 &=& \left( \sfe_G \, \sfa_{G \gamma_2} \, \sfe_{\gamma_2}  \, \sfa_{\gamma_1 \gamma_2}^{-1} \, \sfe_{\gamma_1} \, \sfa_{Y \gamma_1}^{-1} \, \sfe_Y \, , \, \sfe_Y \, \sfa_{Y \gamma_2} \, \sfe_{\gamma_2}  \, \sfa_{\gamma_1 \gamma_2}^{-1} \, \sfe_{\gamma_1} \, \sfa_{G \gamma_1}^{-1}  \, \sfe_G
\right) \ ,
\cr
\sfL_3  &=& \left( \sfe_G \, \sfa_{G \gamma_2} \, \sfe_{\gamma_2} \, \sfa_{\gamma_1 \gamma_2}^{-1} \, \sfe_{\gamma_1} \, \sfa_{\gamma_1 Y} \, \sfe_Y \, , \,
\sfe_Y \sfa_{Y \gamma_2} \, \sfe_{\gamma_2} \, \sfa_{\gamma_1 \gamma_2}^{-1}  \, \sfe_{\gamma_1} \, \sfa_{\gamma_1 G}  \, \sfe_G 
\right) \ ,
\cr
\sfL_4  &=& \left( \sfe_G \, \sfa_{\gamma_2 G }^{-1} \, \sfe_{\gamma_2} \, \sfa_{\gamma_1 \gamma_2}^{-1} \, \sfe_{\gamma_1} \, \sfa_{\gamma_1 Y} \, \sfe_Y \, , \,
 \sfe_Y \, \sfa_{\gamma_2 Y }^{-1} \, \sfe_{\gamma_2} \, \sfa_{\gamma_1 \gamma_2}^{-1} \, \sfe_{\gamma_1} \, \sfa_{\gamma_1 G} \, \sfe_G
\right) \ ,
\cr
\sfL_5  &=& \left( \sfe_G \, \sfa_{\gamma_2 G}^{-1} \, \sfe_{\gamma_2} \, \sfa_{\gamma_2 Y} \, \sfe_Y \, , \,
 \sfe_Y \, \sfa_{\gamma_2 Y}^{-1} \, \sfe_{\gamma_2} \, \sfa_{\gamma_2 G} \, \sfe_G
\right) \ .
\end{eqnarray}
We stress again that writing the trivial paths explicitly is just a notational choice; the same rules would still hold by systematically removing them and keeping track of the starting and terminating nodes, but in a slightly less simple form. Using the aforementioned rules, we obtain the corresponding generating functions (\ref{physgen}) for $q = 1$
\begin{eqnarray}
 F(L_1) &=& P_{GY} \left( D \, M_{\gamma_1} \, U \right) \ P_{YG} \left( D \, M_{\gamma_1} \, U \right) = x_{\gamma_1} \ ,
\cr
 F(L_2) &=& P_{GY} \left( D \, M_{\gamma_2} \, U \, M_{\gamma_1} \, U \right) \ P_{YG} \left(D \, M_{\gamma_2} \, U \, M_{\gamma_1} \, U \right) = x_{\gamma_2} + x_{\gamma_1 + \gamma_2}  \ ,
\cr
 F(L_3)  &=& P_{GY} \left( D \, M_{\gamma_2} \, U \, M_{\gamma_1} \, D  \right) \  P_{YG} \left( D \, M_{\gamma_2} \, U \, M_{\gamma_1} \, D  \right) = x_{-\gamma_1} + x_{\gamma_2} + x_{\gamma_2 - \gamma_1} \ ,
\cr
 F(L_4)  &=& P_{GY} \left( U \, M_{\gamma_2} \, U \, M_{\gamma_1} \, D \right) \    P_{YG} \left( U \, M_{\gamma_2} \, U \, M_{\gamma_1} \, D \right) = x_{- \gamma_1} + x_{-\gamma_1 - \gamma_2}  \ ,
\cr
 F(L_5)  &=& P_{GY} \left( U \, M_{\gamma_2} \, D \right) \ P_{YG} \left( U \, M_{\gamma_2} \, D \right) = x_{-\gamma_2} \ ,
\end{eqnarray}
whose framed BPS spectra agree with \cite{Gaiotto:2010be}. 

\begin{figure}[h]
\centering
\def\svgwidth{11cm}
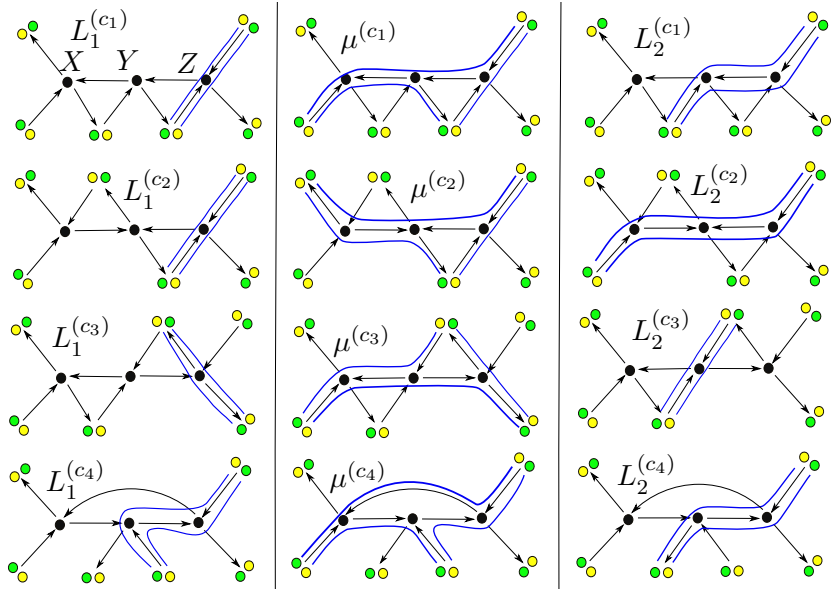
\caption{Line defects in the $A_3$ theory. The three nodes will be labeled by the variables $X$ , $Y$ and $Z$ in this order.}
\label{A3muL1L2}
\end{figure}

We end this section with a few other examples to clarify the above rules in the case of the $A_3$ Argyres-Douglas theory. We will compute the framed spectra of the three line operators illustrated in figure \ref{A3muL1L2}, in different BPS chambers. This discussion has only an illustrative purpose: we will not be careful to identify the correct BPS spectrum in each chamber and will generically label the nodes with the variables $X$, $Y$ and $Z$. This is not correct physically, since in different chambers and for different stability conditions they would correspond to different charges. 	
By using the rules just described we get, for the defect $\mu$
\begin{eqnarray} \nonumber
F \left( \mu^{c_1} \right) &=& P_{GY} \left( D \, M_X \, U \, M_Y \, U \, M_Z \, U \right) \, P_{YG} \left( D \, M_X \, U \, M_Y \, D \right) \, P_{GY} \left( D \, M_Z \, U \right) = X \, Z \ , \\[4pt] \nonumber
F \left( \mu^{c_2} \right) &=& P_{GY} \left( U \, M_X \, D \, M_Y \, U \, M_Z \, U \right) \, P_{YG} \left( U \, M_X \, D \, M_Y \, D \right) \, P_{GY} \left( D \, M_Z \, U \right) = \frac{Z}{X} \ , \\[4pt]  \nonumber
F \left( \mu^{c_3} \right) &=& P_{GY} \left( D \, M_X \, U \, M_Y \, U \right) \, P_{YG} \left( D\, M_X \, U \, M_Y \, D \, M_Z \, D \right) \, P_{YG} \left( U \, M_Z \, D \right) = \frac{X}{Z} \ , \\[4pt] \nonumber
F \left( \mu^{c_4} \right) &=& P_{GY} \left( D \, M_X \, U \, M_Z \, U \right) \, P_{GY} \left( D \, M_Z \, U \, M_Y \, U \right) \, P_{GY} \left( D \, M_Y \, U \, M_X \, U \right) = X \, Y \, Z \ .
\end{eqnarray}
Consider the path in the extended quiver associated with the line operator $\mu^{c_4}$. At first sight it would seem we have an ambiguity, since the top blue line could as well pass through the middle node $Y$. However it is not so; as explained previously since the are no regular punctures, the path is not allowed to go through more that two nodes of the same triangular loop in the quiver. Therefore what is drawn in figure \ref{A3muL1L2} is the only non zero possibility. Similarly we can write down the generating functions for $L_1$
\begin{eqnarray} \nonumber
F \left( L_1^{c_1} \right) &=& P_{GY} \left( D \, M_Z \, U \right) P_{YG} \left( D \, M_Z \, U \right) = Z \ , \\[4pt]  \nonumber
F \left( L_1^{c_2} \right) &=& P_{GY} \left( D \, M_Z \, U \right) P_{YG} \left( D \, M_Z \, U \right) = Z \ , \\[4pt]  \nonumber
F \left( L_1^{c_3} \right) &=& P_{GY} \left( U \, M_Z \, D \right) P_{YG} \left( U \, M_Z \, D \right) = \frac{1}{Z} \ , \\[4pt] 
F \left( L_1^{c_4} \right) &=& P_{YG} \left( D \, M_Y \, D \, M_Z \, U \right) P_{GY} \left( D \, M_Y \, D \, M_Z \, U \right) = Z + Y \, Z \ ,
\end{eqnarray}
and for $L_2$
\begin{eqnarray} \nonumber
F \left( L_2^{c_1} \right) &=& P_{YG} \left( D \, M_Y \, U \, M_Z \, U \right) P_{GY} \left( D \, M_Y \, U \, M_Z \, U \right) = Y (1+Z) \ , \\[4pt] \nonumber
F \left( L_2^{c_2} \right) &=& P_{YG} \left( D \, M_X \, D \, M_Y \, U \, M_Z \, U \right) P_{GY} \left( D \, M_X \, D \, M_Y \, U \, M_Z \, U \right) = (1+X) Y (1+Z) \ , \\[4pt] \nonumber
F \left( L_2^{c_3} \right) &=& P_{GY} \left( D \, M_Y \, U \right) P_{YG} \left( D \, M_Y \, U \right) = Y \ , \\[4pt]
F \left( L_2^{c_4} \right) &=& P_{GY} \left( U \,  M_Y \, D \, M_Z \, U \right)  P_{YG} \left( U \,  M_Y \, D \, M_Z \, U \right) = \frac{1 + Z + Y \, Z}{Y} \ .
\end{eqnarray}
We stress that these results are completely equivalent to the traffic rules of \cite{Gaiotto:2010be}. Given a lamination $\cL \in \scrT (\cC)$ we can construct the corresponding objects in $\sfL \sfA_{\tilde{\sfQ}}$ and then proceed as outlined above to compute the framed spectrum. Of course this procedure can become quite involved if the quiver is complicated. 

We have not addressed the inverse problem, that is which Leavitt path algebra elements corresponds to physically viable laminations. Indeed such a construction would be very useful for quiver theories which do not arise from the triangulation of a surface $\cC$. This problem is very interesting, but a complete algebraic characterization of line defects lies outside of the scope of this work (and furthermore there is no evidence that the relation between Leavitt path algebras and line defects holds for any BPS quiver).

\section{Framed BPS quivers} \label{framedBPS}

As we have seen to a BPS quiver corresponding to the triangulation of a bordered Riemann surface, we can associate an extended BPS quiver. To a lamination on the triangulated surface we can associate certain paths on the extended quiver and then compute the framed BPS spectrum using the matrix rules of Section \ref{matrix}. It can be however tricky to keep track of the paths for complicated quivers and it would be more useful to have a more systematic approach. In this Section we will introduce framed BPS quivers. These quivers have one or more framing nodes which encode all the information about the line operators. Furthermore we will see that they have nice transformation properties under quiver mutations. This fact allows us to use appropriate cluster transformations to compute the framed BPS spectrum. There exists another possibility to associate to a line defect a framed quiver, via the geometric engineering approach of \cite{Chuang:2013wt}. The two approaches should be equivalent since they correspond to string theory dual pictures, but we will not compare them further in this paper.

\subsection{Shear coordinates and framed BPS quivers}

\begin{figure}
\centering
\def\svgwidth{8cm}
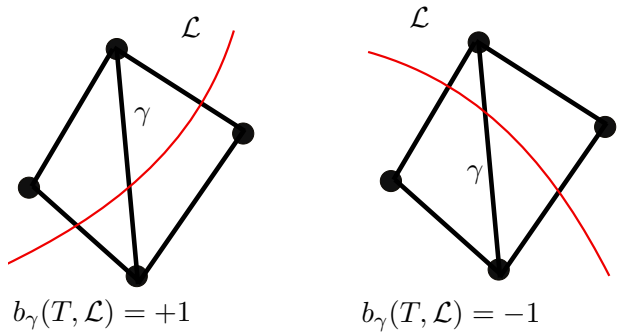
\caption{Shear coordinates on a lamination arc.}
\label{coordonlam}
\end{figure}

\begin{figure}
\centering
\def\svgwidth{8cm}
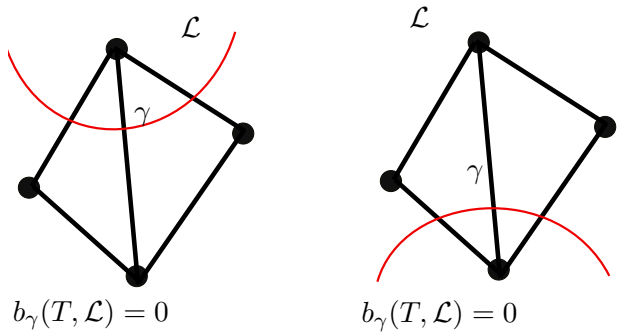
\caption{The shear coordinates associated with the intersection of these arcs with the edge $\gamma$ vanish.}
\label{coordonlam2}
\end{figure}

Given a lamination we  define the \textit{Thurston's shear coordinates} as follows. Fix a triangulation $T$ without self-folded triangles. Consider now a rectangle whose diagonal is labeled by $\gamma$. Assume now that there is a lamination $\cL$ (or the arc of a lamination) which crosses $\gamma$ as shown in figure \ref{coordonlam}. Then to the intersection of $L$ with $\gamma$ we associate an integer as follows. The shear coordinate $b_{\gamma} (T,\cL)$ of $\cL$ with respect to the edge $\gamma \in T$ is $+1$ (respectively $-1$)  if the lamination $\cL$ intersects the edge $\gamma$ in the clockwise (respectively counterclockwise) direction going towards $\gamma$ starting from \textit{each} of the two external edges of the rectangle it crosses. If more then one arc crosses $\gamma$ we extend this definition to each arc. Similarly if an arc $\cP$ has weight $\mathsf{wt} (\cP) = k$, than we count its weight as the number of intersections.
In general to a lamination  $\cL$ on $T$ we associate a set of integers $b_{\gamma} (T,\cL)$ for each edge $\gamma \in T$, defined as the sum of contributions for each intersection of $\cL$ with $\gamma$. Note that the coordinates associated with the arcs of figure \ref{coordonlam2} are identically zero.
These coordinates can be extended to triangulations with self-folded triangles \cite{FT2}, as briefly reviewed in the Appendix \ref{selffolded}.

Now given an arbitrary lamination $\cL$ and a triangulation $T$ we can define a generalized exchange matrix $B^{\cL}$. We simply add an extra row, whose entries are the shear coordinates $b_{\gamma} (T,\cL)$ with respect to the edges of the triangulation. In formulas, assuming the triangulation has $n$ edges $\gamma_i$
\begin{equation}
B^L_{ab} = \left( \begin{matrix}
B_{ij} \ \qquad 1 \le i,j \le n \\  b_{\gamma_j} (T,L) \ \qquad 1 \le j \le n
\end{matrix}
\right)  \ ,
\end{equation}
and similarly if the lamination consists of more curves. More generically we can consider a multi-lamination, that is a collection of laminations $\mathbf{L} = \left( \cL_{n+1} , \cdots , \cL_{m} \right)$ on $\cC$ which corresponds to a set of line defects, and generalize the adjacency matrix by adding a new row for each lamination as follows
\begin{equation} \label{adjacencyBL}
B^{\mathbf{L}}_{ab} = \left( \begin{matrix}
B_{ij} \ \qquad 1 \le i,j \le n \\  b_{\gamma_j} (T,\cL_i) \ \qquad 1 \le j \le n \ , \ n+1 \le i \le m
\end{matrix}
\right) \ .
\end{equation}
Remarkably the generalized adjacency matrix transforms by the mutation transformation rules (\ref{mutB}) under flips of the triangulation $T$ \cite{FT2}. That is, if we flip the triangulation at a certain edge $k$, then the new generalized adjacency matrix is given by $B'^{\mathbf{L}}_{ab} = \mu_k \, B^{\mathbf{L}}_{ab}$.

It is natural to generalize a BPS quiver to include also this information. Recall that for a theory arising from the compactification of two M5 branes on a Riemann surface $\cC$ the BPS quiver can be derived from the triangulation $T$, as we discussed in Section \ref{BPStriang}. Mutations of the quiver, which are used to compute the spectrum of stable BPS states, are associated with flips of the triangulation. As we have seen also a lamination transforms nicely under flips. That is, the generalized adjacency matrix transforms according to the mutation rules. The information about the lamination can be encoded in the BPS quiver by adding an extra framing node \cite{fomin}. This node is connected to the other nodes of the BPS quiver by a number of arrows given by the shear coordinates. That is, if we label by $\gamma_i$ the nodes of the BPS quiver and by $\bullet$ a framing node
\begin{equation}
\Big\{ \text{number of arrows} \ \gamma_i \longrightarrow \bullet \Big\} = b_{\gamma_i} (T,\cL) \ ,
\end{equation}
where if the shear coordinate is negative, the arrows goes from $\bullet$ to $\gamma_i$. Note that if we are given a framed BPS quiver, we can reconstruct both the triangulation (simply neglecting the framing) and the form of the lamination $\cL$ on $\cC$, since the arrow structure of the framing nodes is determined by the shear coordinates. Similarly if we have a multi-lamination, we simply add as many framing nodes as there are laminations and draw arrows accordingly to the generalized adjacency matrix. To keep track of a line defect $L$ associated with the lamination $\cL$ we will label the framing node with $L$ as well.

As an example, the framed BPS quivers corresponding to the line operators of figure \ref{laminA2} are 
\begin{eqnarray} 
& F \left( L_1 \right) = x_{\gamma_1} \  & \begin{matrix}
\xymatrix@C=10mm{  &  \bullet  \ L_1 & \label{L1A2framed}  \\ \label{L2A2framed}
 \ \gamma_2 \ \bullet \  & & \ \bullet \ \gamma_1 \ar[ll] \ar@<-0.5ex>[ul]  \ar@<0.5ex>[ul] 
} \end{matrix} 
\ ,
\\ &
F \left( L_2 \right) =  x_{\gamma_1+\gamma_2} + x_{\gamma_2}  \ &
\begin{matrix}
\xymatrix@C=10mm{  &  \bullet  \ L_2 & \\
 \ \gamma_2 \ \bullet \ \ar@<-0.5ex>[ur]  \ar@<0.5ex>[ur]  & & \ \bullet \ \gamma_1 \ar[ll]
} \end{matrix} \ ,
\\[4pt]  \label{L3A2framed}
& F \left( L_3 \right) = x_{-\gamma_1} + x_{\gamma_2 - \gamma_1} + x_{\gamma_2} \  &
\begin{matrix} \xymatrix@C=10mm{  &  \bullet  \ L_3 \ar@<-0.5ex>[dr]  \ar@<0.5ex>[dr] & \\
 \ \gamma_2 \ \bullet \ \ar@<-0.5ex>[ur] \ar@<0.5ex>[ur]  & & \ \bullet \ \gamma_1 \ar[ll]
} \end{matrix} \ ,
\\[4pt]
& F \left( L_4 \right) = x_{-\gamma_1} + x_{-\gamma_1-\gamma_2}  \  &
\begin{matrix} \xymatrix@C=10mm{  &  \bullet  \ L_4 \ar@<-0.5ex>[dr] \ar@<0.5ex>[dr] & \\
 \ \gamma_2 \ \bullet \  & & \ \bullet \ \gamma_1  \ar[ll]
} \end{matrix}
\vspace{4pt}
\ ,
\label{L4A2framed}
\\  \label{L5A2framed} &
F \left( L_5 \right) =x_{-\gamma_2}
 \  &
\begin{matrix} \xymatrix@C=10mm{  &  \bullet  \ L_5 \ar@<-0.5ex>[dl] \ar@<0.5ex>[dl] & \\
 \ \gamma_2  \ \bullet \  & & \ \bullet \ \gamma_1 \ar[ll]
} \end{matrix} \ .
\end{eqnarray}

The formalism developed so far, has a drawback \cite{FT2}: it cannot accommodate laminations consisting in an arc surrounding a simple puncture. Indeed the shear coordinates of such an arc would be identically zero. For this reason in this Section we will only consider laminations in $\scrT_0 (\cC)$. If such an arc occurs, one has to use the formalism of Section \ref{extended}. Furthermore the shear coordinates act as coordinates for a generic path on the curve $\cC$ \cite{FT2,fomin}. But not every path correspond to what we have called a lamination in $\scrT (\cC)$. For example one can give shear coordinates for a collection of paths which do not satisfy the condition that the sum of their weights vanish at each boundary component. In other words, given a collection of shear coordinates, one needs to impose further conditions to ensure that they correspond to a physical line defect. Not all the framings of BPS quivers are allowed. We will comment further on this problem in Section \ref{admi}.

\subsection{Extended quivers vs framed quivers} \label{sinksource}

Given a certain line defect for a theory of class $\cS [\mathfrak{su} (2)]$ we can derive two objects. One is a certain collection of paths on the extended BPS quiver, given by elements of $\sfL \sfA_{\tilde{Q}}$; the other is a framed BPS quiver. We will now compare briefly the two constructions and show how to pass from one to the other. Associating a line defect with a path on the extended BPS quiver has the obvious advantage of being easy to visualize, since the path is a direct transposition of a lamination $\cL$ on the Riemann surface $\cC$. Furthermore as we have seen the traffic rules used to expand a lamination in the Fock and Goncharov coordinates associated with a triangulation can be directly translated into a set of rules on the extended BPS quiver $\tilde{\sfQ}$. This has to be done ``by hand", drawing paths on the extended quiver which correspond to the lamination, and in particular can become quite cumbersome when dealing with curves of high genus. Another problem is that the drawing of a quiver is largely conventional. While it is straightforward once we are given a lamination on a curve to construct the corresponding path on the extended quiver, it can be difficult to compare two defects since the way we position the nodes of the quiver on a plane is arbitrary. The advantage of the framing is that it is independent on the particular way the quiver is drawn and transforms in a simple way when the triangulation is flipped. These transformations are completely algorithmic and easy to implement. The price to be paid when using framed BPS quivers is the lack of an immediate geometric counterpart to the framing.

\begin{figure}
\centering
\def\svgwidth{10cm}
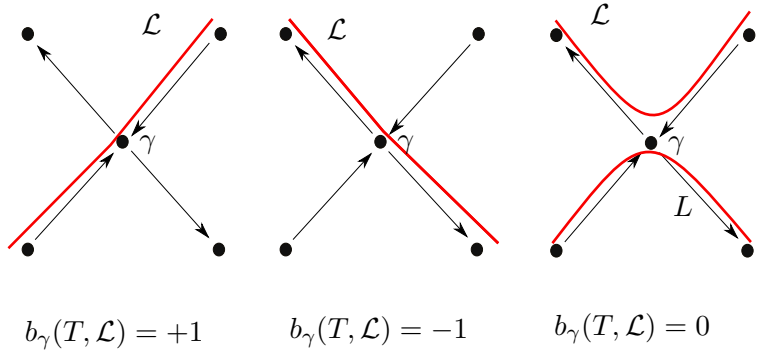
\caption{Shear coordinates associated to a path on a BPS quiver}
\label{shearquiver}
\end{figure}

Nevertheless it is quite easy to pass form one formalism to the other in the case of $\cL \in \scrT_0 (\cC)$. Suppose we are given a certain path corresponding to a line operator on an extended BPS quiver. To derive the corresponding framed BPS quiver we simply have to translate the graphic rules of figures \ref{coordonlam} and \ref{coordonlam2} in quiver language. Indeed figure \ref{coordonlam} simply states that $b_{\gamma} (T,\cL)$ is positive or negative for a path crossing the node $\gamma$ if the path goes through a sink or a source respectively; one the other hand if the path goes upstream or downstream the shear coordinate vanishes. These rules are illustrated in figure \ref{shearquiver}.
The opposite construction proceeds along the same lines. Assume we are given the framed BPS quiver corresponding to a line operator. This quiver can be extended by including the information about the boundary segments. 
We focus on a framing node of the framed BPS quiver and consider the arrows between itself and the BPS quiver. The information we have is that through each node we have to draw a path which goes through a sink, a source or otherwise depending on the shear coordinate. This construction is non local, since we have to consider simultaneously all the nodes. We finally draw the path such that at each node the corresponding conditions are obeyed. While this may not be easy if the quiver is complicated, the fact that the shear coordinates are indeed coordinates on the space of laminations guarantees that this procedure can be always carried on and that the result is unique.

\subsection{Line defects and cluster transformations} \label{linedefcluster}

We will now define certain generating functions $\scrL$ which encode the framed BPS spectrum and introduce certain mutation operators $\mut$ which allow us to generate new framed spectra from known ones. 

Recall that to a line operator we can associate the framed indices $\langle L_{\zeta} \rangle$ and $\langle L_{\zeta} \rangle'$ in (\ref{deftraceL}) and (\ref{deftraceL'}). As we have explained in Section \ref{linedef}, these indices are invariant since the wall-crossing properties of the functions $\mathcal{Y}$ compensate the wall-crossings of the framed degeneracies. For example, crossing the BPS wall corresponding to an hypermultiplet with charge $\gamma_k$ counterclockwise, that is going from a region with $\im (\cZ_{\gamma_k} / \zeta) < 0$ to a region with $\im (\cZ_{\gamma_k} / \zeta) > 0$, the variables $\tilde{\cY}_{\gamma_i}$ transform as
\begin{equation}
\tilde{\cY}_{\gamma_i} \longrightarrow \tilde{\cY}_{\gamma_i} \left( 1 + \tilde{\cY}_{\gamma_k} \right)^{\langle \gamma_i , \gamma_k \rangle} \ ,
\end{equation}
losing or gaining a halo depending on the sign of $\langle \gamma_i , \gamma_k \rangle$. Crossing the same wall clockwise, the inverse transformation applies.

From the framed quiver perspective it is more natural to compose these transformations with quiver mutations. We define formal variables $\{ \sfy_{\gamma} \}$ and $\{ \tilde{\sfy}_{\gamma} \}$ associated with a basis of charges $\{ \gamma \}$ and the transformation rules
\begin{equation} \label{muy}
\mu^{\sfy}_{k,\mp} \ : \ \tilde{\sfy}_{\gamma_i} \longrightarrow \tilde{\sfy}_{\gamma_i} \left( 1 + \tilde{\sfy}_{\gamma_k} \right)^{\pm \langle \gamma_i , \gamma_k \rangle}  \ ,
\end{equation}
where $\mu^{\sfy}_{k,-}$ correspond to the transformation of the $\tilde{\cY}$ variables when crossing a $\gamma_k$ BPS wall in the counterclockwise direction, while $\mu^{\sfy}_{k,+}$ is associated to crossing the same wall in the clockwise direction (and similarly for the variables $\{ \sfy \}$). We will think directly of these variables as being associated with the set of nodes $\sfQ_0$ of a quiver $\sfQ$. The variables $\sfy_{\gamma}$ have a twisted multiplication rule $\sfy_{\gamma_i} \, \sfy_{\gamma_j} = (-1)^{\langle \gamma_i , \gamma_j \rangle} \sfy_{\gamma_i + \gamma_j}$. The variables $\tilde{\sfy}$ obey the untwisted multiplication law $\tilde{\sfy}_{\gamma_i} \, \tilde{\sfy}_{\gamma_j} =  \tilde{\sfy}_{\gamma_i + \gamma_j}$ and are related to the $\sfy$ as $\tilde{\sfy}_{\gamma} = \sigma (\gamma) \, \sfy_{\gamma}$ where $\sigma (\gamma)$ is a quadratic refinement mod 2 of the intersection pairing on the lattice $\Gamma$, 
\begin{equation}
\sigma (\gamma) \ \sigma (\gamma') = \sigma (\gamma + \gamma') \ \sigma (0) (-1)^{\langle \gamma , \gamma'  \rangle} \ .
\end{equation}
In particular for the case of an hypermultiplet $\sigma (\gamma_{\text{hyper}}) = -1$. More formally we can say that the variables $\{ \tilde{\sfy} \}$ are elements of the universal semifield $\IQ_{sf} (\tilde{\sfy})$, the closure of the set of indeterminates $\{ \tilde{\sfy}_{\gamma} \}$ in $\IQ$ under multiplication, addition and division. Since these variables are associated with the nodes of the quiver $\sfQ$, we let quiver mutations act on them as
\begin{equation} \label{muk+}
\mu_{k,+} \left( \tilde{\sfy}_{\gamma_i} \right) = \left\{ \begin{matrix} \tilde{\sfy}_{\gamma_i}^{-1} & i = k \\[4pt] \tilde{\sfy}_{\gamma_i} \, \tilde{\sfy}_{\gamma_k}^{[\langle \gamma_i , \gamma_k \rangle]_+} & i \neq k 
\end{matrix} \right. \ ,
\end{equation}
when crossing a wall $W (\gamma_k)$ in the clockwise order, and 
\begin{equation} \label{muk-}
\mu_{k,-} \left( \tilde{\sfy}_{\gamma_i} \right) = \left\{ \begin{matrix} \tilde{\sfy}_{\gamma_i}^{-1} & i = k \\[4pt] \tilde{\sfy}_{\gamma_i} \, \tilde{\sfy}_{\gamma_k}^{-[\langle \gamma_i , \gamma_k \rangle]_-} & i \neq k 
\end{matrix} \right. \ ,
\end{equation}
when crossing a wall $W (-\gamma_k)$ in the counterclockwise order. Note that these transformations are consistent with (\ref{mu+}) and (\ref{mu-}) in the sense that $\mu_{k,\pm} (\tilde{\sfy}_{\gamma_i}) = \tilde{\sfy}_{\mu_{k,\pm} (\gamma_i)}$.

We define the following transformations acting on the $\{ \tilde{\sfy}_{\gamma} \}$ variables. Assume we are in a fixed chamber, with a certain BPS particle spectrum. This means we have chosen a stability condition $\cZ (u)$ and we draw the corresponding central charge vectors in the upper half plane. Then quiver mutations $\mu_{k,+}$ generate the BPS spectrum going in the clockwise sense in the central charge plane, crossing the BPS rays corresponding to the stable particles. For each mutation corresponding to a particle with charge $\gamma_k$ we define the operator $\mathsf{mut}_{k,\pm} = \mu_{k,\pm}^{\sfy} \circ \mu_{k,\pm}$ as 
\begin{equation} \label{mut+y}
\mathsf{mut}_{k,+} \, \tilde{\sfy}_{\gamma_i}  = \left\{ \begin{matrix} \tilde{\sfy}_{\gamma_i}^{-1} & i = k \\[4pt] \tilde{\sfy}_{\gamma_i} \, \tilde{\sfy}_{\gamma_k}^{[ \langle \gamma_i , \gamma_k \rangle]_+} \left( 1 + \tilde{\sfy}_{\gamma_k} \right)^{- \langle \gamma_i , \gamma_k \rangle} & i \neq k
\end{matrix}
\right. \ .
\end{equation}
These are \textit{cluster transformations} for the coefficients\footnote{We will loosely use the name ``cluster variables" when referring to the coefficients of a cluster algebra. In the literature that name is sometime reserved for the $x$-variables; however we will not use $x$-variables in this paper and no ambiguity should arise.} of a cluster algebra, in the sense of Fomin-Zelevinsky \cite{FZIV}. Similarly going counterclockwise we encounter the rays corresponding to the anti-particles $\cZ_{-\gamma_k} (u)$, and the transformations are
\begin{equation} \label{mut-y}
\mathsf{mut}_{k,-} \, \tilde{\sfy}_{\gamma_i}  = \left\{ \begin{matrix} \tilde{\sfy}_{\gamma_i}^{-1} & i = k \\[4pt] \tilde{\sfy}_{\gamma_i} \, \tilde{\sfy}_{\gamma_k}^{[ \langle \gamma_k , \gamma_i \rangle]_+} \left( 1 + \tilde{\sfy}_{-\gamma_k} \right)^{\langle \gamma_k , \gamma_i \rangle} & i \neq k
\end{matrix}
\right. \ .
\end{equation}
Note that these  auxiliary operators acting on the $\{ \tilde{\sfy}_{\gamma} \}$ do \textit{not} correspond to a wall-crossing; the wall-crossing transformation upon crossing a BPS wall is still given by the halo picture, i.e. the $\{ \tilde{\sfy}_{\gamma} \}$ variables  transform with $\mu^{\sfy}_{k,\mp}$. We will however see that under certain conditions a sequence of operators $\mut_{k,\pm}$ acts as the operator corresponding to a sequence of BPS wall-crossings. Roughly speaking, since the operators $\mut$ are composition of a BPS wall-crossing with a quiver mutation, this happens when a sequence of quiver mutations is trivial.

We introduce the tropical semifield $\mathsf{Trop} (\{ \tilde{\sfy} \})$ or simply $\mathsf{Trop}$, associated with the variables $\tilde{\sfy}$. This is the abelian multiplicative group freely generated by the variables $\{ \tilde{\sfy} \}$ where the addition $\oplus$ is defined as
\begin{equation} \label{tropplus}
\prod_{i \in \sfQ_0} \, \tilde{\sfy}_i^{a_i} \, \oplus \, \prod_{i \in \sfQ_0} \, \tilde{\sfy}_i^{b_i} = \prod_{i \in \sfQ_0} \, \tilde{\sfy}_i^{\min (a_i , b_i)} \ .
\end{equation}
There is a canonical map, the \textit{tropical evaluation}, which to the variables $\{ \tilde{\sfy} \}$ in $\IQ_{\sf} ( \tilde{\sfy} )$  associates the corresponding elements in $\mathsf{Trop}$, which we will denote by $\{ [ \tilde{\sfy} ] \}$. This map amounts to replacing ordinary addition $+$ with the tropical addition $\oplus$. In the following we will also need the result known as the {\it sign-coherence property} of cluster algebras, which states that a Laurent monomial $[ \tilde{\sfy}_i ]$ is either positive or negative \cite{keller,nagao,naka}. A Laurent monomial is said to be positive (or negative) if it is not $1$ and all of its exponents are non-negative (non-positive). This justifies the introduction of the tropical sign $\epsilon (\tilde{\sfy}_i)$ which is $+$ (respectively $-$) if the tropical monomial $[ \tilde{\sfy}_i ]$ is positive (respectively negative). Tropical variables transform under quiver mutations as
\begin{equation}
\mut^t_k \, [  \tilde{\sfy}_{\gamma_i} ] = \left\{ \begin{matrix}  [\tilde{\sfy}_{\gamma_k}]^{-1} & i = k \\[4pt] [ \tilde{\sfy}_{\gamma_i}]  \, [\tilde{\sfy}_{\gamma_k}]^{[ \epsilon (\tilde{\sfy}_k) \langle \gamma_i , \gamma_k \rangle]_+}   & i \neq k
\end{matrix}
\right. \ .
\end{equation}
These transformation rules are obtained from (\ref{mut+y}) and (\ref{mut-y}) by replacing $+$ with $\oplus$ and $\tilde{\sfy}_i$ with $[ \tilde{\sfy}_i ]$. Note that the transformations $\mu_{k,\pm}$ are the tropical limit of the transformations $\mut_{k,\pm}$. 

When composing a number of quiver mutations we will often use the notation $ \tilde{\sfy} [t_m]$ to denote the variables obtained after $m$ quiver mutations, and which are associated to the quiver $\sfQ (t_i)$, in the notation of Section \ref{BPSquivers}. If we have the sequence of quiver mutations ${\mbf \mu} = \left( \mu_{i_1} , \cdots ,  \mu_{i_m} \right)$, then $\tilde{\sfy}_j [t_m] = \mut_{i_m} \cdots \mut_{i_1} \tilde{\sfy}_j$ and $\sfQ (t_m) = \mu_{i_m} \cdots \mu_{i_1} \, \sfQ (t_0)$. A property of cluster algebras is that certain sequence of mutation exhibit periodicity properties. We will call a sequence of mutations $(\mu_{i_1} , \cdots ,  \mu_{i_L})$ a $\Pi$-period for $\sfQ$ if
\begin{equation}
B_{\pi(i) \, \pi(j)} (t_L) = B_{ij} (t_0) \ , \qquad \text{and} \qquad \tilde{\sfy}_{\pi (i)} (t_L) =  \tilde{\sfy}_{i} (t_0) \ ,
\end{equation}
where $\Pi$ acts on the quiver $\sfQ$ as the permutation $\pi$ of its nodes. In plain words we observe periodicity if after a number of quiver mutations, the quiver and its cluster variables are back to their original form and values, eventually up to a permutation. To any sequence of mutations we associate a \textit{tropical sign-sequence} $(\epsilon_1 , \cdots , \epsilon_m)$ where  $\epsilon_t$ is the tropical sign of $\tilde{\sfy}_{i_t} (t)$, the variable corresponding to the node which is mutated. Similarly we define the $\mbf c$-vector  $\beta_t$ of $\tilde{\sfy}_{i_t} (t) $ as the tropical limit
\begin{equation}
[\tilde{\sfy}_{i_t}  (t)] = \tilde{\sfy}^{\beta_t} \ .
\end{equation}
The information about the $\mbf c$-vectors can be packed into the $\mbf c$-matrix $C (t)$, the matrix whose columns are the $\beta_t$. This matrix will play a role below when discussing the BPS spectrum and the quantum dilogarithm identities. We refer the reader to the reviews \cite{keller,naka} for more details.

Having introduced all this formalism, we now return to line defects. As we have discussed in Section \ref{linedef}, a line defect $L$ is completely characterized by its framed BPS spectrum $\underline{\overline{\Omega}} (u , \gamma ; q) $. This information can be conveniently packed into a generating function associated to $L$. For simplicity we now set $q = +1$. We define the generating function
\begin{equation}
\scrL = \sum_{\gamma} \   \underline{\overline{\Omega}} (u , \gamma ; q = + 1 ) \  \tilde{\sfy}_{\gamma} \ ,
\end{equation}
associated to a line defect $L$. This generating function is constructed in terms of the quiver variables $\{ \tilde{\sfy} \}$ and is the direct algebraic counterpart of the indices $\langle L \rangle'$ discussed in Section \ref{linedef}, except that instead of depending on the non trivial functions $\{ \mathcal{Y} \}$, it is written in terms of the formal variables $\{ \tilde{\sfy} \}$. The framed degeneracies undergo wall-crossings as we cross BPS walls. Similarly we let the variables $\{ \tilde{\sfy}_{\gamma} \}$ transform with $\mu^{\sfy}_{k,\mp}$ as we cross the walls. As a result the two wall-crossing transformation compensate and the formal generating function $\scrL$ is invariant, just as the indices $\langle L \rangle'$ are \cite{Gaiotto:2010be}. Note that in general if we would let $\mathsf{mut}_k$ act instead of $\mu^{\sfy}_{k,+}$ upon crossing a wall, then the invariance of $\scrL$ would be lost. We will however see that the operator  $\mathsf{mut}_k$ can be quite useful in studying line defects.

The key observation is that a sequence of quiver mutations acting on the framed quiver can generate the framed quiver corresponding to a new line operator. For this to happen the mutation sequence must act as the identity on the underlying unframed quiver, if we forget the labeling of the nodes, but not on the framed quiver.

Consider now a line defect and the associated framed quiver. To each node of the unframed quiver we associate the formal variables $\{ \tilde{\sfy}_{\gamma} \}$. In particular all the results of the previous sections apply and 
we can construct the generating function
\begin{equation} \label{Lgenfun}
\scrL = \sum_{\gamma} \,  \underline{\overline{\Omega}} (u , \gamma ; q = +1) \ \tilde{\sfy}_{\gamma} \ ,
\end{equation}
computed for example using the rules of Section \ref{matrix}. This generating function is invariant upon crossing BPS walls. Fix a point $u \in \cB$ in the moduli space and assume that the spectrum at $u$ is known and consists of finitely many hypermultiplets. Each BPS state corresponds to a BPS wall, or wall of second kind. Now we claim that a certain sequence of cluster mutation operators $\mut$ has the same effect on (\ref{Lgenfun}) as a sequence of BPS wall-crossing corresponding to all the stable BPS particles in the spectrum. More precisely, consider a sequence of mutations ${\mbf \mu} = \left( \mu_{i_1} , \dots  , \mu_{i_k} \right)$ and assume that the matrix $-C(t_k)$ is a permutation. In plain words this means that after the sequence of mutations, the quiver $\sfQ$ is back to itself up to a permutation of the nodes, and that the charges $\{ \gamma_j \}$ associated with the nodes have all changed sign. If this is the case, then the sequence of cluster mutations $\mut_{\mbf \mu}$ corresponds to the operation of crossing all the BPS walls associated with the stable BPS particles in the chamber identified by the sequence $\mbf \mu$ \cite{Alim:2011ae,Alim:2011kw,Cecotti:2010fi,nagao,keller}. For the time being we will assume this result to be true, and will discuss it in full generality in Section \ref{qmut} using the formalism of quantum cluster algebras. These facts about cluster algebras are at the core of the mutation method discussed in Section \ref{BPSquivers}.

Consider now a sequence of mutations which acts as the identity on the underlying BPS quiver (such that the matrix $- C(t_k)$ is a permutation), but not on the framed quiver. Assume we are given a certain line defect $L$ whose framed BPS spectrum is known and written in terms of the generating function $\scrL$. Since the sequence of mutations $\mbf \mu$ is such that $-C(t_k)$ is a permutation, the corresponding operator $\mut_{\mbf \mu}$ acts as a sequence of wall-crossing on the generating function $\scrL$. However $\scrL$ is invariant under wall-crossing, since the transformations of the framed degeneracies and of the $\{ \tilde{\sfy} \}$ variables compensate each other. Assuming that this happens after $k$ mutations, we have
\begin{equation}
\scrL = \sum_{\gamma} \underline{\overline{\Omega}} (u ; \gamma) \ \tilde{\sfy}_{\gamma} = \sum_{\gamma'} \underline{\overline{\Omega}}^{new} (u ; \gamma') \ \tilde{\sfy}_{\gamma'} \, (t_k) \ .
\end{equation}
The coefficients $\underline{\overline{\Omega}}^{new} (u ; \gamma')$ are such that, when the mutated variables $\tilde{\sfy}_{\gamma'} \, (t_k)$ are expressed in terms of the original variables $\tilde{\sfy}_{\gamma} $, we recover the original expression $\scrL = \sum_{\gamma} \underline{\overline{\Omega}} (u ; \gamma) \ \tilde{\sfy}_{\gamma}$. Since the framed quiver is not invariant under the mutation sequence (only the underlying BPS quiver is), the coefficients $\underline{\overline{\Omega}}^{new} (u ; \gamma') $ are different from the coefficients $\underline{\overline{\Omega}} (u ; \gamma) $. In particular the lamination associated to the framing will now intersect new edges of the triangulations, labelled by the mutated variables $\tilde{\sfy}_{\gamma'} (t_k)$. This is a consequence of the transformation rules under mutations of the generalized adjacency matrix which includes the information about the shear coordinates of the lamination.

The crucial observation is that we can reinterpret the framed degeneracies $ \underline{\overline{\Omega}}^{new} (u ; \gamma') $ as the coefficients of the expansion of a \textit{new} lamination on a \textit{new} basis of  variables $\tilde{\sfy}_{\gamma} (t_k)$. However, since the BPS quiver is invariant under the sequence of mutations (up to a permutation of the nodes) we can relabel the coordinates $\tilde{\sfy}_{\gamma} [i]$ as $\tilde{\sfy}_{\gamma} [i] \longrightarrow \tilde{\sfy}_{\gamma}$, eventually up to a permutation. Then, the object we have constructed is the generating function of a new line operator expressed in the variables of the original BPS quiver. The reason for this is that as we have seen the framing nodes encode the information of the shear coordinates; these being coordinates, they are in one to one correspondence with laminations. Therefore if we relabel the variables corresponding to the mutated quiver, since after the sequence of mutations the BPS quiver is back to its original form but the framed quiver is not, the new framed quiver will correspond to a new line defect. More elegantly, instead of relabeling the coordinates, we will see that we can act on the generating function with the sequence of mutations inverse to the one giving $\tilde{\sfy}_{\gamma} (t_i)$.

More precisely, assume that we are given two line defects $\scrL^{(o)}$ and $\scrL^{(n)}$, and that both defects are described by two framed quivers $\sfQ^{(o)}$ and $\sfQ^{(n)}$ such that the underlying quiver $\sfQ$ is the same. In other words the two framed quivers only differ because of the framing node. Furthermore assume that the two framed quivers are related by a certain sequence of $i$ mutations which acts as minus the identity on the basis of charges, schematically $\mbf{\mu}_{o \rightarrow n}$. In general one would also need a permutation of the nodes; for the time being we consider the simpler situation where this is not the case. Since the sequence of quiver mutations is trivial on the BPS quiver, the corresponding $\mut_{o \rightarrow n}$ operator corresponds to a sequence of BPS wall-crossing, as we will show in Section \ref{qmut}. As a consequence $\scrL^{(o)}$ is invariant. On the other hand after the sequence of mutations $\mbf{\mu}_{o \rightarrow n} $, the framed quiver now corresponds to the line defect $\scrL^{(n)}$, but expanded in the mutated basis $\tilde{\sfy}_{\gamma} [i] = \mut_{\mbf{\mu}_{o \rightarrow n}} \, \tilde{\sfy}_{\gamma}$. Invariance is expressed as the relation
\begin{equation} \label{LoLn}
\scrL^{(o)} \left( \tilde{\sfy}_{\gamma} \right) = \scrL^{(n)} \left( \tilde{\sfy}_{\gamma} [i] \right) =  \scrL^{(n)} \left( \mut_{\mbf{\mu}_{o \rightarrow n}}  \, \tilde{\sfy}_{\gamma} \right) \ ,
\end{equation}
where we have explicitly indicated the set of $\{ \tilde{\sfy} \}$ variables used to compute the generating functions. We can interpret  (\ref{LoLn}) as an equation for the coefficients of the expansion of $\scrL^{(n)}$ on the variables $\{ \tilde{\sfy}_{\gamma} [t_i] \}$. If however we redefine $\{ \tilde{\sfy}_{\gamma} [i] \} \longrightarrow \{ \tilde{\sfy}_{\gamma} \}$, then we find the generating function of a new line defect corresponding to the framed quiver $\sfQ^{(n)}$. Instead of going through the complicated business of solving for the framed degeneracies of $\scrL^{(n)}$ such that (\ref{LoLn}) holds, we can use the fact that the operator $\mut$ is an involution and write
\begin{equation} \label{LnLomut}
\scrL^{(n)} \left( \tilde{\sfy}_{\gamma} \right) = \scrL^{(o)}  \left( \mut_{\mbf{\mu}_{o \rightarrow n}}^{-1}  \tilde{\sfy}_{\gamma} \right) \ ,
\end{equation}
where $\mut_{\mbf{\mu}_{o \rightarrow n}}^{-1}$ is the operator associated to the opposite sequence of mutations $\mbf{\mu}_{o \rightarrow n}$. In the following we will denote these operations by letting the operator $\mut$ act on the generating functions $\scrL$.

Let us clarify this formal argument with a concrete example in the $A_2$ theory. Let us consider a chamber with only two BPS particles, and the line operator given by the quiver
\begin{equation}
\sfQ^{L_5} =  \begin{matrix}  \xymatrix@C=10mm{  &  \bullet  \ L_5  \ar@<-0.5ex>[dl]  \ar@<0.5ex>[dl]  & \\
 \ \tilde{\sfy}_{\gamma_2} \  \bullet \ & & \ \bullet \ \tilde{\sfy}_{\gamma_1} \ar[ll] 
} \end{matrix} \ .
\end{equation}
The framed spectrum is encoded in  $\scrL_5 = \frac{1}{\tilde{\sfy}_{\gamma_2}} $. We now mutate first at $\gamma_1$ and then at $\gamma_2$. Call the variables $\tilde{\sfy}_{\gamma_1} = \sfx$ and $\tilde{\sfy}_{\gamma_2} = \sfy$. After the sequence of mutations the unframed BPS quiver goes back to itself; indeed this sequence corresponds to crossing the BPS walls determined by the charges $\gamma_1$ and $\gamma_2$ in the strong coupling chamber. After the corresponding mutations, the new variables are
\begin{eqnarray}
\sfx[2] &=& \frac{1+\sfy + \sfx \, \sfy}{\sfx} \ , \\
\sfy[2] &=& \frac{1}{\sfy + \sfx \, \sfy} \ ,
\end{eqnarray}
and the line operator has the BPS quiver representation
\begin{equation} \label{exampleL2}
\sfQ'^{L_5} = \mu_2 \, \mu_1 \, \sfQ^{L_5} = \begin{matrix}
\xymatrix@C=10mm{  &  \bullet  \ L_5  \ar@<-0.5ex>[dr] \ar@<0.5ex>[dr] & \\
 \ \sfy[2] \ \bullet \  \ar@<-0.5ex>[ur] \ar@<0.5ex>[ur]  & & \  \ar[ll] \bullet \ \sfx[2]
}
\end{matrix} \ .
\end{equation}
We can compute the generating function for the line defect $L_5$ in the new mutated coordinates $\sfx[2]$ and $\sfy[2]$, for example using the rules of Section \ref{extended}, and indeed see that it is invariant
\begin{equation} \label{exampleL2-2}
\scrL_5  = \frac{1+\sfy[2] + \sfx[2] \, \sfy[2]}{\sfx[2]} =\frac{\sfx}{1 + \sfy+\sfx \, \sfy} + \frac{1}{\sfy + \sfx \, \sfy} \frac{\sfx}{1+ \sfy + \sfx \, \sfy}+\frac{1}{\sfy + \sfx \, \sfy}= \frac{1}{\sfy} \ .
\end{equation}
Note that the quiver (\ref{exampleL2}) has the same form of the framed BPS quiver corresponding to the operator $\scrL_3$ in (\ref{L3A2framed}), except that the nodes correspond to mutated variables; this means that if we forget the labeling of the nodes, we have $\sfQ^{L_3} = \mu_2 \mu_1 \, \sfQ^{L_5}$. Indeed the generating function $\scrL_3 =  \frac{1}{\sfx} + \frac{\sfy}{\sfx}+ \sfy $ appears in the intermediate steps in (\ref{exampleL2-2}) but in the mutated variables. This is obvious: the line defect generating function can be computed by an expansion in the variables corresponding to the nodes, regardless of their labels. On the other hand, since $\scrL_5$ is invariant upon crossing BPS walls, we find a non trivial statement: the rules giving the expansion of the line operator in the mutated framed quiver must be such that in the mutated variables the generating function is the same as in the original quiver variables. In formulas
\begin{equation}
\mut_{2,+}  \, \mut_{1,+} \left(  \frac{1}{\sfx} + \frac{\sfy}{\sfx}+ \sfy \right)  = \mut_{2 \, 1 , +}  \left( \frac{1}{\sfx} + \frac{\sfy}{\sfx}+ \sfy  \right) =\frac{1}{ \sfy} \ .
\end{equation}
Since the mutated framed quiver correspond to a new line operator upon relabeling the node variables, we learn that the generating function of the new operator can be guessed by finding that particular combination of the original cluster variables which, after the sequence of mutations, reproduces the generating function of the framed degeneracies for the line defect we started with. Instead of guessing, we can use the fact that mutations are involutions and simply write down (\ref{LnLomut}) explicitly for this case
\begin{equation}
\scrL_3  =  \frac{1}{\sfx} + \frac{\sfy}{\sfx}+ \sfy  = \mut_{1,+}  \, \mut_{2,+} \left(\scrL_5 \right) = \mut_{1 \, 2 , +}  \left( \frac{1}{\sfy} \right) \ .
\end{equation}

\begin{figure}[h]
\centering
\def\svgwidth{9cm}
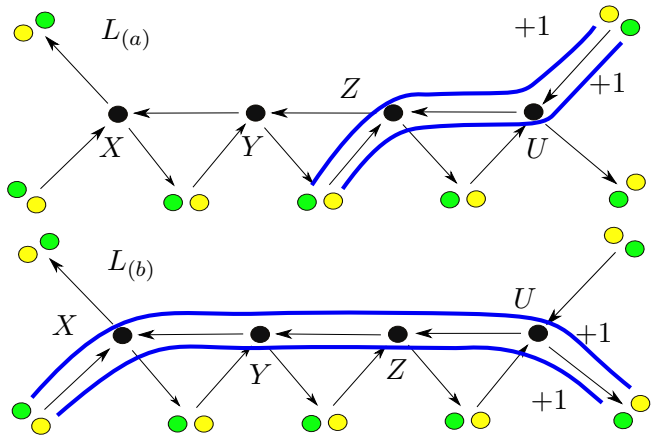
\caption{Paths in the extended quiver corresponding to the line defects $L_{(a)}$ and $L_{(b)}$}
\label{A4example1}
\end{figure}

Let us discuss now a more complicated example. In the $A_4$ Argyres-Douglas theory we can consider the line operator corresponding to the framed BPS quiver
\begin{equation} \label{exampleA4La}
\sfQ^{L_{(a)}} =  \begin{matrix} \xymatrix@C=10mm{  & &  \bullet  \ L_{(a)}  & \\ \ \sfx \ \bullet \  &
 \ \sfy  \ \bullet \ \ar[l] & \ \bullet \ \sfz \    \ar@<-0.5ex>[u] \ar@<0.5ex>[u]  \ar[l] & \ \bullet \ \sfu \ar[l]
} \end{matrix} \ .
\end{equation}
We consider for simplicity a chamber with only four states, corresponding to the nodes of the quiver. By going to the extended quiver shown in figure \ref{A4example1} and using the rules of Section \ref{extended}, one can easily see that the generating function for  the line defect of equation (\ref{exampleA4La}) is
\begin{equation}
 \scrL_{(a)}  = P_{YG} \left( D \, M_{\sfz} \, U  \, M_{\sfu} \, U \right) \ P_{GY}  \left( D \, M_{\sfz} \, U  \, M_{\sfu} \, U \right) = \sfz + \sfu \, \sfz \ .
\end{equation}
Consider now the operator corresponding to the following framed BPS quiver
\begin{equation}
\sfQ^{L_{(b)}} =  \begin{matrix} \xymatrix@C=10mm{  & &  \bullet  \ L_{(b)} \ar@<-0.5ex>[rd] \ar@<0.5ex>[rd]   & \\ \ \sfx \ \bullet \  \ar@<-0.5ex>[urr] \ar@<0.5ex>[urr]  &
 \ \sfy  \ \bullet \ \ar[l] & \ \bullet \ \sfz \    \ar[l] & \ \bullet \ \sfu \ar[l]
} \end{matrix} \ .
\end{equation}
One can easily see using (\ref{adjacencyBL}) and (\ref{mutB}), that $\sfQ^{L_{(b)}} \simeq \left( \mu_4 \, \mu_3 \, \mu_2 \, \mu_1 \right)^5 \ \sfQ^{L_{(a)}} $. Therefore we predict that
\begin{equation}
 \scrL_{(b)}  =   \left( \mut_{1,+} \, \mut_{2,+} \, \mut_{3,+} \, \mut_{4,+} \right)^5 \ \left( \sfz + \sfu \, \sfz \right) \ .
\end{equation}
A straightforward, if boring, computation gives
\begin{eqnarray}
\left( \mut_{1,+} \, \mut_{2,+} \, \mut_{3,+} \, \mut_{4,+} \right)^5 \, \sfz &=& \frac{1+\sfy+\sfy \, \sfz}{\sfy \, \sfz \, \sfu} \ , \\
\left( \mut_{1,+} \, \mut_{2,+} \, \mut_{3,+} \, \mut_{4,+} \right)^5 \, \sfu &=& \sfx + \sfx \, \sfy + \sfx \, \sfy \, \sfz \, + \sfx \, \sfy \, \sfz \, \sfu   \ ,
\end{eqnarray}
and we end up with
\begin{eqnarray} \label{LbA4}
 \scrL_{(b)}  &=& 2 \, \frac{\sfy \, \sfx}{\sfu}+\sfy \, \sfx+\frac{\sfy \, \sfz \, \sfx}{\sfu}+\sfy \, \sfz \,  \sfx+2 \, \frac{\sfx}{\sfu}+\frac{\sfy \, \sfx}{\sfu \, \sfz} \nonumber \\[4pt] && +2 \, \frac{\sfx}{\sfu \, \sfz}+\frac{\sfx}{\sfu \, \sfy \, \sfz}+\sfx+\frac{1}{\sfu}+\frac{1}{\sfu \, \sfz}+\frac{1}{\sfu \, \sfy \, \sfz} \ .
\end{eqnarray}
To check this prediction, we can look at the extended quiver for the line operator $L_{(b)}$ in figure \ref{A4example1}, and indeed one can see that 
\begin{equation}
 \scrL_{(b)}  = P_{YG} \, \left( D \, M_{\sfx} \, U \, M_{\sfy} \, U \, M_{\sfz} \, U \, M_{\sfu} \, D \right) P_{GY} \, \left( D \, M_{\sfx} \, U \, M_{\sfy} \, U \, M_{\sfz} \, U \, M_{\sfu} \, D \right) \ ,
\end{equation}
gives the same result as (\ref{LbA4}).

Therefore given a line operator corresponding to a framed quiver $\sfQ^L$, we can act with a sequence of mutations which leaves the underlying quiver $\sfQ$ invariant to obtain a new framed quiver $\sfQ^{L^{new}}$. In general this will involve a permutation $\Pi$ of the nodes. The new framed quiver corresponds to a new line defect. The generating function of the new defect is obtained by applying the opposite sequence of mutations on the cluster variable expansion of the generating function of the line defect we started with. However in doing this we must also keep track of the permutation of the nodes and of the associated variables. In formulas, if
\begin{equation}
\sfQ^{L^{new}} = \Pi \  \mu_k \, \dots \, \mu_1 \, \sfQ^{L} 
\end{equation}
for a permutation $\Pi$ of the nodes, then we have
\begin{equation}
\scrL^{new} (\tilde{\sfy}_{\gamma}) = \scrL \left( \mut_{\pi(1)} \, \cdots \, \mut_{\pi(k)} \, \Pi  \,  \tilde{\sfy}_{\gamma} \right) \ .
\end{equation}
We seem to have found an organizing principle in the set of line defects, at least when our formalism is valid. We define a \textit{mutation orbit} as the set of line defects generated from a given defect $L$ by applying the same sequence of mutations. The formalism thus discussed gives us a first prediction: \textit{in theories of class $\cS [\mathfrak{su} (2)]$ line defects are organized in mutation orbits}. For example the simple line defects in the $A_2$ Argyres-Douglas theory are listed in (\ref{L1A2framed})-(\ref{L5A2framed}); it is easy to see that they can all be obtained one from another, for example with the $\mut_{12,+}$ operator. In a certain sense, the theory contains only \textit{one} line defect! All the others can be obtained by mutations (and the non simple defects furthermore generated by sums). In general the mutation orbits will not contain a finite number of elements. We will study these families in a series of examples in the next Sections. 

\subsection{Line defects, quantum mutations and quantum dilogarithm identities} \label{qmut}

Now we would like to fill a gap in the previous arguments, and show that for a sequence of mutations which acts as the identity up to a permutation of the nodes and an overall sign change on the basis of charges, the corresponding operator $\mut$ is equivalent to a series of wall-crossings of BPS walls. We will see that this is indeed a consequence of a well known fact, established during the proof of the quantum dilogarithm identities associated with the generalized Donaldson-Thomas invariants. To do this we will need the theory of quantum cluster algebras \cite{FG2}, see \cite{keller,naka} for reviews.

We introduce a set of formal variables $\sfY_{\gamma}$ which generate the quantum torus $\torus_{\Gamma}$ and obey the $q$-commutation relations
\begin{equation}
\sfY_{\gamma_i} \, \sfY_{\gamma_j} = q^{\langle \gamma_i , \gamma_j \rangle} \, \sfY_{\gamma_i + \gamma_j} \ .
\end{equation}
The sets of variables $\sfy_{\gamma}$ and $\tilde{\sfy}_{\gamma}$ are recovered in the limit $q \longrightarrow -1$ and $q \longrightarrow +1$ respectively. Quantum mutations act on the quantum torus $\torus_\Gamma$ variables by adding or removing ``noncommutative" halos. They can be regarded as the composition of a quiver mutation and an automorphism given by the adjoint action of the quantum dilogarithm function. We will also use the notation $\torus_{\sfQ}$ to stress that the pairing is the one associated with the quiver $\sfQ$. In particular, given a quiver $\sfQ$ and its quiver mutation $\sfQ (t_m)$, we can construct the quantum torus $\torus_{\sfQ (t_m)}$. The change of pairing is really just a change in the basis of charges, and all the quantum tori are isomorphic.

Quiver mutations act as
\begin{equation}
\tau_{k,+} \left( \sfY_{\gamma_i} \right) = \left\{ \begin{matrix} \sfY_{\gamma_i}^{-1} & i = k \\[4pt] \sfY_{\gamma_i + \gamma_k \, [\langle \gamma_i , \gamma_k \rangle]_+ } & i \neq k
\end{matrix} \right. \ ,
\end{equation}
when crossing a wall $W (\gamma_k)$ in the clockwise order, and 
\begin{equation}
\tau_{k,-} \left( \sfY_{\gamma_i} \right) = \left\{ \begin{matrix} \sfY_{\gamma_i}^{-1} & i = k \\[4pt] \sfY_{\gamma_i - \gamma_k \, [\langle \gamma_i , \gamma_k \rangle]_-} & i \neq k
\end{matrix} \right. \ ,
\end{equation}
when crossing a wall $W (-\gamma_k)$ in the counterclockwise order. If we have a quiver mutation $\sfQ (t) \xrightarrow{\mu_k} \sfQ(t')$, then $\torus_{\sfQ (t')} \xrightarrow{	\tau_{k, \epsilon}} \torus_{\sfQ (t)}$ is simply the operator which expresses the basis of $\torus_{\sfQ (t')}$ in terms of the basis of $\torus_{\sfQ (t)}$. The two operators $\tau_{k,\pm}$ are not involutions, but their square acts on the variables $\sfY_{\gamma}$ as a change of basis of the form of a Picard-Lefschetz transformation (we still speak of involutions even if these are operators between different quantum tori, since all the quantum tori are isomorphic).

Similarly we consider the following automorphisms, going in the clockwise sense
\begin{equation}
\mathrm{Ad}  \left( \mathbb{E} (\sfY_{\gamma_k}) \right) \ \sfY_{\gamma_i} = \mathbb{E} (\sfY_{\gamma_k}) \, \sfY_{\gamma_i} \, \mathbb{E} (\sfY_{\gamma_k})^{-1} \ ,
\end{equation} 
and in the counterclockwise sense
\begin{equation}
\mathrm{Ad}  \left( \mathbb{E} (\sfY_{-\gamma_k})^{-1} \right) \ \sfY_{\gamma_i} = \mathbb{E} (\sfY_{-\gamma_k})^{-1} \, \sfY_{\gamma_i} \, \mathbb{E} (\sfY_{-\gamma_k}) \ .
\end{equation}
These automorphism act as operators inducing wall crossing transformations of formal line defects across BPS walls, as discussed in Section \ref{linedef}. 
A quantum mutation $\mut^q$ is defined as the composition of these transformations. Remarkably \cite{FG2,keller} the following two compositions coincide
\begin{equation} \label{mutdecomp}
\mut^q_k = \mathrm{Ad}  \left( \mathbb{E} (\sfY_{\gamma_k}) \right) \ \tau_{k,+} = \mathrm{Ad}  \left( \mathbb{E} (\sfY_{-\gamma_k})^{-1} \right) \, \tau_{k,-} \ .
\end{equation}
Quantum mutations are involutions. In particular given a sequence of quiver mutations
\begin{equation}
\sfQ (t_0) \xleftrightarrow{\mu_{i_1}} \sfQ (t_1)  \xleftrightarrow{\mu_{i_2}} \sfQ (t_2) \xleftrightarrow{\mu_{i_3}} \cdots  \xleftrightarrow{\mu_{i_m}} \sfQ (t_m) \ ,
\end{equation}
we have the corresponding operators acting on the quantum tori
\begin{equation}
\torus_{\sfQ (t_0)} \xleftarrow{\mut^q_{i_1}} \torus_{\sfQ (t_1)}  \xleftarrow{\mut^q_{i_2}} \torus_{\sfQ (t_2)} \xleftarrow{\mut^q_{i_3}} \cdots  \xleftarrow{\mut^q_{i_m}} \torus_{\sfQ (t_m)} \ .
\end{equation}

These ideas were made more systematic in \cite{keller,nagao,naka}. In particular the following holds: consider a quiver $\sfQ$ and act with a sequence of mutations ${\mbf \mu} = \left( \mu_{i_1} , \dots , \mu_{i_L} \right)$ to obtain a quiver $\sfQ (t_L)$. Let $C (t_k)$ be the matrix of $\mbf c$-vectors at the step $t_k$, with $k=0$ being the variable associated with the initial quiver $\sfQ$. The columns of $C (t_0)$ give the basis of BPS charges $\{ \gamma_i \}$ associated with the quiver $\sfQ$, i.e. $i \in \sfQ_0$. Consider now the composition of quantum mutations
\begin{equation}
\mut^q_{\mbf \mu} \equiv \mut^q_{i_1}  \cdots  \mut^q_{i_L} \ : \ \torus_{\sfQ (t_L)} \longrightarrow \torus_{\sfQ (t_0)} \ .
\end{equation}
Then the following quantum separation formula holds
\begin{equation}\label{quantumsep}
\mut^q_{\mbf \mu} = \mathrm{Ad} \left( \mathbb{E} ({\mbf \mu}) \right) \ \tau_{i_1 , \epsilon_1} \cdots \tau_{i_L , \epsilon_L} \ ,
\end{equation}
where the product of quantum dilogarithms $\mathbb{E} ({\mbf \mu})$ is defined via the $\mbf c$-vectors at each step $t_k$ as
\begin{equation}
\mathbb{E} ({\mbf \mu}) =  \mathbb{E} (\sfY^{\epsilon_1 \, \beta_1})^{\epsilon_1} \cdots  \mathbb{E} (\sfY^{\epsilon_L \, \beta_L})^{\epsilon_L} \ .
\end{equation}
Consider now the matrix $C (t_L)$ computed at the last step of the mutation sequence. If $C (t_L)$ is a permutation matrix, then we have $\mathbb{E} ({\mbf \mu}) = 1$ and obtain a quantum dilogarithm identity. If on the other hand the matrix $- C (t_L)$ is a permutation, the product of quantum dilogarithms $\mathbb{E} ({\mbf \mu})$ is the Kontsevich-Soibelman operator of Section \ref{genDT}, and contains the information about the spectrum of the BPS states in a given chamber \cite{Alim:2011ae,Alim:2011kw,Cecotti:2010fi,nagao,keller}.

This formalism has a more physical interpretation via line defects. We will give a graphical version of the argument of \cite{Gaiotto:2010be,Andriyash:2010qv} in terms of framed quivers. Assume that a certain theory is characterized by a certain BPS quiver, for example derived from an ideal triangulation via the M5 brane engineering. Consider a certain line defect, physically represented by a very heavy dyon. Ordinary lighter BPS states can bound and result in non-vanishing framed BPS degeneracies. The physical system is an arbitrarily complicated collection of mutually non local particles, bound to a very heavy core state; a rigid analog of the ``supersymmetric galaxy" of \cite{Andriyash:2010qv} in supergravity. This system is graphically represented by a framed quiver $\sfQ^f$. To this system we can associate a generating function in $\torus_{\sfQ}$
\begin{equation}
\scrL^q (\sfQ^f) = \sum_{\gamma} \,  \underline{\overline{\Omega}} (u , \gamma ; q) \, \sfY_{\gamma} \ .
\end{equation}
Assume now that we can find a sequence of quiver mutations $( \mu_{i_1} , \cdots , \mu_{i_L} )$ which leaves the framed quiver (and not only the underlying BPS quiver!) invariant, up to a permutation $\Pi$ of the nodes. That is
\begin{equation}
\mu_{i_L}  \cdots  \mu_{i_1} \, \sfQ^f = \Pi \, \sfQ^f \ .
\end{equation}
Then the corresponding sequence of quantum mutations is the identity on $\torus_{\Gamma}$ up to a permutation. In particular it leaves the generating function $\scrL^q (\sfQ^f)$ invariant up to a permutation:
\begin{equation}
\mut^q_{i_1}  \cdots  \mut^q_{i_L} \, \scrL^q (\sfQ^f) = \Pi \, \scrL^q ( \sfQ^f) \ .
\end{equation}
Now, using the quantum separation formula (\ref{quantumsep}) we find that the adjoint action of an ordered product of quantum dilogarithms acts as the identity on the line defect 
\begin{equation}
\mathrm{Ad} \left( \mathbb{E} (\sfY^{\epsilon_1 \, \beta_1})^{\epsilon_1} \cdots  \mathbb{E} (\sfY^{\epsilon_L \, \beta_L})^{\epsilon_L} \right) \ \scrL^q (\sfQ^f) = \scrL^q ( \sfQ^f)  \ .
\end{equation}
The reason for this is that the composition $ \tau_{i_1 ,\epsilon_1} \cdots \tau_{i_L , \epsilon_L} \, \Pi$ is by assumption the identity map $\torus_{\sfQ (t_L)} \xrightarrow{\simeq} \torus_{\sfQ (t_0)}$. Since the pairing between charges is by assumption non degenerate, the ordered product itself must act as the identity
\begin{equation} 
 \mathbb{E} (\sfY^{\epsilon_1 \, \beta_1})^{\epsilon_1} \cdots  \mathbb{E} (\sfY^{\epsilon_L \, \beta_L})^{\epsilon_L} \ \scrL^q (\sfQ^f) = \scrL^q ( \sfQ^f) \ .
\end{equation}
Since this argument is independent on the particular line defect we are considering, the following must hold
\begin{equation} 
 \mathbb{E} (\sfY^{\epsilon_1 \, \beta_1})^{\epsilon_1} \cdots  \mathbb{E} (\sfY^{\epsilon_L \, \beta_L})^{\epsilon_L}  = 1 \ ,  
\end{equation}
which is indeed a quantum dilogarithm identity \cite{keller}.

For example, the following graphical identity holds for a line operator in the $A_2$ Argyres-Douglas theory:
\begin{eqnarray}
\mu_1 \, \mu_2 \, \mu_1 \, \mu_2 \, \mu_1 \ \begin{matrix}
\xymatrix@C=5mm{  &  \bullet  \ L_1 & \\
 \ 2 \ \bullet \  & & \ \bullet \ 1 \ar[ll] \ar@<-0.5ex>[ul]  \ar@<0.5ex>[ul] 
} \end{matrix} &=& \begin{matrix}
\xymatrix@C=5mm{  &  \bullet  \ L_1 & \\
 \ 1 \ \bullet \  & & \ \bullet \ 2 \ar[ll] \ar@<-0.5ex>[ul]  \ar@<0.5ex>[ul] 
} \end{matrix} 
\nonumber \\
&=& \Pi_{1 \leftrightarrow 2} \begin{matrix}
\xymatrix@C=5mm{  &  \bullet  \ L_1 & \\
 \ 2 \ \bullet \  & & \ \bullet \ 1 \ar[ll] \ar@<-0.5ex>[ul]  \ar@<0.5ex>[ul] 
} \end{matrix} \ .
\end{eqnarray}
If we associate the variables $\tilde{\sfy}_1$ and $\tilde{\sfy}_2$ to the two quiver nodes, it is easy to compute the $\mbf c$-vectors (and done for example in \cite{keller})
\begin{equation}
\beta_1 = \gamma_1 \ , \ \beta_2 = \gamma_2 , \ , \beta_3 = - \gamma_1 , \ , \beta_4 = -\gamma_1 - \gamma_2 , \ , \beta_5 = - \gamma_2 \ ,
\end{equation}
and the tropical sign sequece $(+,+,-,-,-)$. Indeed in this way one recovers the pentagon identity (\ref{pentagon})
\begin{equation}
\bbE (\sfY_{\gamma_1}) \ \bbE (\sfY_{\gamma_2}) \bbE (\sfY_{\gamma_1})^{-1} \ \bbE (\sfY_{\gamma_1 + \gamma_2})^{-1} \ \bbE (\sfY_{\gamma_2})^{-1} = 1 \ .
\end{equation}

On the other hand, consider now the case where the sequence of quiver mutations acts as minus a permutation on the basis of BPS states $\{ \gamma_i \}$. In this case, the same arguments apply almost verbatim, and the result is that $\mathbb{E} (\mbf \mu)$ is the Kontsevich-Soibelman operator. Assume we can find two mutation sequences $\mbf{\mu}$ and $\mbf{\mu'}$ such that both act as minus a permutation, and that their action differ by a permutation. By slight abuse of language, let's say $\mbf{\mu} = \mbf{\mu'} \, \Pi$. Then
\begin{equation}
\mut^q_{\mbf{\mu}} \,  \scrL^q ( \sfQ^f) = \mut^q_{\mbf{\mu'}} \, \Pi \,  \scrL^q ( \sfQ^f)  \ ,
\end{equation} 
for any line defect $ \scrL^q ( \sfQ^f)$. This implies
\begin{equation} \label{AdonL}
\mathrm{Ad} \left( \mathbb{E} (\mbf \mu) \right) \,  \scrL^q ( \sfQ^f) = \mathrm{Ad} \left( \mathbb{E} (\mbf \mu') \right) \,  \scrL^q ( \sfQ^f) \ ,
\end{equation}
or equivalently the Kontsevich-Soibelman wall-crossing formula
\begin{equation}
\mathbb{E} (\mbf \mu) = \mathbb{E} (\mbf \mu') \ .
\end{equation}
Equation (\ref{AdonL}) implies that the sequence of quantum mutations $\mut^q_{\mbf{\mu}}$ acts as conjugation by the Kontsevich-Soibelman operator, up to a permutation and an overall sign change in the basis of charges. In other words this sequence of quantum mutations is the operator which corresponds to the crossing of all the BPS walls corresponding to the BPS stable states in a given chamber. The chamber is identified by the order of the sequence of mutations $\mbf \mu$. Take now the $q \longrightarrow +1$ limit. In this limit the quantum mutation operator $\mut^q_k$ reduces to the classical mutation operator $\mut_{k}$ acting on the commuting variables $\{ \tilde{\sfy}_{\gamma} \}$. In particular conjugation by the quantum dilogarithm $\bbE (\sfY_{\gamma})$ reduces to ordinary wall-crossing transformation (\ref{muy}), adding or removing halos as appropriate. We conclude that acting with the sequence of mutations $\mut_{\mbf \mu}$ on the $\{ \tilde{\sfy} \}$, with $\mbf \mu$ as above, indeed coincides with crossing all the BPS walls corresponding to the stable BPS spectrum. This is what we wanted to show, thus proving the claims of Section \ref{linedefcluster}.

It would be interesting to extend these arguments to string theory on local threefolds. The dynamics of BPS states on singular toric Calabi-Yau is similarly described by framed quivers \cite{Cirafici:2010bd}. This situation is technically more challenging since the relevant quivers are not mutation finite \cite{Cirafici:2011cd}.

Finally one remark about conventions. We could have equivalently defined quantum mutations by composing with $\mathrm{Ad}^{-1}$. In this case the same discussion applies word by word; the result would have given wall-crossing formula with the orientation of the operators reversed. This corresponds to the two different conventions in defining the Kontsevich-Soibelman automorphisms. Similar remarks apply for the operators $\mut_{k,+}$ and $\mut_{k,-}$. In this paper we are using the conventions of \cite{Gaiotto:2010be} for the mutation sequences; to change conventions the reader simply has to find a sequence of quiver mutations which generate the BPS spectrum in her conventions, and then compose quiver mutations with the wall-crossing transformation to obtain the operators $\mut$. 

\subsection{Tropical variables and IR labels}

At this stage we would like to add a brief comment about the IR labels attached to the line defects. In the UV line defects are typically labelled by an appropriate sub-lattice of the co-character lattice $\Lambda_G$ for a lagrangian field theory with structure group $G$. It was argued in \cite{Gaiotto:2010be} that the appropriate IR label can be extracted from the asymptotic behavior of the expansion (\ref{deftraceL}) of the vacuum expectation value $\langle L \rangle$ in the Darboux coordinates associated with the WKB triangulation. More practically one considers the $R \longrightarrow \infty$ and $\zeta \longrightarrow 0$ asymptotics of the functions $\cY_{\gamma}$, in the notation of \cite{Gaiotto:2010be}. The leading term in the expansion of $\langle L \rangle$ is associated with a certain charge $\gamma_t$ which is called the \textit{tropical label}. This label depends on the IR parameters and undergoes wall-crossings upon crossing \textit{anti-BPS walls}, defined as the locus where the leading behavior of $\langle L \rangle$ can change
\begin{equation} 
\overline{W} (\gamma) = \{ (u,\zeta) \, \vert \, \cZ_{\gamma} (u)  / \zeta \in \ii \real_+ \ \text{and} \ \Omega(\gamma;u) \neq 0 \} \ .
\end{equation}
Due to the relation between the triangulation of surfaces and cluster algebras, it is natural to wonder if there is room in our formalism for such labels. Indeed it is so, although the relation is not quite direct. To mimic the correct asymptotic behavior of the $\mathcal{Y}$ functions, we have to introduce a modified tropical sum $\hat{\oplus}$
\begin{equation}
\prod_{i \in \sfQ_0} \, \tilde{\sfy}_i^{a_i} \, \hat{\oplus} \, \prod_{i \in \sfQ_0} \, \tilde{\sfy}_i^{b_i} = \prod_{i \in \sfQ_0} \, \tilde{\sfy}_i^{\max (a_i , b_i)} \ ,
\end{equation}
which involves the $\max$ function and not the $\min$ as in (\ref{tropplus}). This is simply because we have to extract the leading behavior as the functions $\cY \longrightarrow \infty$. The two addition operations are related by
\begin{equation} \label{reloplus}
\prod_{i \in \sfQ_0} \, \tilde{\sfy}_i^{a_i} \, \hat{\oplus} \, \prod_{i \in \sfQ_0} \, \tilde{\sfy}_i^{b_i}  = \left( \prod_{i \in \sfQ_0} \, \tilde{\sfy}_i^{-a_i} \, \oplus \, \prod_{i \in \sfQ_0} \, \tilde{\sfy}_i^{-b_i}  \right)^{-1} \ ,
\end{equation}
and therefore the tropical labels of \cite{Gaiotto:2010be} can indeed be extracted from the tropical limit of our $\tilde{\sfy}$ variables. That the relation is not direct is expected, since tropical labels are not expected to jump at BPS walls, but at anti-BPS walls. Note that in the central charge plane, the location of an anti-BPS wall is at a $\frac{\pi}{2}$ rotation respect to the BPS ray. Similarly their behavior is regulated by a WKB triangulation with angle $\chi = \theta - \frac{\pi}{2}$, as discussed in \cite{Gaiotto:2010be}. 

We can push the formalism a bit further and use an appropriate quiver to study the tropical labels of line defects. One can indeed think of the anti-BPS walls as themselves associated with nodes of a particular quiver. Since the anti-BPS walls are at a $\pi/2$ angle respect to the BPS walls, we get a copy of the original quiver; however the physical interpretation is different. The anti-BPS walls ``spectrum" can be generated by mutations starting from this  quiver. Again we can apply all the formalism of cluster algebras, but this time we take the tropical limit with $\hat{\oplus}$. We call this quiver the tropical quiver $\sfQ^{trop}$. The formalism of Sections \ref{linedefcluster} and \ref{qmut} goes through and we can now study wall-crossing of the tropical charges by studying mutations of $\sfQ^{trop}$. 

Consider now a theory where there is a simple line operator given directly by a cluster coordinate and not by a sum of monomials, such as the theories in the $A_p$ series. We can then obtain other line defects by applying the operators $\mut$ in the appropriate sequence. Then the tropical labels of the new defects, are the $\mbf c$-vectors of the quiver $\sfQ^{trop}$, in the order obtained from the $\mut$-sequence, where now the $\mbf c$-vectors are computed with $\hat{\oplus}$. 

Note that because of (\ref{reloplus}), we could also just invert all the cluster variables $\{ \tilde{\sfy}_{\gamma} \}$, by changing what we mean by particle and anti-particle. The resulting cluster algebras are isomorphic, if we also oppose the quiver. Therefore the tropical labels of  \cite{Gaiotto:2010be}, will appear as $\mbf c$-vectors of $\sfQ^{op}$ up to a sign, where we take the tropical limit with $\oplus$. We have verified this explicitly in a few cases. This support the conclusions of \cite{Gaiotto:2010be} that every stable BPS state $\gamma$ (indeed obtained as a $\mbf c$-vector from $\sfQ$) will appear as a tropical label (since the $\mbf c$-vectors of $\sfQ^{op}$ are the same, up to a sign). On the other hand, we also expect line defects whose generating function cannot be obtained from a single initial cluster variable by applying a sequence of operators $\mut$; this means that all the elements in the mutation orbit are non trivial sums of monomials in the initial cluster variables. Of course, we can always compute the tropical label of these defects directly. We consider this discussion only as an aside comment, and hope to return to the relation between tropical labels of line operators and tropical limits of cluster algebras in the future.

\subsection{Admissible framings} \label{admi}

As we have already explained we are not interested in any lamination. Laminations representing line operators must obey certain conditions. So far we have considered the following problem: given a line defect represented as a lamination on a Riemann surface, how to obtain its framed quiver. Assuming the framed spectrum of the latter is known (computed for example from the Leavitt paths on the extended quiver), we can then generate algorithmically further line defects via mutations. However it would be interesting to consider the opposite problem: which framings correspond to physical line defects. We call these \textit{admissible} framings. The reason one is interested in this problem is that if it were possible to give a purely algebraic characterization of line defects in terms of quiver framing, one could construct line defects for more general theories, which are not associated with an ideal triangulation of a Riemann surface. Indeed, since BPS quivers are not necessarily associated with Riemann curves but are a more general framework, one can hope the same to be true for our framings. Indeed if one thinks of a line defect as an infinitely heavy dyon, it looks like it would always be possible to express the low energy dynamics of the system in term of a framed quiver. For recent work in this direction see \cite{Chuang:2013wt}.

Unfortunately such a characterization is combinatorially very tricky even in the case of simple quivers arising from ideal triangulations. The method of framed quivers is not very efficient in this respect. What we can do very easily is to generate admissible framings, as complicated as they come, starting from a known one and applying quiver mutations. Therefore the best strategy, if a quiver is derived from an ideal triangulation, is to use a bit of geometric intuition to construct a physical line defect on the extended quiver, translate this into an admissible framing using the discussion of Section \ref{sinksource} and then generate new admissible framings by quiver mutations.

While this strategy works for quivers corresponding to theories of class $\cS [\mathfrak{su} (2)]$, one would need a different approach for more general theories. In Sections \ref{gluing} and \ref{boundary} we will discuss a series of gluing rules to generate new framing by cut and join operations on the framed quivers. Again we will derive these rules for theories of class $\cS [\mathfrak{su} (2)]$, but the logic behind them seems to be more general and apply to any quiver. Unfortunately the analysis would have to be done on a case by case basis.

\section{Argyres-Douglas superconformal theories} \label{ADtheories}

In this Section we exemplify the formalism developed so far by looking at line defects in superconformal theories of Argyres-Douglas type. Theories of this type are obtained by a certain scaling limit from $\PP^1$ with regular punctures. The quadratic differential is of the form
\begin{equation}
\lambda^2 = \frac{P^{(1)}_{2n +2} (z)}{(P^{(2)}_{n+3} )^2} \ (\dd z)^2 \ ,
\end{equation}
and has only second order poles at the punctures. Here $P^{(i)}_k$ is a polynomial of degree $k$. In the aforementioned scaling limit, certain zeros of $P^{(2)}_{n+3}$ collide and give rise to irregular singularities. These theories are associated with BPS quivers of the form of a Dynkin graph, and mutations thereof. They are complete and vector-less and have an ADE classification. They always have a minimal chamber, whose stable BPS states correspond to the simple roots of the corresponding Lie algebra, and a maximal chamber whose stable states correspond to the positive roots \cite{Cecotti:2010fi}.

\subsection{Example: $A_3$ theory reprised}

Now we are going to use the results in \ref{matrix} to generate other line defects in the $A_3$ Argyres-Douglas superconformal theory. Recall that in \ref{matrix} we used paths in the extended quiver to write down the framed BPS spectrum of three line defects as an expansion over formal variables associated with the nodes. More precisely we could associate to a lamination an element of the Leavitt path algebra $\sfL \sfA_{\tilde{\sfQ}}$ constructed from the extended quiver $\tilde{\sfQ}$. Now we would like to use the framed BPS quiver formalism to generate other line defects via cluster mutations. 

Consider a chamber with three states $\gamma_1$, $\gamma_2$ and $\gamma_3$ with $\langle \gamma_1 , \gamma_2 \rangle = +1$ and $\langle \gamma_2 , \gamma_3 \rangle= +1$, and write the associated variables $\sfx = \sfy_{\gamma_3}$, $\sfy = \sfy_{\gamma_2}$ and $\sfz=\sfy_{\gamma_1}$. The BPS quiver is 
\begin{equation}
\xymatrix@C=8mm{ 
 \ \sfx  \ \bullet \  & \ \bullet \ \sfy \  \ar[l] & \ \bullet \ \sfz \ar[l]
} \ .
\end{equation}
Consider now the line operator we called $L^{c_1}_1$ in Section \ref{matrix} and whose framed spectrum is $ \scrL^{c_1}_1  = \sfz$. Its framed BPS quiver is
\begin{equation}
\begin{matrix}
\xymatrix@C=8mm{  &  \bullet  \ L_1^{c_1} & \\
 \ \sfx  \ \bullet \  & \ \bullet \ \sfy \  \ar[l] & \ \bullet \ \sfz  \ar[l] \ar@<-0.5ex>[ul] \ar@<0.5ex>[ul]  
} \end{matrix} \ .
\end{equation}
We now mutate on the nodes, starting from $\sfx$, then $\sfy$ and finally $\sfz$
\begin{equation} 
\mu_\sfz \mu_\sfy \mu_\sfx \left( \begin{matrix}
\xymatrix@C=8mm{  &  \bullet  \ L_1^{c_1} & \\
 \ \sfx  \ \bullet \  & \ \bullet \ \sfy \  \ar[l] & \ \bullet \ \sfz  \ar[l] \ar@<-0.5ex>[ul] \ar@<0.5ex>[ul]  
} \end{matrix} \right)
 \simeq \begin{matrix}
\xymatrix@C=10mm{  &  \bullet  \ L_3^{c_1}  \ar@<-0.5ex>[dr] \ar@<0.5ex>[dr] &   \\
 \ \sfx  \ \bullet \  & \ \bullet \ \sfy \    \ar@<-0.5ex>[u] \ar@<0.5ex>[u] 
 \ar[l] & \ \bullet \ \sfz  \ar[l]
 } \end{matrix} \ ,
\end{equation}
and generate the new line operator $L_3^{c_1}$ (we have chosen the labeling to uniformize with \cite{Gaiotto:2010be}). Its framed BPS spectrum its therefore $ \scrL_3^{c_1}  = \mut_{\sfx,+} \, \mut_{\sfy,+} \, \mut_{\sfz,+} \, \scrL_1^{c_1}  = \sfy + \frac{\sfy}{\sfz} + \frac{1}{\sfz} $, as can be confirmed by looking at the corresponding path in the extended quiver. We can act again on the new line defect with the same mutation sequence and generate another line defect, and so on. We find
\begin{eqnarray}
 \begin{matrix}
\xymatrix@C=10mm{  &  \bullet  \ L_3^{c_1}  \ar@<-0.5ex>[dr] \ar@<0.5ex>[dr]  &   \\
 \ \sfx  \ \bullet \  & \ \bullet \ \sfy \    \ar@<-0.5ex>[u] \ar@<0.5ex>[u]   \ar[l] & \ \bullet \ \sfz  \ar[l]
} \end{matrix}
&
\stackrel{\mu_\sfz \, \mu_\sfy \, \mu_\sfx}{\Longrightarrow}
&
\begin{matrix}
\xymatrix@C=10mm{  &  \bullet  \ L_5^{c_1}  \ar@<-0.5ex>[d] \ar@<0.5ex>[d] &   \\
 \ \sfx  \ \bullet \ \ar@<-0.5ex>[ur] \ar@<0.5ex>[ur]   & \ \bullet \ \sfy \   \ar[l] & \ \bullet \ \sfz   \ar[l] 
}
\end{matrix} \\[4pt]
\stackrel{\mu_\sfz \, \mu_\sfy \, \mu_\sfx}{\Longrightarrow}
\begin{matrix}
\xymatrix@C=10mm{  &  \bullet  \ L_7^{c_1} \ar@<-0.5ex>[dl] \ar@<0.5ex>[dl]  &   \\
 \ \sfx  \ \bullet \     & \ \bullet \ \sfy \    \ar[l] & \ \bullet \ \sfz  \ar[l]
}
\end{matrix} 
&
\stackrel{\mu_\sfz \, \mu_\sfy \, \mu_\sfx}{\Longrightarrow}
&
\begin{matrix}
\xymatrix@C=10mm{  &  \bullet  \ L_9^{c_1} &   \\
 \ \sfx  \ \bullet \   \ar@<-0.5ex>[ur] \ar@<0.5ex>[ur]  & \ \bullet \ \sfy \  \ar[l] & \ \bullet \ \sfz  \ar[l]
}
\end{matrix} \\[4pt]
\stackrel{\mu_\sfz \, \mu_\sfy \, \mu_\sfx}{\Longrightarrow}
\begin{matrix}
\xymatrix@C=10mm{  &  \bullet  \ L_{11}^{c_1} \ar@<-0.5ex>[dr] \ar@<0.5ex>[dr]    & \\
 \ \sfx  \ \bullet \  & \ \bullet \ \sfy \  \ar[l] & \ \bullet \ \sfz \ar[l]
} \end{matrix} 
&
\stackrel{\mu_\sfz \, \mu_\sfy \, \mu_\sfx}{\Longrightarrow}
&
\begin{matrix}
\xymatrix@C=10mm{  &  \bullet  \ L_1^{c_1}  & \\
 \ \sfx  \ \bullet \  & \ \bullet \ \sfy \  \ar[l] & \ \bullet \ \sfz  \ar@<-0.5ex>[ul] \ar@<0.5ex>[ul]  \ar[l]
} \end{matrix} \ .
\end{eqnarray}
We see that at the end we have returned to the line defect we started with: indeed we see that line defects come in cluster mutation orbits! Of course all of these statements have a geometrical counterpart from the point of view of the triangulation associated with the BPS quiver. Applying our algorithm to compute the framed BPS spectra, we see
\begin{eqnarray}
 \scrL_5^{c_1}  &=& \mut_{\sfx,+} \, \mut_{\sfy,+} \, \mut_{\sfz,+} \  \scrL_3^{c_1}  = \sfx + \frac{\sfx}{\sfy} + \frac{1}{\sfy} \ , \\
 \scrL_7^{c_1}  &=& \mut_{\sfx,+} \, \mut_{\sfy,+} \, \mut_{\sfz,+}  \ \scrL_5^{c_1}  = \frac{1}{\sfx} \ , \\
 \scrL_9^{c_1}  &=& \mut_{\sfx,+} \, \mut_{\sfy,+} \, \mut_{\sfz,+} \ \scrL_7^{c_1}  = \sfx + \sfx \, \sfy + \sfx \, \sfy \, \sfz \ , \\
 \scrL_{11}^{c_1}  &=& \mut_{\sfx,+} \, \mut_{\sfy,+} \, \mut_{\sfz,+} \  \scrL_9^{c_1}  =  \frac{1}{\sfx \, \sfy \, \sfz} + \frac{1}{\sfy \, \sfz}  + \frac{1}{\sfz} \ .
\end{eqnarray} 
These results agree with Section 10.2 of \cite{Gaiotto:2010be} (we have chosen the same labeling for the line defects). 

To further exemplify our methods, we now turn to a different chamber and a different quiver description. Consider the BPS quiver
\begin{equation}
\xymatrix@C=10mm{  
 \ \sfx  \ \bullet \ & \ \bullet \ \sfy \   \ar[l]  \ar[r]   & \  \bullet \ \sfz
} \ .
\end{equation}
We have chosen a chamber with four states $\gamma_1$, $\gamma_2$, $\gamma_3$ (which again label the three nodes and will again be denoted by $\sfz = \sfy_{\gamma_1}$, $\sfy = \sfy_{\gamma_2}$ and $\sfx = \sfy_{\gamma_3}$) and $\gamma_1 + \gamma_2$, univoquely characterized by  $\arg \cZ_{\gamma_3} < \arg \cZ_{\gamma_2}  < \arg \cZ_{\gamma_1} $. The defect which we called $L_2^{c_3}$ in figure \ref{A3muL1L2} is represented by the framed BPS quiver
\begin{equation}
\begin{matrix}
\xymatrix@C=10mm{  &  \bullet  \ L   & \\
 \ \sfx  \ \bullet  \ & \ \bullet \ \sfy \   \ar[l]  \ar[r]  \ar@<-0.5ex>[u] \ar@<0.5ex>[u]  & \  \bullet \ \sfz 
} \end{matrix} \ .
\end{equation}
Consider now the sequence of mutations $\mu_\sfx \, \mu_\sfz \, \mu_\sfy \, \mu_\sfz $ (where $\mu_\sfz$ acts first), which is the same sequence used to generate the spectrum. The BPS quiver is invariant under this operation, only after a permutation of the labels $\sfz $ and $ \sfy$, which we will call $\Pi_{\sfz \sfy}$. We use this operator to generate new defects
\begin{eqnarray}
&&
 \begin{matrix}
\xymatrix@C=10mm{  &  \bullet  \ L_2^{c_3}   & \\
 \ \sfx  \ \bullet  \ & \ \bullet \ \sfy \   \ar[l]  \ar[r]  \ar@<-0.5ex>[u] \ar@<0.5ex>[u]  & \  \bullet \ \sfz 
}
\end{matrix}
\begin{array}{c}\Pi_{\sfz \sfy} \mu_{\sfx \sfz \sfy \sfz} \\ \Longrightarrow \end{array}
\begin{matrix}
\xymatrix@C=8mm{  &  \bullet  \ L_6^{c_3}   \ar@<-0.5ex>[d] \ar@<0.5ex>[d]  &   \\
 \ \sfx    \ \bullet \   &  \ \bullet \ \sfy \  \ar[l] \ar[r]  &  \ \bullet \ \sfz 
}
\end{matrix} \\[4pt]
&&
\begin{array}{c} \Pi_{\sfz \sfy} \mu_{\sfx \sfz \sfy \sfz} \\ \Longrightarrow \end{array}
 \begin{matrix}
 \xymatrix@C=8mm{  &  \bullet  \ L_4^{c_3}   \ar@<-0.5ex>[d] \ar@<0.5ex>[d]   &   \\
 \ \sfx     \ \bullet \  \ar@<-0.5ex>[ur] \ar@<0.5ex>[ur]   & \  \bullet \ \sfy \  \ar[l] \ar[r] &  \ \bullet \ \sfz   \ar@<-0.5ex>[ul] \ar@<0.5ex>[ul]  
}
\end{matrix}
\begin{array}{c} \Pi_{\sfz \sfy} \mu_{\sfx \sfz \sfy \sfz} \\ \Longrightarrow \end{array}
 \begin{matrix}
\xymatrix@C=10mm{  &  \bullet  \ L_2^{c_3}   & \\
 \ \sfx  \ \bullet  \ & \ \bullet \ \sfy \   \ar[l]  \ar[r]  \ar@<-0.5ex>[u] \ar@<0.5ex>[u]  & \  \bullet \ \sfz 
}
\end{matrix} \ .
\end{eqnarray}
According to our prescription, we can obtain the framed degeneracies by applying the appropriate mutation sequence on $\langle L_2^{c_3} \rangle$. Taking into account the permutation, we find
\begin{eqnarray}
 \scrL_6^{c_3}  &=& \mut_{\pi(\sfx),+} \, \mut_{\pi(\sfz),+} \, \mut_{\pi(\sfy),+} \, \mut_{\pi(\sfz),+} \, \Pi_{\sfz \sfy} \ \sfy = \mut_{\sfy,+} \, \mut_{\sfz,+} \, \mut_{\sfy,+} \, \mut_{\sfx,+} \, \sfz \nonumber \\
&=& \frac{1}{\sfy} + \frac{1}{\sfx \, \sfy} + \frac{1}{\sfy \, \sfz} + \frac{1}{\sfx \, \sfy \, \sfz} \ , \\
 \scrL_4^{c_3}  &=& \mut_{\sfy,+} \, \mut_{\sfz,+} \, \mut_{\sfy,+} \, \mut_{\sfx,+}  \ \left( \scrL_6^{c_3} \vert_{\sfz \leftrightarrow \sfy} \right) \nonumber \\
&=&   \frac{1}{\sfy} + \frac{\sfx}{\sfy} + \sfx + \frac{\sfz}{\sfy} + \frac{\sfx \, \sfz}{\sfy} + \sfz +2 \, \sfx \, \sfz + \sfx \, \sfy \, \sfz \ .
\end{eqnarray}
These results agree with  \cite{Gaiotto:2010be} up to cluster transformations.

The main lesson we have learned is that if we somehow know the framed spectrum of a line defect, we can use the framed BPS quiver formalism to generate other line defects. We see that line defects come in families, which we have called mutation orbits, obtained by repeatedly applying a sequence of mutations on a starting defect. Further ahead in this paper, we will discuss gluing rules for framed BPS quivers, and will learn how to glue together framed sub-quivers to obtain a line operator in a new theory. The combination of gluing rules to construct defects in quantum field theories from known defects in sub-theories (or decoupled limits) and of the framed BPS quiver formalism, allow us in principle to generate a large class of line defects.

\subsection{The $D_4$ superconformal theory}

Consider now the Argyres-Douglas $D_4$ theory. This theory can be engineered by compactifying the $(2,0)$ superconformal theory on a sphere $\cC = \PP^1$ with a regular puncture and an irregular singularity of order $p=2+c=6$. Quivers for theories in the $D$-series have $6g-6+3+c+3=c$ nodes, the number of marked points associated with the irregular singularity. The framed BPS quiver 
\begin{equation}
\begin{matrix}
\xymatrix@C=8mm{   \bullet  \, \sfx_2  \ar[rd] &     &   \\
& \bullet \, \sfx_3  \ar[dl] & \bullet \, \sfx_4 \ar[l] \\
\bullet \, \sfx_1  & &    
} \end{matrix} \ ,
\end{equation}
represents a chamber with four states $\gamma_1$, $\gamma_2$, $\gamma_3$ and $\gamma_4$ with $\arg \cZ_{\gamma_4} >\arg \cZ_{\gamma_2} >\arg \cZ_{\gamma_3} >\arg \cZ_{\gamma_1}$ and to each node we associate the variables $\sfx_i = \sfy_{\gamma_i}$.
\begin{figure}
\centering  
\def\svgwidth{8cm}
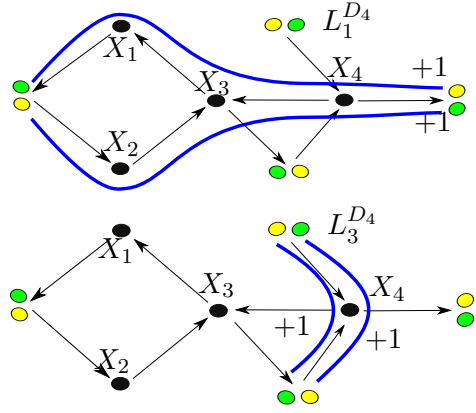
\caption{Paths in the extended quiver corresponding to the line defects $L_1^{D_4}$ and $L_3^{D_4}$.}
\label{D4example}
\end{figure}
We will now consider the two line defects depicted in Figure \ref{D4example}. The associated framed indices are
\begin{eqnarray}
 \scrL_1^{D_4}  &=& P_{YG} \left( D \, M_{\sfx_2}  D \, M_{\sfx_3}  U M_{\sfx_4}  D \right) \, P_{GY} \left( U \, M_{\sfx_1}  U \, M_{\sfx_3}  U \, M_{\sfx_4}  D \right) = \frac{\sqrt{\sfx_2}}{\sqrt{\sfx_1} \, \sfx_4}  \ , \\[4pt]
\scrL_3^{D_4}  &=& P_{GY} \left( D \, M_{\sfx_4} \, U \right) \, P_{YG} \left( D \, M_{\sfx_4} \, U \right) = \sfx_4 \ .
\end{eqnarray}
We can generate other defects using the framed BPS quiver formalism. Firstly we construct the framed quivers associated with the line defects of figure \ref{D4example} using the sink-source rules described in Section \ref{sinksource} and in figure \ref{shearquiver}. Let us begin with $L_1^{D_4}$. The mutation orbit of these defect consists only in another element $L_2^{D_4}$
\begin{equation}
\begin{matrix}
\xymatrix@C=8mm{   \bullet  \, \sfx_2  \ar[rd] &  &  \bullet L_1^{D_4} \ar@<-0.5ex>[d] \ar@<0.5ex>[d]   \\
& \bullet \, \sfx_3 \ar[ur]  \ar[dl] & \bullet \, \sfx_4 \ar[l] \\
\bullet \, \sfx_1  & &    
}
\end{matrix}
\begin{array}{c}  \mu_{\sfx_1} \mu_{\sfx_3} \mu_{\sfx_2}  \mu_{\sfx_4} \\ \Longrightarrow \end{array}
\begin{matrix}
\xymatrix@C=8mm{   \bullet  \, \sfx_2  \ar[rd] &  &  \bullet L_2^{D_4}  \ar[ll] \\
& \bullet \, \sfx_3   \ar[dl] & \bullet \, \sfx_4 \ar[l] \ar[u] \\
\bullet \, \sfx_1 \ar@/_1.5pc/[uurr]  & &    
}
\end{matrix} \ ,
\end{equation}
and from the framed BPS quiver we see
\begin{equation}
 \scrL_2^{D_4}  = \mut_{\sfx_4,+} \mut_{\sfx_2,+} \mut_{\sfx_3,+} \mut_{\sfx_1,+} \, \left( \frac{\sqrt{\sfx_2}}{\sqrt{\sfx_1}} \, \frac{1}{\sfx_4} \right) = \left( \frac{\sqrt{\sfx_1}}{\sqrt{\sfx_2}} \, \sfx_4 \right) \ .
\end{equation}
The cluster orbit of $L_3^{D_4}$ has period $4$
\begin{eqnarray} \nonumber
\begin{matrix}
\xymatrix@C=8mm{   \bullet  \, \sfx_2  \ar[rd] &  &  \bullet L_3^{D_4}    \\
& \bullet \, \sfx_3 \ar[dl] & \bullet \, \sfx_4 \ar[l] \ar@<-0.5ex>[u] \ar@<0.5ex>[u] \\
\bullet \, \sfx_1  & &    
}
\end{matrix}
\begin{array}{c}  \mu_{\sfx_1} \mu_{\sfx_3} \mu_{\sfx_2}  \mu_{\sfx_4} \\ \Longrightarrow \end{array}
\begin{matrix}
\xymatrix@C=8mm{   \bullet  \, \sfx_2  \ar[rd] &  &  \bullet L_4^{D_4} \ar@<-0.5ex>[d] \ar@<0.5ex>[d]   \\
& \bullet \, \sfx_3  \ar[dl] & \bullet \, \sfx_4 \ar[l] \\
\bullet \, \sfx_1  & &    
}
\end{matrix}
\\[4pt]  \nonumber
\begin{array}{c}  \mu_{\sfx_1} \mu_{\sfx_3} \mu_{\sfx_2}  \mu_{\sfx_4} \\ \Longrightarrow \end{array}
\begin{matrix}
\xymatrix@C=8mm{  \bullet  \, \sfx_2  \ar[rd] &  &   \\
\bullet L_5^{D_4} \ar@<-0.5ex>[u] \ar@<0.5ex>[u]   & \bullet \, \sfx_3  \ar[dl] & \bullet \, \sfx_4 \ar[l] \\
\bullet \, \sfx_1  \ar@<-0.5ex>[u] \ar@<0.5ex>[u]  & &    
}
\end{matrix}
\begin{array}{c}  \mu_{\sfx_1} \mu_{\sfx_3} \mu_{\sfx_2}  \mu_{\sfx_4} \\ \Longrightarrow \end{array}
\begin{matrix}
\xymatrix@C=8mm{   \bullet  \, \sfx_2  \ar[rd] &  &  \bullet L_6^{D_4} \ar@<-0.5ex>[d] \ar@<0.5ex>[d]   \\
& \bullet \, \sfx_3 \ar@<-0.5ex>[ur] \ar@<0.5ex>[ur]   \ar[dl] & \bullet \, \sfx_4 \ar[l] \\
\bullet \, \sfx_1  & &    
}
\end{matrix} \ ,
\end{eqnarray}
and the framed BPS spectrum is obtained by mutations from $L_3^{D_4}$
\begin{eqnarray}
 \scrL_4^{D_4}  &=& \mut_{\sfx_4,+} \mut_{\sfx_2,+} \mut_{\sfx_3,+} \mut_{\sfx_1,+} \, \sfx_4 = \frac{1}{\sfx_4} + \frac{1}{\sfx_3 \, \sfx_4} + \frac{1}{\sfx_4 \, \sfx_3 \, \sfx_1} \ , \\ \nonumber
 \scrL_5^{D_4}  &=& \left( \mut_{\sfx_4,+} \mut_{\sfx_2,+} \mut_{\sfx_3,+} \mut_{\sfx_1,+} \right)^2 \, \sfx_4 \\  \nonumber
&=& \sfx_1 + \frac{1}{\sfx_2} + 2 \frac{\sfx_1}{\sfx_2} + \sfx_1 \, \sfx_3 + \frac{\sfx_1 \, \sfx_3}{\sfx_2} + \frac{1}{\sfx_2 \, \sfx_3} + \frac{\sfx_1}{\sfx_2 \, \sfx_3} + \frac{\sfx_1 \, \sfx_4}{\sfx_2}  \\ && + \sfx_1 \, \sfx_3 \, \sfx_4 + \frac{\sfx_1 \, \sfx_3 \, \sfx_4}{\sfx_2} \ ,  \\
 \scrL_6^{D_4}  &=& \left( \mut_{\sfx_4,+} \mut_{\sfx_2,+} \mut_{\sfx_3,+} \mut_{\sfx_1,+} \right)^3 \, \sfx_4 = \sfx_3 + \sfx_2 \, \sfx_3 + \frac{1}{\sfx_4} + \frac{\sfx_3}{\sfx_4} + \frac{\sfx_2 \, \sfx_3}{\sfx_4} \ .
\end{eqnarray}

\section{$SU(2)$ gauge theories} \label{SU2gauge}

We will now discuss the framed BPS quivers for $SU(2)$ with matter, reproducing and extending the results of \cite{Gaiotto:2010be}. Again we aim to show how to use our methods, and not towards a complete classification of defects. In the simplest case of pure super Yang-Mills, the curve $\cC$ is a $\PP^1$ with two ideal boundaries, each with a marked point. The quadratic differential looks like
\begin{equation}
\lambda^2 = \left( \frac{\Lambda^2}{z^3} + \frac{2 \, u}{z^2} + \frac{\Lambda^2}{z} \right) \ (\dd z)^2 \ ,
\end{equation}
$\Lambda$ being the strong coupling scale. Adding massive flavors corresponds to regular singularities on $\cC$, whose residue of the Seiberg-Witten differential is precisely the mass. The appropriate curves and triangulations were discussed at length in \cite{Gaiotto:2009hg}.

The corresponding BPS quiver have all the form \cite{Cecotti:2011rv,Alim:2011ae,Alim:2011kw}
\begin{equation}
\begin{matrix}
\xymatrix@C=8mm{\bullet  \gamma_{i_1} \ar[dr] & \cdots  &  \cdots & \cdots & \bullet \gamma_{i_{N_f}}  \ar[dlll]  \\ &
 \bullet \, \gamma_1    \ar@<-0.5ex>[rr] \ar@<0.5ex>[rr]  & & \bullet \, \gamma_2  \ar[ulll] \ar[ur]
}
\end{matrix} \ ,
\end{equation}
and mutations thereof. The number nodes $\gamma_{i_k}$ corresponds to the number of flavors. It will be sometime useful to use a mutation equivalent form of this quivers.

\subsection{Pure $SU(2)$}

The case of pure $SU(2)$ was studied in detail in  \cite{Gaiotto:2010be} and already discussed in Section \ref{genDT}. We will use it as a warm up, to exemplify our methods. The idea is to start with a very simple line defect, for example such that its framed BPS spectrum is given by a single $\{ \tilde{\sfy}_{\gamma} \}$ variable, and then obtain new line defects by applying appropriate sequences of mutations. The $SU(2)$ quiver has only two nodes, labelled by $\gamma_1$ and $\gamma_2$ and we label the corresponding variables as $\tilde{\sfy}_{\gamma_1} = \sfx_1$ and $\tilde{\sfy}_{\gamma_2} = \sfx_2$. It is  easy to see, using the formalism developed in Section \ref{matrix}, that the generating function of the defect $L_{0,-2}^{SU(2)}$ is simply given by $\frac{1}{\sfx_2}$. Therefore we can start from this defect, and start applying the mutation sequence $\mu_{\sfx_2 \, \sfx_1}$. We obtain 
\begin{eqnarray}
\begin{matrix}
\xymatrix@C=8mm{   &  \bullet L_{0,-2}^{SU(2)}  \ar[dr]^2 &   \\
 \bullet \, \sfx_1  \ar@<-0.5ex>[rr] \ar@<0.5ex>[rr]  & & \bullet \, \sfx_2 
}
\end{matrix}
\begin{array}{c}  \mu_{\sfx_2} \mu_{\sfx_1} \\ \Longrightarrow \end{array}
\begin{matrix}
\xymatrix@C=8mm{   &  \bullet L_{-4,2}^{SU(2)}  \ar[dl]^4 &   \\
 \bullet \, \sfx_1  \ar@<-0.5ex>[rr] \ar@<0.5ex>[rr]  & & \bullet \, \sfx_2 \ar[ul]_2
}
\end{matrix}
\\[4pt]
\begin{array}{c}  \mu_{\sfx_2} \mu_{\sfx_1} \\ \Longrightarrow \end{array}
\begin{matrix}
\xymatrix@C=8mm{   &  \bullet L_{-8,6}^{SU(2)}  \ar[dl]^8 &   \\
 \bullet \, \sfx_1  \ar@<-0.5ex>[rr] \ar@<0.5ex>[rr]  & & \bullet \, \sfx_2  \ar[ul]_6
}
\end{matrix}
\begin{array}{c}  \mu_{\sfx_2} \mu_{\sfx_1} \\ \Longrightarrow \end{array}
\begin{matrix}
\xymatrix@C=8mm{   &  \bullet L_{-12,10}^{SU(2)} \ar[dl]^{12}  &   \\
 \bullet \, \sfx_1  \ar@<-0.5ex>[rr] \ar@<0.5ex>[rr]  & & \bullet \, \sfx_2 \ar[ul]_{10}
}
\end{matrix} \ .
\end{eqnarray}
The framed BPS spectrum can be generated via the operator $\mut_{\sfx_1 \, \sfx_2,+}$
\begin{eqnarray}
 \scrL_{0,-2}^{SU(2)}  &=& \frac{1}{\sfx_2} \ ,
\\
 \scrL_{-4,2}^{SU(2)}  &=& \frac{\left(\sfx_2^2 (\sfx_1+1)+2 \, \sfx_2+1\right)^2}{\sfx_2^3 \, \sfx_1^2} \ ,
\\
 \scrL_{-8,6}^{SU(2)}  &=&\frac{\left(\sfx_2^4 (\sfx_1+1)^3+2 \, \sfx_2^3 \left(\sfx_1^2+3 \, \sfx_1+2\right)+3 \, \sfx_2^2 (\sfx_1+2)+4 \, \sfx_2+1\right)^2}{\sfx_2^5 \,
   \sfx_1^4}
\\ \nonumber
 \scrL_{-12,10}^{SU(2)}  &=& \frac{1}{\sfx_2^7 \, \sfx_1^6} \Big(
\sfx_2^6 (\sfx_1+1)^5+2 \, \sfx_2^5 (\sfx_1+1)^3 (\sfx_1+3)+3 \, \sfx_2^4 \left(\sfx_1^3+6 \, \sfx_1^2+10\,  \sfx_1+5 \, \right) \\ && +4
   \, \sfx_2^3 \left(\sfx_1^2+5 \, \sfx_1+5\right)+5 \, \sfx_2^2 (\sfx_1+3)+6  \, \sfx_2+1
\Big)^2 \ .
\end{eqnarray}
Consider now the sequence of mutations
\begin{eqnarray}
\begin{matrix}
\xymatrix@C=8mm{   &  \bullet L_{0,-2}^{SU(2)}  \ar[dr]_2 &   \\
 \bullet \, \sfx_1  \ar@<-0.5ex>[rr] \ar@<0.5ex>[rr]  & & \bullet \, \sfx_2 
}
\end{matrix}
\begin{array}{c}  \mu_{\sfx_1} \mu_{\sfx_2} \\ \Longrightarrow \end{array}
\begin{matrix}
\xymatrix@C=8mm{   &  \bullet L_{0,2}^{SU(2)}  &   \\
 \bullet \, \sfx_1  \ar@<-0.5ex>[rr] \ar@<0.5ex>[rr]  & & \bullet \, \sfx_2 \ar[ul]^2 
}
\end{matrix}
\\[4pt]
\begin{array}{c}  \mu_{\sfx_1} \mu_{\sfx_2} \\ \Longrightarrow \end{array}
\begin{matrix}
\xymatrix@C=8mm{   &  \bullet L_{-4,6}^{SU(2)}  \ar[dl]^4 &   \\
 \bullet \, \sfx_1  \ar@<-0.5ex>[rr] \ar@<0.5ex>[rr]  & & \bullet \, \sfx_2   \ar[ul]_6
}
\end{matrix}
\begin{array}{c}  \mu_{\sfx_1} \mu_{\sfx_2} \\ \Longrightarrow \end{array}
\begin{matrix}
\xymatrix@C=8mm{   &  \bullet L_{-8,10}^{SU(2)} \ar[dl]^{8}  &   \\
 \bullet \, \sfx_1  \ar@<-0.5ex>[rr] \ar@<0.5ex>[rr]  & & \bullet \, \sfx_2  \ar[ul]_{10} 
}
\end{matrix} \ .
\end{eqnarray}
In this sequence we have inverted the order of the mutations. If we choose to remain in the strong coupling chamber, then this sequence corresponds to a mutation operator which crosses BPS walls in the counterclockwise direction\footnote{However it turns out that this is immaterial due to the symmetries of the quiver and we could have used $\mut_{\sfx_2 \, \sfx_1,+}$ }. The corresponding line operators are obtained by applying to $ \scrL_{0,-2}^{SU(2)}$ the mutation  $\mut_{\sfx_2 \, \sfx_1,-}$, giving
\begin{eqnarray}
 \scrL_{0,2}^{SU(2)}  &=& \sfx_2 \, (\sfx_1+1)^2 \ ,
\\
 \scrL_{-4,6}^{SU(2)}  &=& \frac{\left(\sfx_2^2 (\sfx_1+1)^3+2 \, \sfx_2 (\sfx_1+1)+1\right)^2}{\sfx_2 \, \sfx_1^2} \ ,
\\ \nonumber
 \scrL_{-8,10}^{SU(2)}  &=& \frac{1}{\sfx_2^3 \, \sfx_1^4}  \Big(\sfx_2^4 (\sfx_1+1)^5+4 \, \sfx_2^3 (\sfx_1+1)^3+3 \, \sfx_2^2 \left(\sfx_1^2+3 \, \sfx_1+2\right) \\ &&+2 \, \sfx_2
   (\sfx_1+2)+1\Big)^2 \ .
\end{eqnarray}
Similarly we can start with the line defect $\scrL_{2,0}^{SU(2)}$
\begin{eqnarray}
\begin{matrix}
\xymatrix@C=8mm{   &  \bullet L_{2,0}^{SU(2)}  &   \\
 \bullet \, \sfx_1 \ar[ur]^2  \ar@<-0.5ex>[rr] \ar@<0.5ex>[rr]  & & \bullet \, \sfx_2 
}
\end{matrix}
\begin{array}{c}  \mu_{\sfx_2} \mu_{\sfx_1} \\ \Longrightarrow \end{array}
\begin{matrix}
\xymatrix@C=8mm{   &  \bullet L_{-2,0}^{SU(2)}  \ar[dl]^2 &   \\
 \bullet \, \sfx_1  \ar@<-0.5ex>[rr] \ar@<0.5ex>[rr]  & & \bullet \, \sfx_2 
}
\end{matrix}
\\[4pt]
\begin{array}{c}  \mu_{\sfx_2} \mu_{\sfx_1} \\ \Longrightarrow \end{array}
\begin{matrix}
\xymatrix@C=8mm{   &  \bullet L_{-6,4}^{SU(2)}  \ar[dl]^6 &   \\
 \bullet \, \sfx_1  \ar@<-0.5ex>[rr] \ar@<0.5ex>[rr]  & & \bullet \, \sfx_2  \ar[ul]_4
}
\end{matrix}
\begin{array}{c}  \mu_{\sfx_2} \mu_{\sfx_1} \\ \Longrightarrow \end{array}
\begin{matrix}
\xymatrix@C=8mm{   &  \bullet L_{-10,8}^{SU(2)} \ar[dl]^{10}  &   \\
 \bullet \, \sfx_1  \ar@<-0.5ex>[rr] \ar@<0.5ex>[rr]  & & \bullet \, \sfx_2 \ar[ul]_{8}
}
\end{matrix} \ ,
\end{eqnarray}
and obtain
\begin{eqnarray} 
 \scrL_{2,0}^{SU(2)}  &=& \sfx_1 \ ,
\\
 \scrL_{-2,0}^{SU(2)}  &=& \frac{1+2 \, \sfx_2 + \, \sfx_2^2}{\sfx_2^2 \, \sfx_1} \ ,
\\
\scrL_{-6,4}^{SU(2)}  &=&\frac{\left(\sfx_2^3 (\sfx_1+1)^2+\sfx_2^2 (2 \, \sfx_1+3)+3 \, \sfx_2+1\right)^2}{\sfx_2^4 \, \sfx_1^3} \ ,
\\ \nonumber
\scrL_{-10,8}^{SU(2)}  &=&\frac{1}{\sfx_2^6 \, \sfx_1^5} \Big(\sfx_2^5 (\sfx_1+1)^4+\sfx_2^4 (\sfx_1+1)^2 (2 \, \sfx_1+5)+\sfx_2^3 \left(3 \, \sfx_1^2+12 \, \sfx_1+10\right) \\ && +2 \, \sfx_2^2 (2 \,
   \sfx_1+5)+5 \, \sfx_2+1\Big)^2 \ .
\end{eqnarray}
via the operator $\mut_{\sfx_1 \, \sfx_2,+}$. Mutatis mutandis, the sequence
\begin{eqnarray}
\begin{matrix}
\xymatrix@C=8mm{   &  \bullet L_{2,0}^{SU(2)}  &   \\
 \bullet \, \sfx_1 \ar[ur]^2  \ar@<-0.5ex>[rr] \ar@<0.5ex>[rr]  & & \bullet \, \sfx_2 
}
\end{matrix}
\begin{array}{c}  \mu_{\sfx_1} \mu_{\sfx_2} \\ \Longrightarrow \end{array}
\begin{matrix}
\xymatrix@C=8mm{   &  \bullet L_{-2,4}^{SU(2)}  \ar[dl]^2 &   \\
 \bullet \, \sfx_1  \ar@<-0.5ex>[rr] \ar@<0.5ex>[rr]  & & \bullet \, \sfx_2 \ar[ul]_4
}
\end{matrix}
\\[4pt]
\begin{array}{c}  \mu_{\sfx_1} \mu_{\sfx_2} \\ \Longrightarrow \end{array}
\begin{matrix}
\xymatrix@C=8mm{   &  \bullet L_{-6,8}^{SU(2)}  \ar[dl]^6 &   \\
 \bullet \, \sfx_1  \ar@<-0.5ex>[rr] \ar@<0.5ex>[rr]  & & \bullet \, \sfx_2  \ar[ul]_8
}
\end{matrix}
\begin{array}{c}  \mu_{\sfx_1} \mu_{\sfx_2} \\ \Longrightarrow \end{array}
\begin{matrix}
\xymatrix@C=8mm{   &  \bullet L_{-10,12}^{SU(2)} \ar[dl]^{10}  &   \\
 \bullet \, \sfx_1  \ar@<-0.5ex>[rr] \ar@<0.5ex>[rr]  & & \bullet \, \sfx_2 \ar[ul]_{12}
}
\end{matrix} \ ,
\end{eqnarray}
produces via the operator $\mut_{\sfx_2 \, \sfx_1 , -}$ the framed spectrum 
\begin{eqnarray}
 \scrL_{-2,4}^{SU(2)}  &=&\frac{\left(\sfx_2 (\sfx_1+1)^2+1\right)^2}{\sfx_1} \ ,
\\
 \scrL_{-6,8}^{SU(2)}  &=& \frac{\left(\sfx_2^3 (\sfx_1+1)^4+3 \, \sfx_2^2 (\sfx_1+1)^2+\sfx_2 (2 \, \sfx_1+3)+1\right)^2}{\sfx_2^2  \, \sfx_1^3} \ ,
\\ \nonumber
 \scrL_{-10,12}^{SU(2)}  &=& \frac{1}{\sfx_2^4\,  \sfx_1^5} \Big( 
\sfx_2^5 (\sfx_1+1)^6+5 \, \sfx_2^4 (\sfx_1+1)^4+2 \, \sfx_2^3 (2 \, \sfx_1+5) (\sfx_1+1)^2 \\ && + \sfx_2^2 \left(3 \, \sfx_1^2+12 \,
   \sfx_1+10\right)+\sfx_2 (2 \, \sfx_1+5)+1
\Big)^2 \ .
\end{eqnarray}
These results agree with Section 10.3 of \cite{Gaiotto:2010be} upon relabeling $\sfx_2 = Y$ and $\sfx_1 = X$. Note that the procedure is completely algorithmic.

From the extended quiver we see that there is another type of line defect that we can study, namely the one corresponding to the closed path $\sfe_{\sfx_1} \, \sfa_{\sfx_1 \sfx_2}^{(1)} \, \sfe_{\sfx_2} ( \sfa_{\sfx_1 \sfx_2}^{(2)})^{-1}  \sfe_{\sfx_1}$, where the distinction between the two arrows  $\sfa_{\sfx_1 \sfx_2}^{(1)}$ and $\sfa_{\sfx_1 \sfx_2}^{(2)}$ is immaterial. Since the path is closed, we are computing the trace
\begin{equation}
\scrL^{SU(2)}_{\circ} = \Tr (M_{\sfx_1} \, D \, M_{\sfx_2} \, U) = \sqrt{\sfx_1 \, \sfx_2} + \frac{1}{\sqrt{\sfx_1 \, \sfx_2}} + \sqrt{\frac{\sfx_2}{\sfx_1}} \ .
\end{equation}
The corresponding framed BPS quiver is
\begin{equation}
\begin{matrix}
\xymatrix@C=8mm{   &  \bullet L_{\circ}^{SU(2)} \ar[dl]  &   \\
 \bullet \, \sfx_1  \ar@<-0.5ex>[rr] \ar@<0.5ex>[rr]  & & \bullet \, \sfx_2 \ar[ul]
}
\end{matrix} \ ,
\end{equation}
and it is the unique element in its mutation class.

\subsection{$SU(2)$ with $N_f=1$}

We will now discuss other examples, in the case when $SU(2)$ Yang-Mills is coupled to a massive flavor particle. We start by considering a chamber in the moduli space where the theory is described by the quiver
\begin{equation}
\begin{matrix}
\xymatrix@C=8mm{   &  \bullet \, \gamma_3 \ar[dr]  &   \\
 \bullet \, \gamma_1  \ar[rr] \ar[ur]  & & \bullet \, \gamma_2 
}
\end{matrix} \ .
\end{equation}
If we choose the ordering $\arg \cZ_{\gamma_1} > \arg \cZ_{\gamma_3} > \arg \cZ_{\gamma_2}$ then the only three stable BPS states are the simple representations associated with the charges $\gamma_1$, $\gamma_2$ and $\gamma_3$. The spectrum in this chamber is generated by $\mu_{231,+}$. Again we choose the labels $\sfx_i = \tilde{\sfy}_{\gamma_i}$, with $i=1,2,3$.

Let us begin by considering the defect
\begin{equation}
\begin{matrix}
\xymatrix@C=8mm{ \bullet L^{N_f=1}_{2,0,0} &   &  \bullet \, \sfx_3 \ar[dr]  &   \\
& \ar[ul]_2  \bullet \, \sfx_1  \ar[rr] \ar[ur]  & & \bullet \, \sfx_2 
}
\end{matrix} \ .
\end{equation}
Again we label the defects by their arrow structure as $L^{N_f=1}_{b_{\gamma_1},b_{\gamma_2},b_{\gamma_3}}$. The corresponding framed BPS generating function is $\scrL^{N_f=1}_{2,0,0} = \sfx_1$. This is easy to see by using the dictionary of Section \ref{sinksource} to express the line defect as an element of the Leavitt path algebra associated with the extended quiver. 

Now starting from this generating function, we will generate other line defects. Consider iterated actions of the mutation sequence $\mu_{231} = \mu_{\sfx_2} \, \mu_{\sfx_3} \, \mu_{\sfx_1}$ beginning at node $\sfx_1$
\begin{eqnarray}
\begin{matrix}
\xymatrix@C=8mm{ \bullet L^{N_f=1}_{2,0,0} &   &  \bullet \, \sfx_3 \ar[dr]  &   \\
& \ar[ul]_2  \bullet \, \sfx_1  \ar[rr] \ar[ur]  & & \bullet \, \sfx_2 
}
\end{matrix}
\begin{array}{c}  \mu_{231} \\ \Longrightarrow \end{array}
\begin{matrix}
\xymatrix@C=8mm{ \bullet L^{N_f=1}_{-2,0,0} \ar[dr]^2 &   &  \bullet \, \sfx_3 \ar[dr]  &   \\
&  \bullet \, \sfx_1  \ar[rr] \ar[ur]  & & \bullet \, \sfx_2 
}
\end{matrix}
\cr \begin{array}{c}  \mu_{231} \\ \Longrightarrow \end{array}
\begin{matrix}
\xymatrix@C=8mm{ \bullet L^{N_f=1}_{-4,4,-2} \ar[rr]^2 \ar[dr]_4 &   &  \bullet \, \sfx_3 \ar[dr]  &   \\
&  \bullet \, \sfx_1  \ar[rr] \ar[ur]  & & \bullet \, \sfx_2 \ar[lllu]_4
}
\end{matrix}
\begin{array}{c}  \mu_{231} \\ \Longrightarrow \end{array}
\begin{matrix}
\xymatrix@C=8mm{ \bullet L^{N_f=1}_{-8,6,0}  \ar[dr]_8 &   &  \bullet \, \sfx_3 \ar[dr]  &   \\
&  \bullet \, \sfx_1  \ar[rr] \ar[ur]  & & \bullet \, \sfx_2 \ar[lllu]_6
}
\end{matrix} \ .
\end{eqnarray}
This mutation orbit contains infinite elements. The first ones are obtained by applying the operator $\mut_{\sfx_1 \sfx_3 \sfx_2,+}$ and read
\begin{eqnarray}
\scrL^{N_f=1}_{-2,0,0} &=& \frac{1}{\sfx_1 \sfx_2^2 \, \sfx_3}+\frac{2}{\sfx_1 \sfx_2 \sfx_3}+\frac{1}{\sfx_1 \sfx_2}+\frac{1}{\sfx_1 \sfx_3}+\frac{1}{\sfx_1} \ ,
\cr
\scrL^{N_f=1}_{-4,4,-2} &=& \frac{1}{ {\sfx_1^2 \, \sfx_2^3 \, \sfx_3^3}} (\sfx_2 (\sfx_3+1) ((\sfx_1+1) \sfx_2 \, \sfx_3+\sfx_2+2)+1) \cr & & \times  \left(\sfx_2 \left(\sfx_2 \left(\sfx_2 (\sfx_1
   \sfx_3+\sfx_3+1)^2+\sfx_3 (\sfx_1 (\sfx_3+2)+\sfx_3+4)+3\right)+2 \, \sfx_3+3\right)+1\right)
\cr
\scrL^{N_f=1}_{-8,6,0} &=& \frac{1}{\sfx_1^4 \, \sfx_2^5 \, \sfx_3^4}    \Big[  \sfx_2 (\sfx_3+1) \Big( \sfx_2 \Big(\sfx_2 (\sfx_1 \sfx_3+\sfx_3+1) \Big(\sfx_2 (\sfx_1 \sfx_3+\sfx_3+1)^2 \ ,
\cr & & +2
   (\sfx_1+2) \sfx_3+4\Big)+3 (\sfx_1+2) \sfx_3+6\Big)+4 \Big)+1\Big] 
   \cr & & 
   \times  \Big[ \sfx_2 \Big(\sfx_2 \Big(\sfx_2^2 (\sfx_1
   \sfx_3+\sfx_3+1)^3+\sfx_2 (\sfx_3 (\sfx_1 (\sfx_3+2)+\sfx_3+5)+4) (\sfx_1 \sfx_3+\sfx_3+1) \cr & & + \sfx_3 (2 \, \sfx_1 \sfx_3+3 \,
   \sfx_1+3 \, \sfx_3+9)+6\Big)+3 \, \sfx_3+4\Big)+1\Big] \ .
\end{eqnarray}
Similarly we can consider the defect
\begin{equation}
\begin{matrix}
\xymatrix@C=8mm{   &  \bullet \, \sfx_3 \ar[dr]  &  & \bullet \, L^{N_f=1}_{0,-2,0}  \ar[dl]_2 \\
 \bullet \, \sfx_1  \ar[rr] \ar[ur]  & & \bullet \, \sfx_2 &
}
\end{matrix} \ ,
\end{equation}
whose generating function is simply $\scrL^{N_f=1}_{0,-2,0} = \frac{1}{\sfx_2}$, and follow the mutation sequence
\begin{eqnarray}
\begin{matrix}
\xymatrix@C=8mm{   &  \bullet \, \sfx_3 \ar[dr]  &  & \bullet \, L^{N_f=1}_{0,-2,0}  \ar[dl]_2 \\
 \bullet \, \sfx_1  \ar[rr] \ar[ur]  & & \bullet \, \sfx_2 &
}
\end{matrix}
\begin{array}{c}  \mu_{231} \\ \Longrightarrow \end{array}
\begin{matrix}
\xymatrix@C=8mm{   &  \bullet \, \sfx_3 \ar[dr]  &  & \bullet \, L^{N_f=1}_{-2,2,-2}  \ar[ll]_2 \ar[llld]_2 \\
 \bullet \, \sfx_1  \ar[rr] \ar[ur]  & & \bullet \, \sfx_2 \ar[ur]_2 &
}
\end{matrix}
\cr \begin{array}{c}  \mu_{231} \\ \Longrightarrow \end{array}
\begin{matrix}
\xymatrix@C=8mm{   &  \bullet \, \sfx_3 \ar[dr]  &  & \bullet \, L^{N_f=1}_{-6,4,0}  \ar[dlll]_6 \\
 \bullet \, \sfx_1  \ar[rr] \ar[ur]  & & \bullet \, \sfx_2 \ar[ur]_4 &
}
\end{matrix}
\begin{array}{c}  \mu_{231} \\ \Longrightarrow \end{array}
\begin{matrix}
\xymatrix@C=8mm{   &  \bullet \, \sfx_3 \ar[dr]  &  & \bullet \, L^{N_f=1}_{-8,8,-2}  \ar[ll]_2 \ar[llld]_8 \\
 \bullet \, \sfx_1  \ar[rr] \ar[ur]  & & \bullet \, \sfx_2  \ar[ur]_8 &
}
\end{matrix} \ .
\end{eqnarray}
Also this mutation orbit contains infinite elements. By using again the operator $\mut_{132,+}$ we obtain
\begin{eqnarray}
\scrL^{N_f=1}_{-2,2,-2} &=& \frac{(\sfx_2 \sfx_3+\sfx_2+1) \left(\sfx_2 \sfx_3 (\sfx_1 \sfx_2+\sfx_2+1)+(\sfx_2+1)^2\right)}{\sfx_1 \, \sfx_2^2 \, \sfx_3^2}
\cr
\scrL^{N_f=1}_{-6,4,0} &=& \frac{1}{\sfx_1^3 \, \sfx_2^4 \, \sfx_3^3}  \left(\sfx_2 (\sfx_3+1) \left(\sfx_2 \left(\sfx_2 (\sfx_1 \sfx_3+\sfx_3+1)^2+(2 \, \sfx_1+3) \sfx_3+3\right)+3\right)+1\right) \cr & & \times
   \left(\sfx_2 \left(\sfx_2 \left(\sfx_2 (\sfx_1 \sfx_3+\sfx_3+1)^2+\sfx_3 (\sfx_1 (\sfx_3+2)+\sfx_3+4)+3\right)+2 \,
   \sfx_3+3\right)+1\right)
\cr 
\scrL^{N_f=1}_{-8,8,-2} &=& \frac{1}{\sfx_1^4 \, \sfx_2^5 \, \sfx_3^5} 
\cr & &
\Big[
\sfx_2 (\sfx_3+1) \Big(\sfx_2 \Big(\sfx_2 (\sfx_1 \sfx_3+\sfx_3+1) \Big(\sfx_2 (\sfx_1 \sfx_3+\sfx_3+1)^2+2 (\sfx_1+2)
   \sfx_3+4\Big) \cr & & +3 (\sfx_1+2) \sfx_3+6\Big)+4\Big)+1
\Big]
   \cr & & \Big[
\sfx_2 \Big(\sfx_2 \Big(\sfx_2^3 (\sfx_1 \sfx_3+\sfx_3+1)^4+\sfx_2^2 (\sfx_3 (\sfx_1 (\sfx_3+2)+\sfx_3+6)+5) (\sfx_1
   \sfx_3+\sfx_3+1)^2 \cr & & +\sfx_2 \Big(2 (\sfx_1+1) (\sfx_1+2) \sfx_3^3+3 (\sfx_1 (\sfx_1+6)+6) \sfx_3^2+12 (\sfx_1+2)
   \sfx_3+10\Big) \cr & & +3 (\sfx_1+2) \sfx_3^2+4 (\sfx_1+4) \sfx_3+10\Big)+4 \, \sfx_3+5\Big)+1 \Big] \ .
\end{eqnarray}
Computing these operators by direct means would have been quite challenging; in particular $\scrL^{N_f=1}_{-8,8,-2}$ contains over a thousand terms. Our algorithm can be easily implemented on any symbolic computation software and these results are immediate.

As a further example, consider the two line operators, connected by the mutation sequence $\mu_{231}$
\begin{equation}
\begin{matrix}
\xymatrix@C=8mm{   &  \bullet \, \sfx_3 \ar[dr]  &  & \bullet \, L^{N_f=1}_{0,1,-1}  \ar[ll] \\
 \bullet \, \sfx_1  \ar[rr] \ar[ur]  & & \bullet \, \sfx_2 \ar[ur] &
}
\end{matrix}
\begin{array}{c}  \mu_{231} \\ \Longrightarrow \end{array}
\begin{matrix}
\xymatrix@C=8mm{ \bullet \, L^{N_f=1}_{-1,0,+1}  \ar[dr] &   &  \bullet \, \sfx_3 \ar[dr]  \ar[ll] &  \\ &
 \bullet \, \sfx_1  \ar[rr] \ar[ur]  & & \bullet \, \sfx_2 
}
\end{matrix} \ .
\end{equation}
By using the dictionary of Section \ref{sinksource}, one can easily see that $L^{N_f=1}_{0,1,-1}$ corresponds to the Leavitt path algebra element $\sfe_{Y} \sfa_{Y \gamma_1} \sfe_{\gamma_1} \sfa_{\gamma_1 \gamma_2} \sfe_{\gamma_2} \sfa^{-1}_{\gamma_2 \gamma_3} \sfe_{\gamma_3} \sfa_{\gamma_3 G} \sfe_G$ (this operator corresponds to an open lamination which encircles one of the boundaries of the annulus and has both ends on the other) out of which we compute the generating function $\scrL^{N_f=1}_{0,1,-1} = \sqrt{\frac{\sfx_1 \, \sfx_2}{\sfx_3}}$. Then
\begin{equation}
\scrL^{N_f=1}_{-1,0,+1}  = \mut_{\sfx_1 \sfx_3 \sfx_2,+} \scrL^{N_f=1}_{0,1,-1}  = \sqrt{\frac{\sfx_3}{\sfx_1 \, \sfx_2}} \ ,
\end{equation}
and these operators are the only elements in the mutation orbit. As a check of our formalism, we can consider a different mutation sequence, starting again from the same operator $L^{N_f=1}_{0,1,-1} $. For example
\begin{equation}
\begin{matrix}
\xymatrix@C=8mm{   &  \bullet \, \sfx_3 \ar[dr]  &  & \bullet \, L^{N_f=1}_{0,1,-1}  \ar[ll] \\
 \bullet \, \sfx_1  \ar[rr] \ar[ur]  & & \bullet \, \sfx_2 \ar[ur] &
}
\end{matrix}
\begin{array}{c}  \mu_{2321} \\ \Longrightarrow \end{array}
\begin{matrix}
\xymatrix@C=8mm{   &  \bullet \, \sfx_3 &  & \bullet \, \hat{L}^{N_f=1}_{-1,1,0}  \ar[llld] \\
 \bullet \, \sfx_1  \ar[rr] \ar[ur]  & & \bullet \, \sfx_2 \ar[ur] \ar[ul]&
}
\end{matrix} \ .
\end{equation}
The second unframed quiver has the same form as the first one if we exchange the labels of the nodes $\sfx_2$ and $\sfx_3$. If we do this exchange, we find that $ \hat{L}^{N_f=1}_{-1,1,0} \vert_{2 \leftrightarrow 3} = L^{N_f=1}_{-1,0,+1}$. Consistency of the formalism developed in Section \ref{linedefcluster} requires that
\begin{equation}
\scrL^{N_f=1}_{-1,0,+1} = \mut_{1323,+} \ \left( \scrL^{N_f=1}_{0,1,-1} \right) \vert_{\sfx_2 \leftrightarrow \sfx_3} = \sqrt{\frac{\sfx_3}{\sfx_1 \, \sfx_2}} \ ,
\end{equation}
which is indeed true. In other words as long as the unframed quiver goes back to itself up to a permutation of the labels of the nodes, our formalism produces the same results. This is clear physically: changing the mutation sequence simply corresponds to changing chamber. For example the mutation sequence $\mu_{231}$ is associated with a stability condition $\cZ$ which produces a chamber with the three states $(\gamma_1,\gamma_2,\gamma_3)$. On the other hand if the mutation sequence is $\mu_{2321}$, we find the states $(\gamma_1 , \gamma_2 , \gamma_2+\gamma_3 , \gamma_3)$ (in order of decreasing $\arg \cZ$). The fact that we find the same result in both chambers is just a manifestation of the wall-crossing invariance of the generating functions $\scrL$. In particular the number of line defects in a given mutation orbit is independent on where we are in the moduli space, in accordance with our expectations from physical reasoning.

\subsection{$SU(2)$ with $N_f=2$}

We will now show some other examples in the case where we couple Yang-Mills to two flavors. We choose a chamber were the theory is described by the BPS quiver
\begin{equation}
\begin{matrix}
\xymatrix@C=8mm{   &  \bullet \, \gamma_3 \ar[dl]  &   \\
 \bullet \, \gamma_1  \ar@<-0.5ex>[rr] \ar@<0.5ex>[rr]   & & \bullet \, \gamma_2 \ar[ul] \ar[dl] \\
 & \bullet \, \gamma_4 \ar[ul] &
}
\end{matrix} \ .
\end{equation}
We fix the chamber by choosing an appropriate stability condition, corresponding to the mutation sequence $\mu_{4313424321,+}$ ($\mu_{1,+}$ acts first). In this chamber the Donaldson-Thomas invariant is
\begin{eqnarray}
\bbE (\sfY_{\gamma_1}) \, 
\bbE (\sfY_{\gamma_2}) \, 
\bbE (\sfY_{\gamma_1+\gamma_2+\gamma_3}) \, 
\bbE (\sfY_{\gamma_1+\gamma_2+\gamma_4}) \, 
\bbE (\sfY_{2 \gamma_1+\gamma_2+\gamma_3 + \gamma_4}) \, 
\bbE (\sfY_{\gamma_1+\gamma_3}) \cr \times \,
\bbE (\sfY_{\gamma_1 + \gamma_4}) \, 
\bbE (\sfY_{\gamma_1+\gamma_3+\gamma_4}) \, 
\bbE (\sfY_{\gamma_3}) \, 
\bbE (\sfY_{\gamma_4}) \ . 
\end{eqnarray}
We have chosen a complicated sequence on purpose, to illustrate the power of the formalism. 

We start with the operator $L^{N_f=2}_{+2,0,0,0}$
\begin{equation}
\begin{matrix}
\xymatrix@C=8mm{  \bullet \, L^{N_f=2}_{+2,0,0,0} & &  \bullet \, \sfx_3 \ar[dl]  &   \\
 & \bullet \, \sfx_1 \ar[ul]_2  \ar@<-0.5ex>[rr] \ar@<0.5ex>[rr]   & & \bullet \,\sfx_2 \ar[ul] \ar[dl] \\
 & & \bullet \, \sfx_4 \ar[ul] &
}
\end{matrix} \ ,
\end{equation}
where again we use the notation $L^{N_f=2}_{b_{\gamma_1},b_{\gamma_2},b_{\gamma_3},b_{\gamma_4}} $ to denote a generic line defect. By using the rules of Section \ref{sinksource} we associate to this defect the Leavitt path algebra element
\begin{equation}
\left( 
\sfe_{Y} \sfa_{Y \gamma_3} \sfe_{\gamma_3} \sfa_{\gamma_3 \gamma_1} \sfe_{\gamma_1} \sfa_{\gamma_1 \gamma_4}^{-1} \sfe_{\gamma_4} \sfa_{G \gamma_4}^{-1} \sfe_G
\, , \,
\sfe_{G} \sfa_{G \gamma_3} \sfe_{\gamma_3} \sfa_{\gamma_3 \gamma_1} \sfe_{\gamma_1} \sfa_{\gamma_1 \gamma_4}^{-1} \sfe_{\gamma_4} \sfa_{Y \gamma_4}^{-1} \sfe_Y
\right) \ ,
\end{equation}
and using the rules of Section \ref{matrix}, the generating function
\begin{equation}
\scrL^{N_f=2}_{+2,0,0,0} = \sfx_1 \, (1+\sfx_3) \, (1+\sfx_4) \ .
\end{equation}
Now we begin to generate new line defects by applying the aforementioned sequence of mutations
\begin{eqnarray}
\begin{matrix}
\xymatrix@C=8mm{  \bullet \, L^{N_f=2}_{+2,0,0,0}  &  \bullet \, \sfx_3 \ar[dl]  &   \\
  \bullet \, \sfx_1 \ar[u]_2  \ar@<-0.5ex>[rr] \ar@<0.5ex>[rr]   & & \bullet \, \sfx_2 \ar[ul] \ar[dl] \\
  & \bullet \, \sfx_4 \ar[ul] &
}
\end{matrix}
\begin{array}{c}  \mu_{4313424321} \\ \Longrightarrow \end{array}
\begin{matrix}
\xymatrix@C=8mm{  \bullet \, L^{N_f=2}_{-2,0,0,0} \ar[d]_2  &  \bullet \, \sfx_3 \ar[dl]  &   \\
  \bullet \, \sfx_1  \ar@<-0.5ex>[rr] \ar@<0.5ex>[rr]   & & \bullet \, \sfx_2 \ar[ul] \ar[dl] \\
  & \bullet \, \sfx_4 \ar[ul] &
}
\end{matrix}
\cr
\begin{array}{c}  \mu_{4313424321} \\ \Longrightarrow \end{array}
\begin{matrix}
\xymatrix@C=8mm{  \bullet \, L^{N_f=2}_{-2,0,2,2}  \ar[d]_2 &  \bullet \, \sfx_3 \ar[dl]  \ar[l]^2 &   \\
  \bullet \, \sfx_1   \ar@<-0.5ex>[rr] \ar@<0.5ex>[rr]   & & \bullet \, \sfx_2 \ar[ul] \ar[dl] \\
  & \bullet \, \sfx_4 \ar[ul] \ar[uul]^2 &
}
\end{matrix}
\begin{array}{c}  \mu_{4313424321} \\ \Longrightarrow \end{array}
\begin{matrix}
\xymatrix@C=8mm{  \bullet \, L^{N_f=2}_{-6,4,0,0} \ar[d]^6  &  \bullet \, \sfx_3 \ar[dl]  &   \\
  \bullet \, \sfx_1   \ar@<-0.5ex>[rr] \ar@<0.5ex>[rr]   & & \bullet \, \sfx_2 \ar[ull]_4 \ar[ul] \ar[dl] \\
  & \bullet \, \sfx_4 \ar[ul] &
}
\end{matrix} \ .
\end{eqnarray}
Now we compute the corresponding generating functions by applying the mutation operator $\mut_{\sfx_1 \sfx_2 \sfx_3 \sfx_4 \sfx_2 \sfx_4 \sfx_3 \sfx_1 \sfx_3 \sfx_4,+}$ iteratively
\begin{eqnarray}
L^{N_f=2}_{-2,0,0,0} &=&  \frac{(\sfx_2 \sfx_3+\sfx_3+1) (\sfx_2 \sfx_4+\sfx_4+1)}{\sfx_1\,  \sfx_2^2 \, \sfx_3 \, \sfx_4} \ ,
\cr
L^{N_f=2}_{-2,0,2,2} &=& \frac{(\sfx_2 \sfx_3 (\sfx_1 \sfx_2+\sfx_2+2)+\sfx_2+\sfx_3+1) (\sfx_2 \sfx_4 (\sfx_1
   \sfx_2+\sfx_2+2)+\sfx_2+\sfx_4+1)}{\sfx_1 \, \sfx_2^2} \ ,
\cr
L^{N_f=2}_{-6,4,0,0} &=& \frac{1}{\sfx_1^3 \, \sfx_2^4 \, \sfx_3 \, \sfx_4} 
\left(\sfx_2 \left(\sfx_3 \left(\sfx_2 \left((\sfx_1+1)^2 \sfx_2+2 \, \sfx_1+3\right)+3\right)+\sfx_1
   \sfx_2+\sfx_2+2\right)+\sfx_3+1\right) \cr & &  \left(\sfx_2 \left(\sfx_4 \left(\sfx_2 \left((\sfx_1+1)^2 \sfx_2+2 \,
   \sfx_1+3\right)+3\right)+\sfx_1 \sfx_2+\sfx_2+2\right)+\sfx_4+1\right) \ .
\end{eqnarray}
Similarly we can consider the following sequence of defects
\begin{eqnarray}
\begin{matrix}
\xymatrix@C=8mm{   &  \bullet \, \sfx_3 \ar[dl]  &   \bullet \, L^{N_f=2}_{0,-2,0,0} \ar[d]_2 \\
  \bullet \, \sfx_1  \ar@<-0.5ex>[rr] \ar@<0.5ex>[rr]   & & \bullet \, \sfx_2 \ar[ul] \ar[dl] \\
  & \bullet \, \sfx_4 \ar[ul] &
}
\end{matrix}
\begin{array}{c}  \mu_{4313424321} \\ \Longrightarrow \end{array}
\begin{matrix}
\xymatrix@C=8mm{   &  \bullet \, \sfx_3 \ar[dl]   \ar[r]^2 &   \bullet \, L^{N_f=2}_{0,-2,2,2} \ar[d]_2 \\
  \bullet \, \sfx_1  \ar@<-0.5ex>[rr] \ar@<0.5ex>[rr]   & & \bullet \, \sfx_2 \ar[ul] \ar[dl] \\
  & \bullet \, \sfx_4 \ar[ul] \ar[uur]_2 &
}
\end{matrix}
\cr
\begin{array}{c}  \mu_{4313424321} \\ \Longrightarrow \end{array}
\begin{matrix}
\xymatrix@C=8mm{   &  \bullet \, \sfx_3 \ar[dl]  &   \bullet \, L^{N_f=2}_{-4,2,0,0} \ar[lld]^4  \\
  \bullet \, \sfx_1  \ar@<-0.5ex>[rr] \ar@<0.5ex>[rr]   & & \bullet \, \sfx_2 \ar[ul] \ar[dl]  \ar[u]_2 \\
  & \bullet \, \sfx_4 \ar[ul] &
}
\end{matrix}
\begin{array}{c}  \mu_{4313424321} \\ \Longrightarrow \end{array}
\begin{matrix}
\xymatrix@C=8mm{   &  \bullet \, \sfx_3 \ar[dl]  \ar[r]^2 &   \bullet \, L^{N_f=2}_{-4,2,0,2} \ar[lld]^4  \\
  \bullet \, \sfx_1  \ar@<-0.5ex>[rr] \ar@<0.5ex>[rr]   & & \bullet \, \sfx_2 \ar[ul] \ar[dl] \ar[u]_2 \\
  & \bullet \, \sfx_4 \ar[ul]  \ar[uur]_2 &
}
\end{matrix} \ .
\end{eqnarray}
To these line defects we associate the following generating functions
\begin{eqnarray}
\scrL^{N_f=2}_{0,-2,0,0}  &=& \frac{(1+ \sfx_3) (1+\sfx_4)}{\sfx_2 \, \sfx_3 \, \sfx_4} \ ,
\cr
\scrL^{N_f=2}_{0,-2,2,2} &=& \frac{(\sfx_2 \sfx_3+\sfx_3+1) (\sfx_2 \sfx_4+\sfx_4+1)}{\sfx_2} \ ,
\cr
\scrL^{N_f=2}_{-4,2,0,0} &=& \frac{(\sfx_2 \sfx_3 (\sfx_1 \sfx_2+\sfx_2+2)+\sfx_2+\sfx_3+1) (\sfx_2 \sfx_4 (\sfx_1
   \sfx_2+\sfx_2+2)+\sfx_2+\sfx_4+1)}{\sfx_1^2 \, \sfx_2^3 \, \sfx_3 \, \sfx_4} \ ,
\cr
\scrL^{N_f=2}_{-4,2,2,2} &=& \frac{1}{\sfx_1^2 \,  \sfx_2^3} \left(\sfx_2 \left(\sfx_3 \left(\sfx_2 \left((\sfx_1+1)^2 \sfx_2+2 \, \sfx_1+3\right)+3\right)+\sfx_1
   \sfx_2+\sfx_2+2\right)+\sfx_3+1\right) \cr & & \left(\sfx_2 \left(\sfx_4 \left(\sfx_2 \left((\sfx_1+1)^2 \sfx_2+2 \,
   \sfx_1+3\right)+3\right)+\sfx_1 \sfx_2+\sfx_2+2\right)+\sfx_4+1\right) \ ,
\end{eqnarray}
where the first one has been computed directly by using the rules of Section \ref{matrix}, from the Leavitt path algebra element 
\begin{equation}
\left( \sfe_Y \sfa_{\gamma_4 Y}^{-1} \sfe_{\gamma_4} \sfa_{\gamma_2 \gamma_4}^{-1} \sfe_{\gamma_2} \sfa_{\gamma_2 \gamma_3} \sfe_{\gamma_3} \sfa_{\gamma_3 G} \sfe_{G} 
\,
,
\, 
 \sfe_G \sfa_{\gamma_4 G}^{-1} \sfe_{\gamma_4} \sfa_{\gamma_2 \gamma_4}^{-1} \sfe_{\gamma_2} \sfa_{\gamma_2 \gamma_3} \sfe_{\gamma_3} \sfa_{\gamma_3 Y} \sfe_{Y} 
\right) \ ,
\end{equation}
and the other have been generated using the operator $\mut_{\sfx_1 \sfx_2 \sfx_3 \sfx_4 \sfx_2 \sfx_4 \sfx_3 \sfx_1 \sfx_3 \sfx_4,+}$ iteratively.

As in the previous case, we find a mutation orbit with two elements
\begin{equation}
\begin{matrix}
\xymatrix@C=8mm{   &  \bullet \, \sfx_3 \ar[dl]  &   \bullet \, L^{N_f=2}_{0,0,-1,0} \ar[l] \\
  \bullet \, \sfx_1  \ar@<-0.5ex>[rr] \ar@<0.5ex>[rr]   & & \bullet \, \sfx_2 \ar[ul] \ar[dl] \\
  & \bullet \, \sfx_4 \ar[ul] &
}
\end{matrix}
\begin{array}{c}  \mu_{4313424321} \\ \Longrightarrow \end{array}
\begin{matrix}
\xymatrix@C=8mm{   &  \bullet \, \sfx_3 \ar[dl] \ar[r] &   \bullet \, L^{N_f=2}_{0,0,1,0}  \\
  \bullet \, \sfx_1  \ar@<-0.5ex>[rr] \ar@<0.5ex>[rr]   & & \bullet \, \sfx_2 \ar[ul] \ar[dl] \\
  & \bullet \, \sfx_4 \ar[ul] &
}
\end{matrix} \ ,
\end{equation}
with generating functions
\begin{eqnarray}
\scrL^{N_f=2}_{0,0,-1,0} &=& \frac{1}{\sqrt{\sfx_1 \,  \sfx_2} \, \sfx_3} \ , \cr
\scrL^{N_f=2}_{0,0,1,0} &=& \sqrt{\sfx_1 \, \sfx_2} \, \sfx_3 \ .
\end{eqnarray}

This time, arguing purely from the symmetries of the quiver, we can conjecture the existence of a second mutation orbit with two elements
\begin{equation}
\begin{matrix}
\xymatrix@C=8mm{   &  \bullet \, \sfx_3 \ar[dl]  &   \bullet \, L^{N_f=2}_{0,0,0,-1} \ar[ldd] \\
  \bullet \, \sfx_1  \ar@<-0.5ex>[rr] \ar@<0.5ex>[rr]   & & \bullet \, \sfx_2 \ar[ul] \ar[dl] \\
  & \bullet \, \sfx_4 \ar[ul] &
}
\end{matrix}
\begin{array}{c}  \mu_{4313424321} \\ \Longrightarrow \end{array}
\begin{matrix}
\xymatrix@C=8mm{   &  \bullet \, \sfx_3 \ar[dl] &   \bullet \, L^{N_f=2}_{0,0,0,1}  \\
  \bullet \, \sfx_1  \ar@<-0.5ex>[rr] \ar@<0.5ex>[rr]   & & \bullet \, \sfx_2 \ar[ul] \ar[dl] \\
  & \bullet \, \sfx_4 \ar[ul] \ar[ruu] &
}
\end{matrix} \ ,
\end{equation}
with spectra
\begin{eqnarray}
\scrL^{N_f=2}_{0,0,0,-1} &=& \frac{1}{\sqrt{\sfx_1 \,  \sfx_2} \, \sfx_4} \ , \cr
\scrL^{N_f=2}_{0,0,0,1} &=& \sqrt{\sfx_1 \, \sfx_2} \, \sfx_4 \ .
\end{eqnarray}
A direct computation confirms that it is so. It is interesting that we can guess the framed BPS spectrum of new line defects solely from the symmetries of the framed quiver.

\section{Gluing and surgery with line defects} \label{cutjoin}

So far we have discussed two algebraic formalisms to study line defects in a given quantum field theory, via extended and framed BPS quivers. Now we would like to change our perspective and investigate what happens to defects when we change the underlying theory, for example by coupling it with another quantum field theory, or decoupling it from a  subsystem, or simply gauging some flavor symmetry. Indeed all these operations, and others, have a simple interpretation in term of quiver ``cut and join" rules, at least for complete theories \cite{Cecotti:2011rv}. In this Section we will investigate certain gluing and surgery rules in the case where the quantum field theories are defined with line defects. Indeed, we expect the behavior of defects to be ``functorial" in an appropriate sense \cite{Kapustin:2010ta}; we expect the rules we will find in this Section to be part of a broader categorical picture.

\subsection{Gluing defects via extended and framed quivers}  \label{gluing}

So far we have studied quantum field theories which admit a BPS quiver description and investigated the consequences of this description on the framed BPS spectrum. The BPS quiver, or the equivalent triangulation of the curve $\cC$ whenever available, gives also a rather elegant description of the physical processes of coupling together various subsystems or taking decoupling limits \cite{Gaiotto:2009we,Cecotti:2011rv}. For example an arc of a triangulation passing through a puncture corresponding to a mass parameter, and dividing a curve $\cC_{g,n}$ in two, can be removed disconnecting the curve into two separate curves. This surgery is the topological description of the process of decoupling the flavor BPS state by sending its mass to infinity. The algebraic counterpart is obtained by removing a single node from a quiver, resulting in two disconnected quivers. Similarly the inverse process describes two distinct subsystems which are coupled via a massive flavor particle; physically, adding a new BPS state whose charge is the only one having non vanishing pairing with other charges from either of the two subsystems. Other more complicate surgeries are possible. If one is given two $\cN=2$ quantum field theories such that the corresponding curves have both a self-folded triangle (up to mutation equivalence), then the two curves can be glued together: firstly one removes the self-folded triangles leaving two boundaries with a marked point on each, and then joins the two curves with an annulus ending on those two boundaries. As we have seen the annulus with a puncture on each boundary describes an $SU(2)$ gauge theory, and the whole gluing process corresponds to coupling the two original field theories by gauging their two $SU(2)$ global symmetries \cite{Cecotti:2011rv}.

In summary the description of supersymmetric quantum field theories in terms of Riemann surfaces or quivers gives as a byproduct a set of simple graphical rules to couple two systems, which corresponds to a set of gluing rules for the respecting quivers. Similarly it is easy to study the decoupling process, which corresponds to cutting or splitting Riemann surfaces. It is an interesting problem to explore how these rules are generalized for quantum field theories in the presence of line defects; this means study surgery or gluing rules for Riemann surfaces with laminations, and the equivalent procedure on the extended and framed quivers. Naturally, if we have two distinct quantum field theories and we couple them in one of the ways outlined above, we expect the set of line defects of the new theory to contain the set of defects of the two subsystems in some limit (although their framed BPS spectra may change). Indeed by consistency it should always be possible to recover either set by the inverse decoupling procedure. But the new quantum field theory will allow also new type of line defects, associated for example to the new particles added in the spectrum or to the gauged symmetries. From the framed BPS spectrum point of view, since the new states exist as one particles states, in principle they can also bound to new line operators. This is easy to understand graphically, since gluing together two Riemann surfaces allows for the possibility of new laminations starting from one surface and ending on the other. We would like to give a more algebraic description, in terms of (framed) BPS quivers. We will do so in a few cases, without any pretension at being exhaustive. In particular we will focus on Riemann surfaces with boundary, whose associated quantum field theories are asymptotically free, or conformal Argyres-Douglas models.

Consider now two Riemann surfaces with boundaries. They correspond to two distinct supersymmetric field theories. Their respective BPS quivers are obtained from an ideal triangulation. The two quantum field theories can be coupled by gluing together the two curves. A generic gluing prescription was devised in \cite{Alim:2011ae}. The idea is to glue two boundary components of the two surfaces as two sides of a triangle. To be more definite call $\sfQ^{(1)}$ and $\sfQ^{(2)}$ the relevant quivers and denote with $\widetilde{\sfQ^{(1)}}$ and $\widetilde{\sfQ^{(2)}}$ their extended quivers. Consider two nodes in the extended set $\sfa \in \mathsf{B}_0^{(1)}$ and $\mathsf{b}\in \mathsf{B}_0^{(2)}$. Then the respective boundary edges can be glued as two sides of a triangle whose third side we will denote by $\mathsf{c}$. In quiver language
\begin{equation} \label{gluingquivers}
\begin{matrix}
\xymatrix@C=8mm{ &  &  \circ  \ \mathsf{c}   \ar[dl]  & & \\
\sfQ^{(1)}  \ar@{.}[r] &  \  \mathsf{a}  \ar[rr]  \ \bullet \  &  & \ar[ul] \ \bullet \ \mathsf{b}  & \sfQ^{(2)} \ar@{.}[l]
} 
\end{matrix} \ .
\end{equation}
Note the both $\mathsf{a}$ and $\mathsf{b}$ are now ordinary nodes of the glued quiver $\sfQ^{(1)}  \oplus_{\mathsf{a}}^{\mathsf{b}}  \sfQ^{(2)}$ while $\mathsf{c}$ is an extended node. The notation $\sfQ^{(1)}  \oplus_{\mathsf{a}}^{\mathsf{b}}  \sfQ^{(2)}$ for the glued quiver stresses the fact that the gluing is done according to (\ref{gluingquivers}). This gluing rule guarantees that if $\sfQ^{(1)}$ and $\sfQ^{(2)}$ have finite chambers, also $\sfQ^{(1)}  \oplus_{\mathsf{a}}^{\mathsf{b}}  \sfQ^{(2)}$ has a finite chamber. In \cite{Alim:2011ae} this gluing was broken down to the reiteration and composition of four minimal operation. Topologically these minimal operations correspond to adding a marked point, adding a puncture, adding a boundary component or increasing the genus of a given Riemann surface. We will discuss them in Section \ref{boundary} and for the moment restrict ourselves to the general abstract case.

We wish to understand what happens when the Riemann surfaces are equipped with a line defect. Assume that two multi laminations are given, $\cL^{(1)}$ on $\cC^{(1)}$ and $\cL^{(2)}$ on $\cC^{(2)}$. 
The only three relevant cases are when or $\cL^{(1)}$ ends at the boundary segment $\mathsf{a}$, or $\cL^{(2)}$ ends at the boundary segment $\mathsf{b}$, or both happen; if none of these happens then there is simply no gluing of defects to be considered. Indeed if we have a quantum field theory characterized by a set of line defects and we couple it to another system, for example by adding matter or by gauging symmetries, there will be defects which remain unaffected by the coupling. Defects of this sort are represented by laminations on the curve $\cC$ which are not involved in the gluing process. Physically they correspond to line defects which do not bound to the new BPS states introduced by the coupling with another quantum field theory. Such a bound could be for example forbidden by certain symmetries or energetically disfavored.  Note however that these arguments strictly speaking only apply in the chamber where the gluing is made and that the situation becomes  quite complicated when one is free to tune all the parameters of the theory.

\begin{figure}
\centering
\def\svgwidth{12cm}
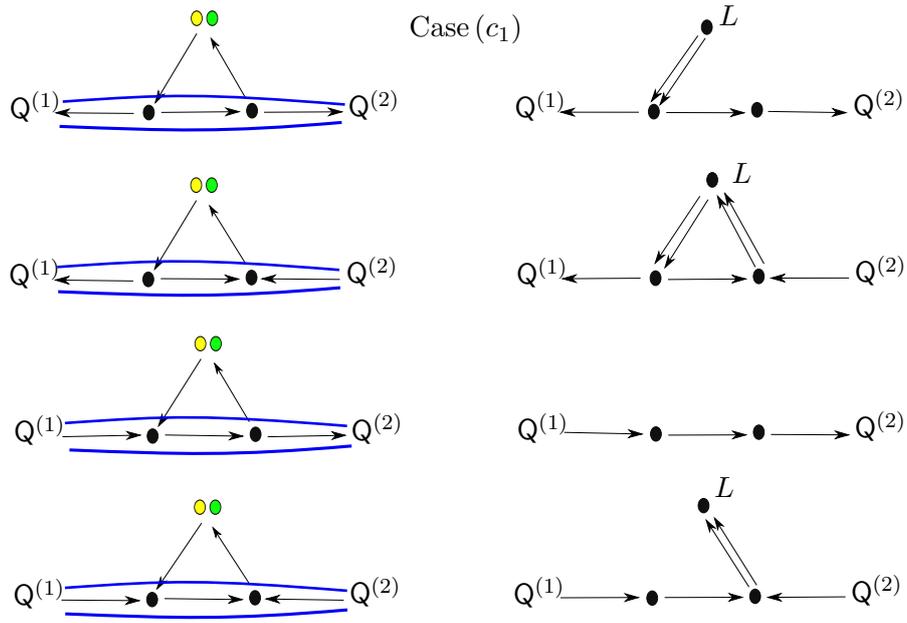
\caption{Gluing line operators in case ($c_1$) and their framed quivers. As before we only show the new arrows connecting the framing node which are induced by the gluing. We assume the multiplicities of the curves to be the same and for simplicity omit them.}
\label{gluingC}
\end{figure}

\begin{figure}
\centering
\def\svgwidth{12cm}
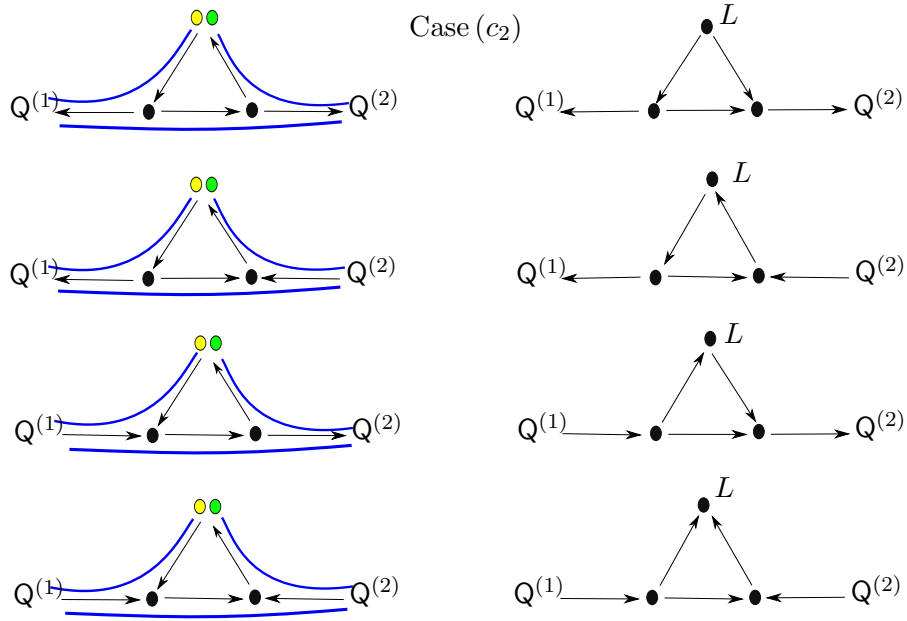
\caption{Gluing line operators in case ($c_2$) and their framed quivers. The same specifications of figure \ref{gluingC} apply.}
\label{gluingC2}
\end{figure}

To simplify our construction we will assume that the gluing is made in a chamber where both $\sfQ^{(1)}$ and $\sfQ^{(2)}$ have a finite spectrum consisting of finitely many hypermultiplets with multiplicity one. We will consider now the gluing exemplified in (\ref{gluingquivers}) both from the point of view of the paths on the extended quiver, which is a straightforward rendering of the laminations on the Riemann surface $\cC$, and from the point of view of the framed quivers. For simplicity we will consider one defect at the time, consisting of a lamination made of two paths, each with multiplicity one, and ending at the same boundary (the weight at the boundary is cancelled by a special curve; we will not discuss special curves since they can be safely removed and reinstated as necessary). More general cases can be treated similarly. Furthermore when drawing the  quivers, we will draw only one arrow between $\sfQ^{(1)}$ and $\sfQ^{(2)}$ and the nodes $\sfa$ and $\sfb$: the arrows connecting $\sfa$ with the node in $\sfQ^{(1)}$ which corresponds to the last edge of the triangulation crossed by the lamination before arriving at the edge represented by $\sfa$ (and similarly for $\sfb$ and $\sfQ^{(2)}$). The framing depends on the orientation of this arrow. In general other arrows will be present but we omit them for simplicity, confident that no confusion should arise. Similarly we will not draw the arrows connecting $\sfQ^{(1)}$ and $\sfQ^{(2)}$ with the framing node, even if in general they will be present; we will only draw the new arrows connecting the framing node with $\sfa$ or $\sfb$. Let us consider the following cases in turn:
\begin{figure}
\centering
\def\svgwidth{12cm}
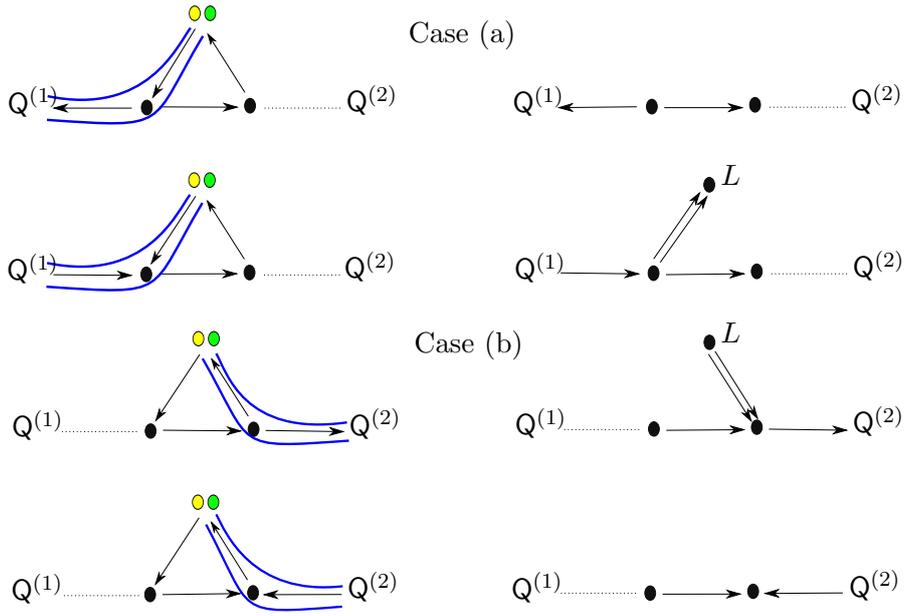
\caption{Gluing line operators in cases (a) and (b) and their framed quivers. We draw only the arrows and paths involved in the gluing. On the left are shown the extended quivers equipped with paths, while on the right the relevant framed quivers. The framing node in general will be connected with $\sfQ^{(1)}$ or $\sfQ^{(2)}$ by several arrows independently on the gluing process. We omit these arrows and only show the new ones, if any, introduced during the gluing.} 
\label{gluingA}
\end{figure}
\begin{itemize}
\item[(a)] The lamination $\cL^{(1)}$ on $\cC^{(1)}$ ends on the boundary node $\mathsf{a}$ while no lamination ends on the boundary node $\mathsf{b}$. There is no lamination on $\cC^{(2)}$. The node $\mathsf{a}$ is a boundary node of $\sfQ^{(1)}$ but an internal node of the glued quiver $\sfQ^{(1)} \oplus_{\mathsf{a}}^{\mathsf{b}} \sfQ^{(2)}$. Therefore if we want the gluing of the line operator to be consistent, the path ending at $\mathsf{a}$ of $\sfQ^{(1)}$ must now end at the boundary node $\mathsf{c}$ (since no lamination is allowed on $\cC^{(2)}$ by assumption). We illustrate this in figure \ref{gluingA}, where we draw both the extended quiver and the framed quiver. Note that the framed BPS spectrum associated with the defect will depend on the direction of the arrow joining the quiver $\sfQ^{(1)}$ with the node called $\sfa$ in (\ref{gluingquivers}). Depending on this direction the lamination can pass through a sink or through a simple arrow concatenation, affecting or not the framing node. Note that there can be other arrows connecting the framing node with the quiver $\sfQ^{(1)}$ which are associated with the particular form of $\cL^{(1)}$ on $\sfQ^{(1)}$; as we have already said, these are omitted from the drawing for simplicity.
\item[(b)] The lamination $\cL^{(2)}$ on $\cC^{(2)}$ ends on the boundary node $\mathsf{b}$ while no lamination ends on the boundary node $\mathsf{a}$. This case is similar to case (a), with the role played by the quiver $\sfQ^{(1)}$ and $\sfQ^{(2)}$ interchanged. The differences are illustrated in figure \ref{gluingA}. Now the only contribution to the framing comes from the case where the arrows are oriented to form a source at the relevant node.
\item[(c)] The lamination $\cL^{(1)}$ on $\cC^{(1)}$ ends on the boundary node $\mathsf{a}$ and the lamination $\cL^{(2)}$ on $\cC^{(2)}$ ends on the boundary node $\mathsf{b}$. In this case the two laminations have to be glued together, otherwise there would be an intersection\footnote{To be clear, it is certainly possible that a lamination on $\cC^{(1)}$ gets extended in the interior of $\cC^{(2)}$. However this case is topologically (but not physically) equivalent to the case where two independent laminations on $\cC^{(1)}$ and $\cC^{(2)}$ are glued together. In other words for classification purposes we do not need to consider it as a separate case.}. There are two ways of doing so, by gluing all the curves together (Case $c_1$) or by letting some end on the boundary (Case $c_2$). Examples of the relevant paths on the extended quivers and their framed quiver counterpart are drawn in figures \ref{gluingC} and \ref{gluingC2}. Note that for the gluing to be possible it is necessary that each curve has the same multiplicity. We exclude the case when one of the lamination is glued back to itself, since it would be possible to further contract it back to its original curve of provenience.
\end{itemize}

\subsection{Line defects and surgery on surfaces with boundary} \label{boundary}

We will now show some elementary examples of surgeries of curves with laminations. In this section we will assume that a certain surface $\cC$ with boundary is given together with an ideal triangulation and the associated quiver $\sfQ$. The associated gauge theories are complete and furthermore the associated BPS quiver contains a chamber with a finite number of BPS hypermultiplets. Theories associated with Riemann surfaces with boundaries correspond to asymptotically free theories with gauge group a product of $SU(2)$, or to conformal Argyres-Douglas theories. There is a set of elementary operations using which one can construct a different surface that still has a finite BPS chamber. These operations were studied in \cite{Alim:2011ae}. A surface with boundary is characterized by its genus $g$, the number of punctures $n$ and the number of boundary components $b$ each with $c_i \ge 1$ marked points, $i=1 , \dots , b$. Given such a surface we can construct another one by the following operations: by adding a marked point to a boundary component, by adding a puncture, by adding a boundary component with a marked point and by increasing the genus. To each of these operations there is associated a quiver gluing rule \cite{Alim:2011ae}.

We will now extend these rules to the case when there is a lamination on the surface $\cC$. Again the only relevant case is when the lamination is somewhat involved in the surgery process. The case where the lamination does not pass through the boundary involved in the surgery and is only a spectator is trivial and will not be considered. We will show how framed BPS quivers behave under these gluing rules with a series of examples. To simplify the rules we will assume that there is a single lamination $\cL$ on $\cC$ and that only two paths intersect the boundary segment upon which we perform the surgery, each with multiplicity one. Furthermore we will only consider the case where both paths arrive at the boundary component after having crossed the same edge, which precedes or follows the boundary segment in clockwise direction. Locally the quiver $\sfQ$ will look like
\begin{equation}
\mathsf{Q} \longrightarrow \bullet \ \mathsf{a} \ , \qquad \text{or} \qquad \mathsf{Q} \longleftarrow \bullet \ \mathsf{a} \ ,
\end{equation}
where $\mathsf{a}$ is the node of the quiver corresponding to the boundary segment on which we perform the surgery (and therefore is a boundary node for $\sfQ$ but an internal node for the quiver resulting from the surgery). 
We only draw the arrow connecting $\sfa$ with the node of $\sfQ$ corresponding to the edge of the triangulation last crossed by the lamination before arriving to the edge corresponding to $\sfa$. There could be other arrows from $\sfa$ to other nodes of $\sfQ$, but we will not show them explicitly to simplify the drawings. Of course other cases are possible: for example the two paths might come from two distinct internal edges, or the lamination might consists of more that two paths. All these cases can be dealt with with obvious modifications of the formalism and we will leave them to the reader.

\subsection*{Adding a marked point}

\begin{figure}[h]
\centering
\def\svgwidth{ 
10cm}
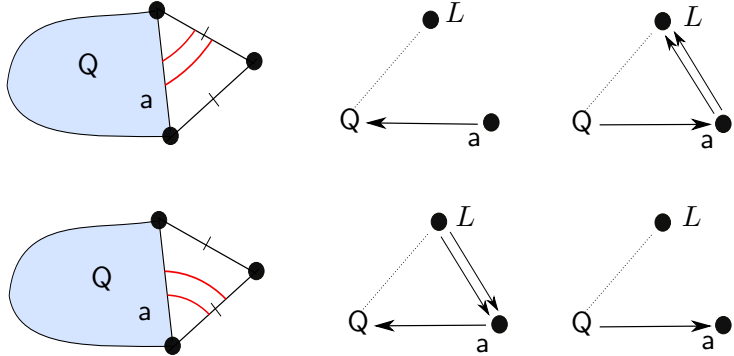
\caption{Adding a marked point.}
\label{addmarked}
\end{figure}

The first example we will consider is when we add a single marked point to a boundary component of $\cC_{g,n,b,c}$. Let us consider a boundary segment labelled by $b_i$, with $c_i$ marked points. We increase the number of marked points by one, that is $c_i \longrightarrow c_i + 1$. This is equivalent to gluing an unpunctured triangle to the boundary edge $\mathsf{a}$, located between two marked points of $b_i$. Now the boundary edge $\sfa$ becomes an internal edge of the new quiver. Since we have glued a triangle to the boundary edge, a lamination crossing $\sfa$ has only two possibilities: it can end on the boundary edge on its left or on the one on its right. We do not consider the case where the lamination is glued back to itself, since at least a part of it would become contractible. The possible cases, within the aforementioned conditions, and their description in terms of framed quivers  are shown in figure \ref{addmarked}. We omit special curves from the discussion, which are however needed to ensure that the sum of the weights of all the paths ending at the same boundary edge is zero.

\subsection*{Adding a puncture}

\begin{figure}
\centering
\def\svgwidth{ 
14cm}
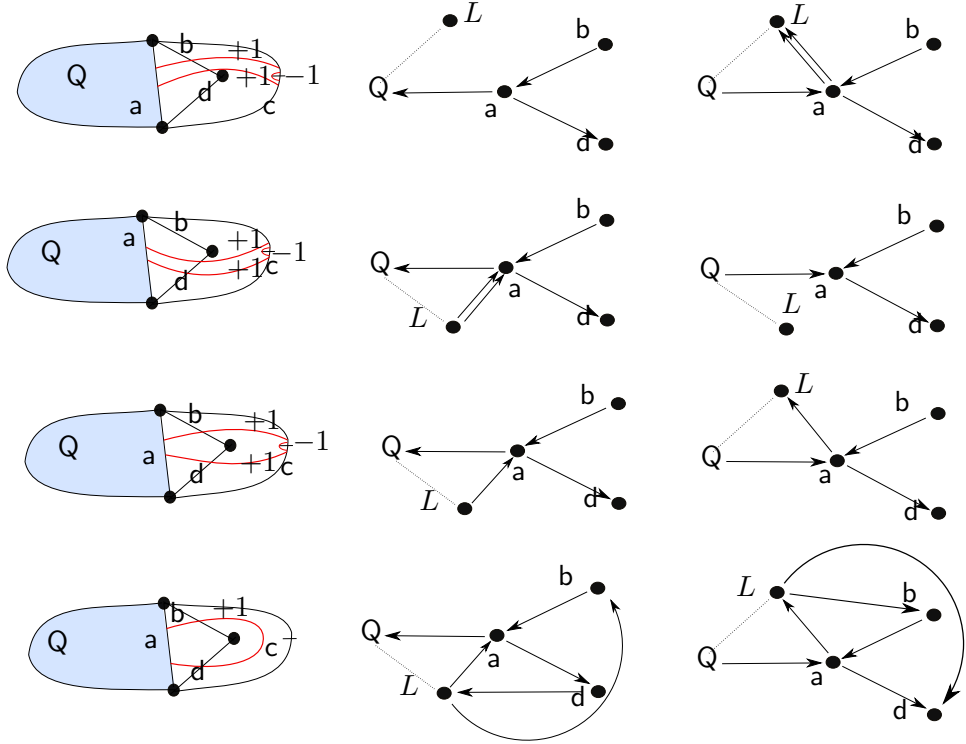
\caption{Adding a puncture. For simplicity we only consider the case where the paths have multiplicity $+1$.}
\label{addpuncture}
\end{figure}

Consider now adding a puncture to $\cC_{g,n,b,c}$. The new curve is of the form $\cC_{g,n+1,b,c}$ and in particular the field theory has an extra flavor charge by (\ref{flavornum}). In the language of the previous section $\sfQ^{(2)}$ corresponds to a once-punctured monogon whose boundary edge is $\mathsf{b}$. This corresponds to a triangulation with a self-folded triangle. As discussed in Appendix \ref{selffolded}, this is not a problem, since the rules to determine the shear coordinates extend to triangulations with self-folded triangles (and more generically to tagged triangulations). However it is simpler to dispose of the self-folded triangle by flipping the edge corresponding to $\mathsf{b}$. Indeed it is always possible to reduce a triangulation with self-folded triangles to a triangulation without \cite{FT1}. The relevant cases and their associated framed quivers are drawn in figure \ref{addpuncture}. Note that in this case it is possible to glue the lamination back to itself without producing a contractible path. As is shown in the last case of figure \ref{addpuncture}, the path in the lamination now encircles the new puncture. This is the only case qualitatively different from the previous gluing, due to the fact that $\cC_{g,n+1,b,c}$ has a more complicated topology allowing for new non trivial homotopy paths. 

\subsection*{Adding a boundary component}

\begin{figure}[h]
\centering
\def\svgwidth{ 
12cm}
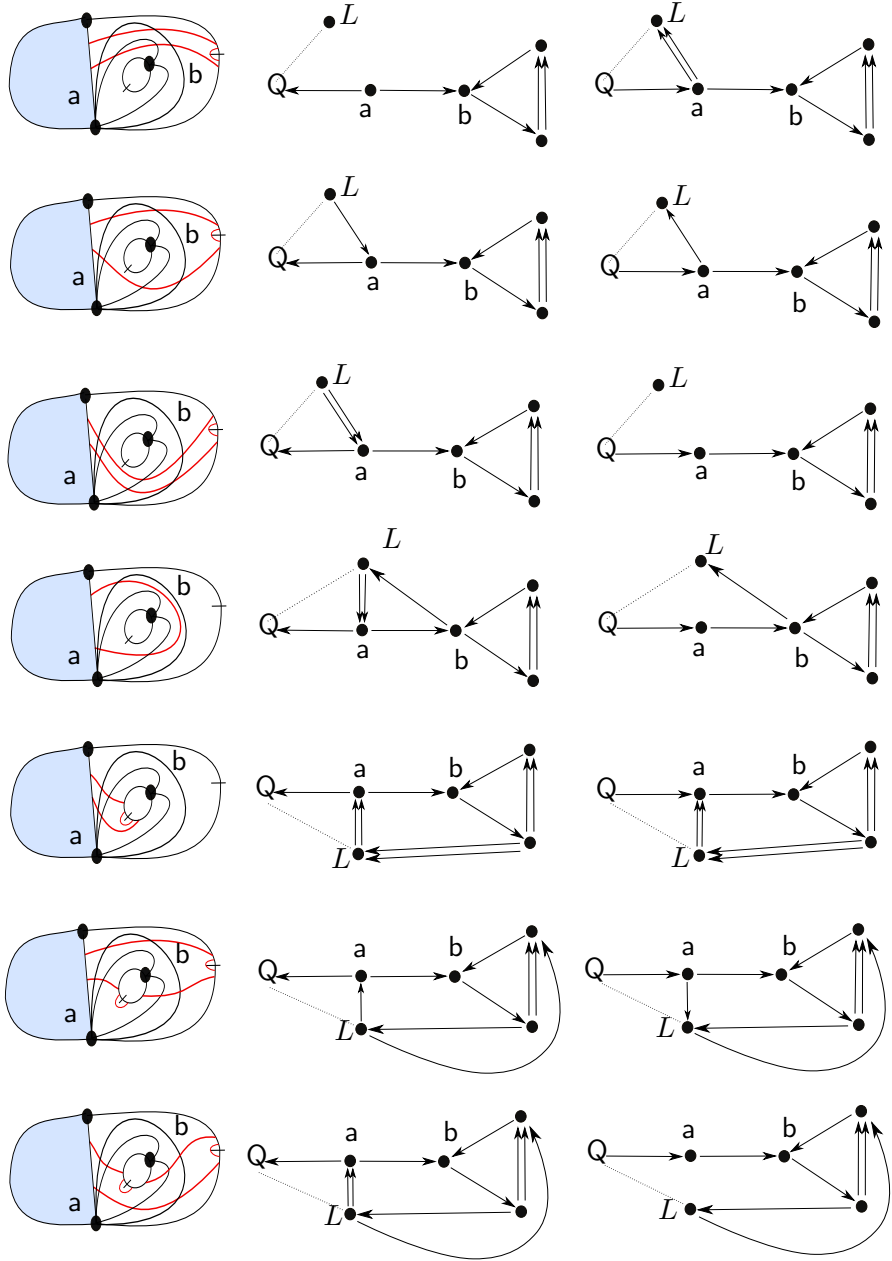
\caption{Adding a boundary component. Now $\sfQ^{(2)}$ is an annulus with one marked point in each boundary. The boundary $\sfb$ is now an internal node of the new quiver. Note that one can see a new $SU(2)$ sector corresponding to the annulus (and whose edges are not labelled). This case is topologically much richer, allowing for paths encircling or ending on the new boundary.}
\label{addboundary}
\end{figure}

To add a boundary component one can glue to the original curve $\cC_{g,n,b,c}$ an annulus with one marked point on each boundary component. In this case the quiver $\sfQ^{(2)}$ is the quiver representing the annulus. One of the boundary segments $\mathsf{b}$ of the annulus will become an internal vertex of the new quiver, while the other boundary segment will become the new boundary edge. The resulting curve $\cC_{g,n,b+1,c'}$ has indeed $b+1$ boundary components and the newly added one has a marked point. As in the previous case, now it is possible for a path in a lamination to circle the new boundary and go back to the quiver $\sfQ^{(1)}$. The qualitatively new feature is that now there are new open laminations which can end on the new boundary edge. We exemplify few cases in figure \ref{addboundary}, for the case where all the laminations have multiplicity $+1$.

\subsection*{Increasing the genus}

\begin{figure}[h]
\centering
\def\svgwidth{ 
11cm}
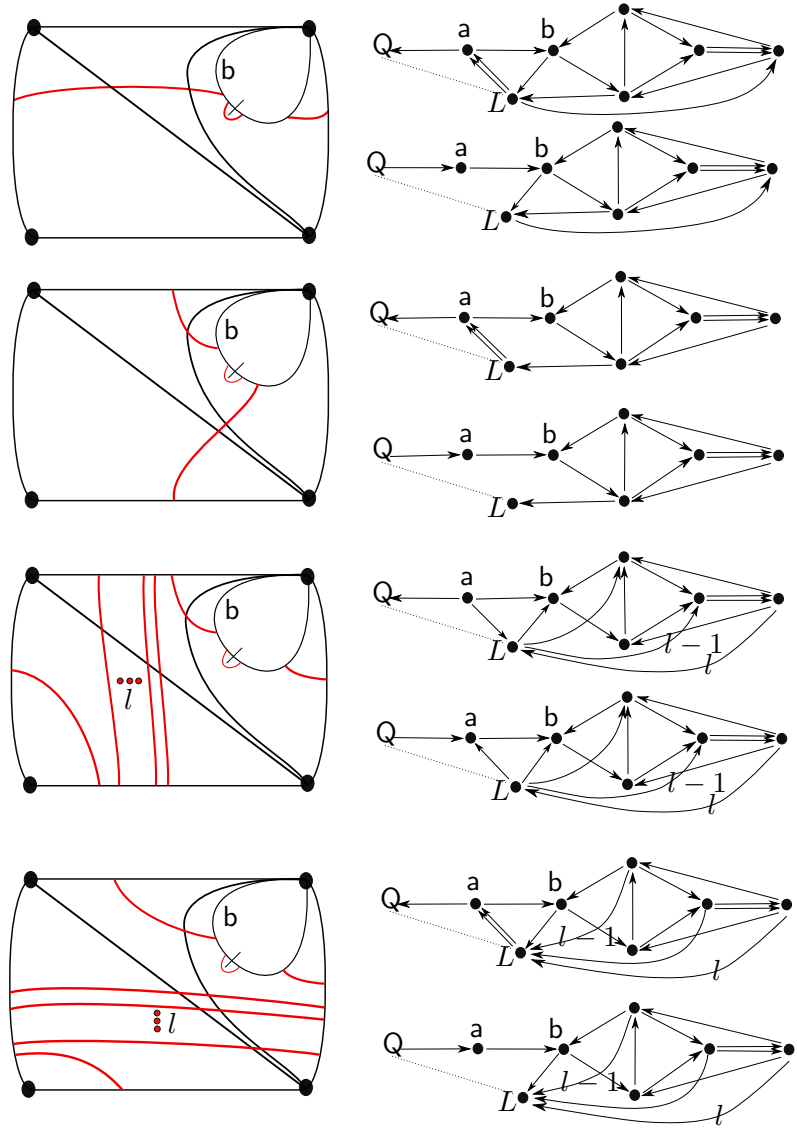
\caption{Increasing the genus by adding a torus with one boundary component with a single marked point. In the figure of the curve we only draw the torus with boundary. Conventionally we let the path wind along one of the homology cycle of the torus $l$ times in the last two figures.}
\label{addgenus} 
\end{figure}

To increase the genus by one unit we add to $\cC_{g,n,b,c}$ a torus with one boundary component with a single marked point. In this case the torus is $\sfQ^{(2)}$ and the boundary component is the edge $\sfb$ folded on itself. The new curve has the form $\cC_{g+1,n,b,c}$. Adding a torus has added two generators to the homology of $\cC$. Correspondingly the new set of laminations is quite complicated and involves curves winding along one of the new homology generators. The framed BPS quivers are shown in figure \ref{addgenus}, where we only draw the torus with its boundary $\sfb$ which has to be glued to $\sfQ$. Again we only consider paths with multiplicity $+1$, which can however wind an indefinite number of time along the two non-trivial homology cycles of the torus.

\subsection{Argyres-Douglas, again}

Finally we put all that we have learned to good use in a simple example. We will now take a line defect in the Argyres-Douglas superconformal theory of type $A_2$ and use it to generate new defects in the $A_4$ theory. We are actually going to consider the most simple case available. We pick the line defect $L_5$
\begin{equation}
\begin{matrix} \xymatrix@C=10mm{  &  \bullet  \ L_5 \ar@<-0.5ex>[dl] \ar@<0.5ex>[dl] & \\
 \ \gamma_2  \ \bullet \  & & \ \bullet \ \gamma_1 \ar[ll]
} \end{matrix} \ ,
\end{equation}
which we have already discussed in Sections \ref{linedefcluster} and \ref{matrix}. If we label $\tilde{\sfy}_{\gamma_2} = \sfy$ and $\tilde{\sfy}_{\gamma_1} = \sfx$, then $\scrL_{5}^{A_2} = \frac{1}{\sfy}$. We can use the quiver gluing rules of \cite{Cecotti:2011rv,Alim:2011ae} to obtain the $A_4$ theory, whose BPS quiver we take to be
\begin{equation}
\begin{matrix} \xymatrix@C=10mm{  
 \ \sfy  \ \bullet \  & & \ \bullet \ \sfx \ar[ll]  & & \ar[ll] \ \sfu  \ \bullet \  & & \ \bullet \ \sfv \ar[ll]
} \end{matrix} \ ,
\end{equation}
in the appropriate chamber. From the rules of Section \ref{gluing} and \ref{boundary}, we see that there exists a defect 
\begin{equation}
\begin{matrix} \xymatrix@C=10mm{  &  \bullet  \ L_5^{A_4} \ar@<-0.5ex>[dl] \ar@<0.5ex>[dl] & & & & &  \\
 \ \sfy  \ \bullet \  & & \ \bullet \ \sfx \ar[ll] & & \ar[ll] \ \sfu  \ \bullet \  & & \ \bullet \ \sfv \ar[ll]
} \end{matrix} \ ,
\end{equation}
with the property $\scrL_5^{A_4} = \frac{1}{\sfy}$. Indeed in this case the gluing rules are trivial, since the lamination does not take part in the gluing. Equivalently, the relevant Leavitt path algebra elements only involve the node $\sfy$, before and after the gluing. We can however start generating new line defects using quiver mutations. Since the theory is complete, we can pick a stability condition corresponding to the mutation sequence $\mu_{\sfy \, \sfx \, \sfu \, \sfv}$ and start to generate new framed quivers:
\begin{eqnarray}
\begin{matrix} \xymatrix@C=1.8mm{  &  \bullet  \ L_5^{A_4} \ar@<-0.5ex>[dl] \ar@<0.5ex>[dl] & & & & &  \\
 \ \sfy  \ \bullet \  & & \ \bullet \ \sfx \ar[ll] & & \ar[ll] \ \sfu  \ \bullet \  & & \ \bullet \ \sfv \ar[ll]
} \end{matrix} 
\begin{array}{c} \mu_{\sfy \, \sfx \, \sfu \, \sfv} \\ \Longrightarrow \end{array}
\begin{matrix} \xymatrix@C=1.8mm{  &  \bullet  \ L_6^{A_4} \ar@<-0.5ex>[dr] \ar@<0.5ex>[dr] & & & & &  \\
 \ \sfy \ar@<-0.5ex>[ur] \ar@<0.5ex>[ur] \ \bullet \  & & \ \bullet \ \sfx \ar[ll] & & \ar[ll] \ \sfu  \ \bullet \  & & \ \bullet \ \sfv \ar[ll]
} \end{matrix} 
\\
\begin{array}{c} \mu_{\sfy \, \sfx \, \sfu \, \sfv} \\ \Longrightarrow \end{array}
\begin{matrix} \xymatrix@C=1.8mm{  & & &  \bullet  \ L_7^{A_4} \ar@<-0.5ex>[dr] \ar@<0.5ex>[dr]  & & &  \\
 \ \sfy \ \bullet \  & & \ar@<-0.5ex>[ur] \ar@<0.5ex>[ur]  \ \bullet \ \sfx \ar[ll] & & \ar[ll] \ \sfu  \ \bullet \  & & \ \bullet \ \sfv \ar[ll]
} \end{matrix} 
\begin{array}{c} \mu_{\sfy \, \sfx \, \sfu \, \sfv} \\ \Longrightarrow \end{array}
\begin{matrix} \xymatrix@C=1.8mm{  & & & & &  \bullet  \ L_8^{A_4} \ar@<-0.5ex>[dr] \ar@<0.5ex>[dr]   &  \\
 \ \sfy \ \bullet \  & & \ \bullet \ \sfx \ar[ll] & & \ar@<-0.5ex>[ur] \ar@<0.5ex>[ur]  \ar[ll] \ \sfu  \ \bullet \  & & \ \bullet \ \sfv \ar[ll]
} \end{matrix} 
\\
\begin{array}{c} \mu_{\sfy \, \sfx \, \sfu \, \sfv} \\ \Longrightarrow \end{array}
\begin{matrix} \xymatrix@C=1.8mm{  & & & & &  \bullet  \ L_9^{A_4} &  \\
 \ \sfy \ \bullet \  & & \ \bullet \ \sfx \ar[ll] & &  \ar[ll] \ \sfu  \ \bullet \  & & \ \bullet \ \sfv \ar[ll] \ar@<-0.5ex>[ul] \ar@<0.5ex>[ul]   
} \end{matrix} 
\begin{array}{c} \mu_{\sfy \, \sfx \, \sfu \, \sfv} \\ \Longrightarrow \end{array}
\begin{matrix} \xymatrix@C=1.8mm{  & & & & &  \bullet  \ L_{10}^{A_4} \ar@<-0.5ex>[dr] \ar@<0.5ex>[dr]   &  \\
 \ \sfy \ \bullet \  & & \ \bullet \ \sfx \ar[ll] & & \ar[ll] \ \sfu  \ \bullet \  & & \ \bullet \ \sfv \ar[ll]
} \end{matrix} 
\\
\begin{array}{c} \mu_{\sfy \, \sfx \, \sfu \, \sfv} \\ \Longrightarrow \end{array}
\begin{matrix} \xymatrix@C=2mm{  &   \bullet  \ L_{11}^{A_4} & & & & &  \\
 \ \sfy \   \ar@<-0.5ex>[ur] \ar@<0.5ex>[ur]  \bullet \  & & \ \bullet \ \sfx \ar[ll] & &  \ar[ll] \ \sfu  \ \bullet \  & & \ \bullet \ \sfv \ar[ll]  
} \end{matrix} 
\end{eqnarray}

The framed spectrum will now be given applying iteratively the operator $\mut_{\sfv \, \sfu \, \sfx \, \sfy,+}$
\begin{eqnarray}
\scrL_6^{A_4} &=& \mut_{\sfv \, \sfu \, \sfx \, \sfy,+} \scrL_5^{A_4} = \frac{1}{\sfx} + \frac{\sfy}{\sfx} + \sfy
\\
\scrL_7^{A_4} &=& \mut_{\sfv \, \sfu \, \sfx \, \sfy,+} \scrL_6^{A_4} = \frac{1}{\sfu} + \frac{\sfx}{\sfu} + \sfx
\\
\scrL_8^{A_4} &=& \mut_{\sfv \, \sfu \, \sfx \, \sfy,+} \scrL_7^{A_4} = \sfu + \frac{1}{\sfv} + \frac{\sfu}{\sfv}
\\
\scrL_9^{A_4} &=& \mut_{\sfv \, \sfu \, \sfx \, \sfy,+} \scrL_8^{A_4} = \sfv
\\
\scrL_{10}^{A_4} &=& \mut_{\sfv \, \sfu \, \sfx \, \sfy,+} \scrL_9^{A_4} = \frac{1}{\sfx \, \sfy \, \sfu \, \sfv} +  \frac{1}{\sfx  \, \sfu \, \sfv} + \frac{1}{\sfu \, \sfv} + \frac{1}{\sfv} 
\\
\scrL_{11}^{A_4} &=& \mut_{\sfv \, \sfu \, \sfx \, \sfy,+} \scrL_{10}^{A_4} = \sfy + \sfx \, \sfy + \sfu \, \sfx \, \sfy  + \sfu \, \sfv \, \sfx \, \sfy
\end{eqnarray}
Remarkably starting from a defect in a known quantum field theory, we have obtained \textit{new} defects in a \textit{new} quantum field theory (the fact that both theories are complete is here crucial). We have shown this only in a simple example, but the method is completely general and gives a new strategy to find line defects and determine their framed degeneracies. If we can find a defect which is not involved in the gluing process, this method gives a very efficient tool to study new defects in the new theory by using quiver mutations.  

Overall a clever combination of the techniques developed so far, extended and framed quivers and the respective gluing rules, give a rather powerful formalism to compute framed BPS spectra.

\section{Discussion} \label{discussion}

In this paper we have provided an algebraic perspective in terms of quivers and certain algebras on line defects in supersymmetric field theories, especially in connection with their geometrical description in terms of laminations in Teichm\"uller theory \cite{Gaiotto:2010be}. Certain aspects of this algebraic perspective are rather powerful and in particular give rise to easily implementable algorithms. The price to be paid is a loss of geometrical intuition. While this is expected on general grounds, we have encountered many tricky combinatorial problems; the approach we have taken is to proceed on a case by case analysis, and leave more general classification problems for future work. This paper roughly speaking is divided into three main parts. Let us in turn summarize the results obtained. 

Given a line defect, represented as a lamination on a curve $\cC$, we can construct certain elements of the Leavitt path algebra of the associated extended BPS quiver. Out of these elements we can compute the framed BPS spectrum directly with a set of matrix rules. This method is a direct translation in the quiver context of the techniques exploited in  \cite{Gaiotto:2010be}. However in the process we have discovered an interesting connection between laminations in  Teichm\"uller theory and Leavitt algebras associated with extended quivers. Our definition of the Leavitt path algebra is slightly different from the one usually found in the literature, as we had to include extra elements, namely doubling the boundary nodes, to accommodate for all kinds of line defects. On the extended quiver we can identify a series of paths associated with a lamination, determine the corresponding Leavitt path algebra elements and from them compute the framed BPS spectrum.

Using the theory of quantum cluster algebras we have devised an algorithm which constructs new line defects from known ones. The algorithm is based on a rule which associates to a lamination a series of shear coordinates. These coordinates can be used to frame the BPS quiver, by adding a new node corresponding to a line defect: the structure of the arrows connecting the framing node with the BPS quiver is determined by the shear coordinates. The framed BPS quiver transforms nicely under quiver mutations, meaning that the shear coordinates transform with the same rules as the adjacency matrix of the quiver. In particular if we find a sequence of quiver mutations such that the BPS quiver goes back to itself, but the \textit{framed} BPS quiver does not, we have generated a new line defect. Not only that, but it is even possible to compute its framed BPS spectrum from the line defect we started with. This is done by associating a set of variables $\{ \tilde{\sfy}_{\gamma} \}$ to the nodes of the unframed BPS quiver, which are commutative limits of quantum cluster variables. From the sequence ${\mbf \mu}$ of quiver mutations used to determine the new line defects, we have constructed an operator $\mut_{\mbf \mu}$ which gives the framed BPS spectrum of the new line defect. More formally our approach constructs line defects out of coefficients of cluster algebras, selecting among all the cluster variables the appropriate ones and combinations thereof which are physical (respect the positivity property). Line defects come in families, which we have dubbed mutation orbits, obtained by applying mutation sequences on a given starting defect. Certain orbits have only one element, others have infinite elements; the ones with finite elements exhibit a periodicity property inherited by the underlying cluster algebra. From a representative of the orbit we can in principle reconstruct all the others via the mutation algorithm.

Finally we have given a series of graphical rules to perform surgery and gluing of line defects between different theories, as operations on the framed BPS quivers. These operations generalize to the case of line defects the construction of \cite{Cecotti:2011rv,Alim:2011ae}. In general gluing and surgery operations on the curve $\cC$ can be used to classify complete supersymmetric theories by iterating certain operations, such as coupling or gauging. These operations correspond to a series of quiver gluing rules starting from elementary building blocks, which were used in the classification of mutation finite quivers \cite{FT1}. We have discussed how a certain set of gluing rules naturally extends to curves $\cC$ with line defects and to framed BPS quivers. These rules turn out to be combinatorially more involved, due to the large number of cases which are possible. However, limiting ourselves to elementary line defects, we have obtained a simple set of rules. These gluing rules, combined with the aforementioned approaches to compute framed BPS spectra, give us a tool to compute new framed spectra in new quantum field theories, starting from a line defect in a given quantum field theory. We have exemplified this in a simple case, and hope to return to this method in the future for more complicated quantum field theories.

In this paper we have made no attempt in classifying defects or in exploring systematically the results aforementioned; we have taken a more pragmatic approach and applied our formalism to a series of examples. Some of these computations reproduce known results in the literature; others are new. We hope to return to a more systematic study in a following publication. In particular it would be very interesting to give a set of necessary and sufficient conditions on a framing for it to correspond to a physical line defect.

Recently other works have appeared using algebraic methods to study line defects. In \cite{Chuang:2013wt}, framed BPS quivers were derived using geometric engineering for magnetic line defects. A careful analysis of framed stability conditions allows the computation of generalized Donaldson-Thomas invariants associated with the framed quivers and their identification with the framed BPS states. It is possible that similar representation theory techniques can be adapted to the formalism used in this paper and it would be interesting to make the connection more precise. In  \cite{Xie:2013lca} cluster algebras were used as well to study line defects. Our approach is very similar but, in a sense, complementary. The characterization of line defects in  \cite{Xie:2013lca} uses tropical coordinates and $X$ cluster variables, while we use shear coordinates and coefficient $\sfy$ cluster variables. Needless to say, it would be quite interesting to connect both approaches more in detail; we leave this problems for the future. 

In our approach the generating functions of framed degeneracies corresponding to simple defects are written in terms of certain cluster variables. Which cluster variables have the right to appear in the generating function of a  physical line defect is determined by a certain sequence of mutations. Undoubtedly, this is connected with the periodicity property of cluster algebras and with the Laurent phenomenon; yet this connection should be made more precise.  Furthermore there is clearly a relation between formal line operators and quantum cluster algebras, which we have observed only implicitly in Section \ref{qmut} and which deserves to be investigated further. Another interesting aspect of defects that we haven't considered is the study of their algebras. Line defects can be multiplied and this operation characterized by a set of fusion coefficients \cite{Kapustin:2005py,Kapustin:2007wm,Gaiotto:2010be}. It would be very interesting to recast this statement as certain operations on the framed quivers. 

In this paper we have only considered complete theories. Yet, precisely as for BPS quivers, the framing procedure is more general. Unfortunately, if the theory is not engineered from a curve $\cC$, we do not have any algebraic  intuition on how to attach a specific framing to a given defect. It is possible that our algebraic gluing rules could be used to construct line defects in non complete theories, or at least as a starting point. In general it would be interesting to apply our formalism to other quiver models, such as those studied in  \cite{Cecotti:2011gu,DelZotto:2011an,Cecotti:2012jx,Cecotti:2013lda,Cecotti:2013sza,Cecotti:2012gh,Cecotti:2012sf}, for which very few results on defects are available.
 
There are two other important aspects of line defects that we haven't considered: their connection with the geometry of the Hitchin systems, and their categorification. We believe for example that the gluing rules we have discussed could play a role in constructing the appropriate morphisms in higher categories \cite{Kapustin:2010ta}. It would also be quite interesting to extend the categorical methods of \cite{Cecotti:2012va} to line defects.

Finally this paper, together with its companion \cite{Cirafici:2013nja}, is part of a more broad project which aims to understand the modifications to Donaldson-Thomas theory induced by defects. Also there, in certain simple cases, we found a connection between defects and quivers (namely between representations of a certain quiver and moduli of parabolic sheaves corresponding to certain divisor defects). It seems likely that this connection is quite general. In particular it would be interesting to study the necessary modifications on the BPS quivers to incorporate surface defects, using quiver methods to recover the results of \cite{Gaiotto:2011tf}. We hope to return to all of these issues soon.

\textbf{Note added:} Another open issue to clarify is the relation between our algorithms and representation theory. In this paper the framed BPS degeneracies are \textit{not} derived from the representation theory of our framed quivers. Indeed a direct representation theory analysis of our quivers would give the wrong result in many cases, for example already in the Argyres-Douglas cases studied in Sections \ref{framedBPS} and \ref{ADtheories}. The reason for this is that our framed quivers are constructed out of the shear coordinates on the UV curve $\cC$; the framing node does not correspond (directly) to any charge associated with the line defects (for example by thinking of the line defect as an infinitely heavy dyon) but is constructed in terms of local data on $\cC$. The advantage of using the shear coordinates is that they behave nicely under mutations. Once the generating functional associated with a framed quiver is known (for example computed via the connection with Leavitt path algebras) all the generating functionals in its mutation class can be computed explicitly using the mutation operators of Section \ref{framedBPS}. Similarly cutting and surgery rules are described rather easily in terms of shear coordinates. Recently a different proposal was made in \cite{Cordova:2013bza} to characterize line defects using framed quivers. In their proposal the framing is obtained in terms of IR data using the defect renormalization group flow map. As a consequence it has a more clear physical meaning, and the corresponding techniques to derive the framed BPS spectra are more directly related to representation theory. 

A simple connection between the two proposals could be obtained as follows: once the generating functionals have been computed using our algorithms, one can easily obtain an IR label for the line defect using the renormalization group flow map of \cite{Cordova:2013bza}. A preliminary analysis of the Argyres-Douglas quivers of Sections \ref{framedBPS} and \ref{ADtheories} reveals that this operation precisely reproduces the quivers of \cite{Cordova:2013bza}. It remains to be seen if a general combinatorial rule at the quiver level exists or not. In particular it would be very interesting to investigate how the cutting and surgery rules of Section \ref{cutjoin} extend to the quivers of \cite{Cordova:2013bza}.

\bigskip

\acknowledgments

I thank M. Del Zotto, N. Orantin and A.-L. Thiel for discussions. A very preliminary version of these results was presented during a series of informal seminars at CMAGDS, Lisbon in May 2012; I wish to thank the participants for the many remarks which helped this paper taking shape. The author was supported in part by the Funda\c{c}\~{a}o para a Ci\^{e}ncia e Tecnologia (FCT/Portugal) via the Ci\^{e}ncia2008 program and via the grants PTDC/MAT/119689/2010 and EXCL/MAT-GEO/0222/2012, and by the Center for Mathematical Analysis, Geometry and Dynamical Systems, a unit of the LARSyS laboratory.

\appendix

\section{Self-folded triangles} \label{selffolded}

In all this paper we have assumed for simplicity that our triangulations do not include self-folded triangles. In this Appendix we will briefly explain how to extend our results to this case, referring the reader to the original literature for more details \cite{FT1,FT2}. To deal with self-folded triangles it is useful to introduce the concept of \textit{tagged triangulation}. A tagged arc is obtained by considering an arc of the triangulation which does not cut out a once punctured monogon, and tagging each of its ends. There are two ways to tag an end, plain or notched. The only restrictions are that the endpoints ending on the boundary must be tagged plain, and both ends of a loop must be tagged in the same way. Upon imposing appropriate compatibility conditions on the arcs, one can define a tagged triangulation.

In particular, an ordinary triangulation can be seen as a tagged triangulation by mapping the ordinary arcs into tagged arcs. If an ordinary arc does not cut out a once-punctured monogon, then it can be represented by a tagged arc, both ends of which are tagged plain. On the other hand, if the ordinary arc cuts out a once-punctured monogon it can be replaced by a tagged arc as follows. The ordinary arc is a loop based at some marked point. We replace it with an arc connecting the marked point with the puncture inside the monogon, tagged plain at the marked point and notched at the puncture. The notion of flip extends to tagged triangulations, by simply replacing a tagged arc by a different tagged arc. In particular one can define a signed adjacency matrix for any tagged triangulation, which transforms with a mutation upon flipping a tagged arc. Indeed adjacency matrices for tagged triangulation are defined precisely by consistency with mutations starting from an ordinary triangulation.

What makes tagged triangulations useful in our formalism, is that the shear coordinates of a lamination (and therefore the framing of a BPS quiver) can be extended also to the tagged case. Here we quote Definition 13.1 of \cite{FT2}. Shear coordinates are uniquely defined by the rules
\begin{itemize}
\item Consider two tagged triangulations $T_1$ and $T_2$ which coincide, except that at a certain puncture $p$ the tags of the arcs in $T_1$ are all different from the tags of the arcs of $T_2$ at the same puncture $p$. Suppose also that we have two lamination $\cL_1$ and $\cL_2$, which coincide except for the fact that each curve in $\cL_1$ which spirals into $p$ is replaced in $\cL_2$ by a curve which spirals into $p$ in the opposite direction. Then $b_{\gamma_1} (T_1 , \cL_1) = b_{\gamma_2} (T_2 , \cL_2)$ for each tagged arc $\gamma_1 \in T_1$ and its counterpart $\gamma_2 \in T_2$.
\item By applying appropriate tag-changing transformations we can convert any tagged triangulation into a tagged triangulation not containing any notches, except maybe inside once-punctured digons. This triangulation corresponds to an ordinary triangulation. Then if an arc is not contained inside a self-folded triangle, we define the shear coordinate as usual, as the shear coordinate of the corresponding arc in the ordinary triangulation. If it is inside a self-folded triangle enveloping a puncture $p$, we can apply a tag-changing transformation at $p$ as in the previous point, and then determine the shear coordinate as the one of the corresponding ordinary arc.
\end{itemize}
In the paper we have used a more pragmatic approach, by mutating appropriately any ordinary triangulation until it does not contain any self-folded triangle. Operationally, this is more or less equivalent to the previous definition, by reducing the problem to the definition of the shear coordinate with respect to an ordinary arc, and then requiring consistent transformations under mutations.

\end{document}